\documentclass[a4paper]{article}
\usepackage{graphicx}
\usepackage{natbib}
\usepackage[colorlinks=true,linkcolor=blue,citecolor=green, unicode=true, pdfcreator={David F. Crawford}]{hyperref}
\usepackage[all]{hypcap}
\begin{document}

\newcommand{\tablenotemark}[1]{${}^{#1}$}
\newcommand{\tablecomments}[1]{}
\newcommand{\tablenotetext}[2]{${}^{#1}${#2}\\}
\newlength{\figurewidth}

\setlength{\figurewidth}{12.4 cm}

\def\slantfrac#1#2{\hbox{$\,^#1\!/_#2$}}
\newcommand{\arcdeg}{\ensuremath{^{\circ}}}
\newcommand{\arcmin}{\hbox{$^\prime$}}
\newcommand{\arcsec}{\hbox{$^{\prime\prime}$}}
\newcommand{\hour}{\hbox{$^{\rm h}$}}
\newcommand{\minute}{\hbox{$^{\rm m}$}}
\newcommand{\secnd}{\hbox{$^{\rm s}$}}
\newcommand{\farcs}{\mbox{\ensuremath{.\!\!^{\prime\prime}}}}

\newcommand{\sun}{\ensuremath{\odot}}
\newcommand{\aj}{Astronomical J.}
\newcommand{\ajp}{Aust. J. Phys.}
\newcommand{\araa}{Annual Review of Astronomy and Astrophysics}
\newcommand{\apj}{Astrophysical Journal}
\newcommand{\apjl}{Astrophysical Journal Letters}
\newcommand{\apjs}{Astrophysical Journal Supplement}
\newcommand{\aap}{Astronomy and Astrophysics}
\newcommand{\aaps}{Astronomy and Astrophysics Supplement}
\newcommand{\pasa}{Publications of the Astronomical Society of Australia}
\newcommand{\mnras}{MNRAS} 
\newcommand{\nat}{Nature} 
\newcommand{\jcos}{Journal of Cosmology} 
\newcommand{\CC}{{\rm \scriptscriptstyle CC}}
\newcommand{\BB}{{\rm \scriptscriptstyle BB}}

\setlength{\fboxrule}{0.4mm}

\begin{center}
{\large Observational evidence favors a static universe}\\
\vspace{1 cm}

\begin{quote}
\parbox{8cm}{
David F. Crawford\\
Sydney Institute for Astronomy,\\
School of Physics, University of Sydney.\\
Correspondence: 44 Market St, Naremburn, 2065,\\
NSW, Australia\\
email: davdcraw@bigpond.net.au}
\end{quote}
\end{center}

\vspace{1.0 cm}

\begin{abstract}
The common attribute of all Big Bang  cosmologies  is that they are based on the assumption that the universe is expanding. However examination of  the evidence for this expansion clearly favors a static universe.  The major topics considered are: Tolman surface brightness, angular size, type 1a supernovae, gamma ray bursts, galaxy distributions, quasar distributions, X-ray background radiation, cosmic microwave background radiation,  radio source counts, quasar variability and the Butcher--Oemler effect. An analysis of the best raw data for these topics shows that they are consistent with expansion only if there is evolution that cancels the effects of expansion. An alternate cosmology, curvature cosmology, is a tired-light cosmology that predicts a well defined static and stable universe and is fully described. It not only predicts accurate values for the Hubble constant and the temperature of cosmic microwave background radiation but  shows good agreement with most of the topics considered.  Curvature cosmology also predicts the deficiency in solar neutrino production rate and can explain the anomalous acceleration of {\it Pioneer} 10.\\

\noindent
Note:\\
The editor of the Journal of Cosmology required this paper to be divided into three parts. Except for the abstracts, introductions and conclusions the text of the three papers is identical to this version. For convenience a table of contents is included in this version. The correspondence between the sections in this version and those in the Journal of Cosmology are:

\begin{tabular}{l l c l}\\
Part I & sections \ref{s2}, \ref{s3}, \ref{s4} & 72pp & 2011, \href{http://journalofcosmology.com/crawford1.pdf}{JCos, 13, 3875-3946}\\
Part II & section \ref{s5} & 53pp & 2011, \href{http://journalofcosmology.com/crawford2.pdf}{JCos, 13, 3947-3999}\\
Part III & sections \ref{s6} \& \ref{s7} & 58pp & 2001, \href{http://journalofcosmology.com/crawford3.pdf}{JCos, 13, 4000-4057} \\
\end{tabular}
\end{abstract}
\vspace{5 mm}
cosmology: observations, large-scale structure of universe, theory
\pagebreak
\tableofcontents

\section{Introduction}
\label{s1}
The common attribute of all Big Bang  cosmologies (BB) is that they are based on the assumption that the universe is expanding \citep{Peebles93}. An early alternative was the steady-state theory of Hoyle, Bondi and Gold (described with later extensions by \citet{Hoyle00}) that required continuous creation of matter. However steady-state theories have serious difficulties in explaining the cosmic microwave background radiation. This left BB as the dominant cosmology but still subject to criticism. Recently \citet{Lal10} and \citet{Joseph10} have continued major earlier criticisms of Big Bang cosmologies \citep{Ellis84,Lerner91,Disney00,Flandern02}. Whereas most of theses criticisms have been of a theoretical nature this paper concentrates on whether observational data supports BB or a static cosmological model, curvature cosmology (CC), described below.

Expansion produces two distinct effects. The first effect of expansion is the increasing redshift with distance as described by Hubble's law. This could be due to either a genuine expansion or resulting from a tired-light phenomenon. The second effect of expansion is time dilation resulting from the slowing down of the arrival times of the photons as the source gets further away. Section~\ref{s4} concentrates on the evidence for expansion as shown by time dilation and shows that the evidence is only consistent with BB if there is evolution in either luminosity or in angular size that closely cancels the effects of time dilation.  To illustrate that a static cosmology can explain the data, a particular model, curvature cosmology (CC), is used.  Curvature cosmology is based on the hypothesis of curvature redshift and the hypothesis of curvature pressure. Curvature redshift arises from the principle that any localized wave travelling in curved space time will follow geodesics and be subject to geodesics focussing.  Since this will alter the transverse properties of the wave some of its properties such as angular momentum will be altered which is contrary to quantum mechanics. For a photon the result is an interaction that results in three new photons. One with almost identical energy and momentum as the original and two extremely low energy secondary photons. In effect the photon loses energy via an interaction with curved space-time. The concept of curvature pressure arises from the idea that the density of  particles produce curved space-time. Then as a function of their velocity there will be a reaction pressure that acts to decrease the local space-time curvature.

It is the evaluation of evidence for this expansion that forms the major basis of this paper. Consequently minor differences between different expansion cosmologies are not particularly important here; it is the broad brush approach that is relevant. Nevertheless to provide appropriate numerical quantities the evaluation is based on a particular BB cosmology, the ($\Lambda$CDM) model, which is defined in Section~\ref{s2} by the equations for angular size, volume and distance modulus. Since the major difference between BB and CC arises from the expansion in BB, the major results of the comparison are applicable to other static cosmologies and do not depend on the validity of CC. A problem in evaluating a  well established cosmology like  Big Bang cosmology  is that all of the observations have been analyzed within the BB paradigm. Thus there can be subtle effects that may lead to a  possible bias. In order to avoid this  bias and wherever possible comparisons are made using original observations.

Section~\ref{s3} discusses the theoretical justification for the basic and additional hypotheses that have been incorporated into the cosmologies. For BB these includes inflation, dark matter and dark energy. For CC the hypotheses are curvature redshift and curvature pressure.

The  evaluation of BB  and its comparison with CC using observational evidence is divided into four major parts. The first in Section~\ref{s4} concentrates on those observations that have measurements that are strongly dependent on expansion. It is found that in all cases where there is direct evidence for evolution this evolution is close to what is required to cancel the expansion term in the BB equations. This is an extraordinary coincidence. The simplest conclusion is that the universe is not expanding. Section~\ref{s5} looks at observations that have different explanations in CC from what they have in BB. It is found that not only does CC gave better agreement with most of the observations but it does so without requiring extra ad hoc parameters or hypotheses.  The next section is a description of CC and possible experimental tests of its validity. Finally Section~\ref{s7} examines additional important topics that are relevant to CC.

The test for Tolman surface brightness in Section~\ref{s4} is through the expected variation of apparent surface brightness with redshift. The results strongly favor a static universe but could be consistent with BB provided there is luminosity evolution.

The relationship between angular size and linear size in BB includes a aberration factor of $(1+z)$ that does not occur in static cosmologies. However the available data does not support the inclusion of this factor and is more consistent with a static universe.

Next it is argued that the apparent time dilation of the supernova light curves is  not due to expansion. The analysis is complex and is based on the premise that the most constant characteristic of the supernova explosion is its total energy and not its peak magnitude. If this is correct, then selection effects can account for the apparent time dilation. The CC analysis is in complete agreement with the known correlation between  peak luminosity and light curve duration. Furthermore the analysis overcomes a serious problem with the current redshift distribution of supernovae.  Finally using CC the distribution of the total energy for each supernova as a function of ($1+z$) has an exponent of $0.047\pm0.089$. This shows that there is no redshift dependence that occurs in the BB analysis. Thus there is no need for dark energy.

The raw data of various time measures taken from the light curves for gamma ray bursts (GRB) show no evidence of the time dilation that is expected in BB. Since it can be argued that evolutionary and other effects that may have cancelled the expected time dilation in BB are unlikely a reasonable conclusion is that there is no time dilation in GRB.

It is shown for galaxies with types E--S$_{\rm a}$ that have a well a defined peak in their luminosity distribution
the magnitude of  this peak is independent of redshift when the analysis was done using  a static cosmology.

Analysis of  quasar distributions in BB shows that luminosity evolution is required to explain the observations. A novel method is used to analyst the quasar distribution. Because the quasar distribution is close to an exponential distribution in absolute magnitude (power law in luminosity) then for a small redshift range it is also an exponential distribution in apparent magnitude. Then for a small redshift range  it is possible to use statistical averages to get an estimate of the distance modulus directly from the raw data. The only input required from the cosmological model is the variation of volume with redshift. The results are shown in Fig.~\ref{quaf1}. They show much better agreement with CC than with BB.

An analysis of the distribution of radio sources is included in this section not because of explicit evidence of time dilation but because it has been generally accepted that the distribution can only be explained by having strong evolution. It is shown that a distribution of radio sources in a static universe can have a good fit to the observations.

\citet{Hawkins10,Hawkins01, Hawkins03} has been monitoring quasar variability using a Fourier method since about 1975 and finds no variation in their time scales with redshift. Although it is generally accepted that the variations are intrinsic to the quasar there is a possibility that they may be due to micro-lensing which could place their origin to modulation effects in our own galaxy.

The Butcher--Oemler effect of the increasing proportion of blue galaxies in clusters at higher redshift is shown to be non-existent or at least greatly exaggerated.

Apart from its lack of expansion, curvature cosmology  makes further specific predictions that can be compared with BB.  These are considered in Section~\ref{s5}. Whereas the conclusions about expansion and evolution in Section~\ref{s4} would be applicable for any reasonable static cosmology this section considers topics that are specific to CC and BB.

The topic of X-ray background radiation is very important for CC. Not only can CC explain the radiation from 10--300 keV but the results enable estimates for the temperature and density of the cosmic gas (the gas external to clusters of galaxies).
For the measured density of $N=1.55\pm 0.01\,\mbox{m}^{-3}$ the calculated value of the Hubble constant is $H=64.4\pm 0.2\mbox{\,km\,s}^{-1}\mbox{\,Mpc}^{-1}$ (c.f. equation~\ref{ce63}) whereas the value estimated from the type 1a supernova data (Section~\ref{s4.3}) is $63.8\pm0.5\,$kms$^{-1}$ Mpc$^{-1}$ and the result from the Coma cluster (Section~\ref{s5.14}) is $65.7\,$kms$^{-1}$ Mpc$^{-1}$. In CC the theoretical temperature for the cosmic gas is $2.56\times 10^9\,$K  and the temperature estimated from fitting the X-ray data is $(2.62\pm 0.04)\times10^9\,$K.

In CC the CMBR is produced by very high energy electrons via curvature-redshift radiation in the cosmic plasma. The predicted temperature of the CMBR is 3.18\,K to be compared with an observed value of 2.725\,K \citep{Mather90}.  The prediction does depend on the nuclei mix in the cosmic gas and could vary from this value by several tenths of a degree. It is argued that in CC the apparent larger CMBR temperature at large redshifts could be explained by the effects of curvature redshift on the width of spectral lines. Evidence for correlations between CMBR intensity and galaxy density is consistent with CC.

Regarding dark matter not only does CC have a quite different explanation for the velocity dispersion in clusters of galaxies but it can make a good estimate, without any free parameters, of its value for the Coma cluster. In BB it is assumed that the redshift dispersion is a genuine velocity dispersion and the mass of a cluster of galaxies is determined by using the virial theorem. In CC the redshift dispersion is due to curvature redshift produced by the intra-cluster gas.

The Sunyaev--Zel'dovich effect, gravitational lens, the Lyman-$\alpha$ forest and the Gunn--Peterson trough can be explained by or are fully consistent with CC.
BB offers a good explanation for the primordial abundances of the light nuclei, albeit with some uncertainty of the density of the early universe at their time of formation. In CC the distribution of light elements is determined by nuclear reactions in the very high temperature cosmic gas. This explanation needs a quantitative analysis.

Galactic rotation curves are a problem for both cosmologies. BB requires an extensive halo of dark matter around the galaxy while CC requires a  reasonable halo of normal matter to produce the apparent rotation via curvature redshift. Its problem is getting the required asymmetry in the halo distribution.

Anomalous redshifts are the controversial association of high redshift quasars with much lower redshift galaxies. Although they are inexplicable in BB, CC could offer a partial explanation for some observations.

Finally voids and other large scale structures in the redshift distribution of quasars and galaxies is easily explained in CC by the extra redshift due to curvature redshift in higher density gas clouds. In BB it is a complicated result of the evolution of these objects.

Section~\ref{s6} provides a complete description of CC and its two major hypotheses: curvature redshift and curvature pressure. Although it is not a new idea it is argued that gravitation is an acceleration and not a force. This idea is used to justify the averaging of accelerations rather than forces in deriving curvature pressure.

The next Section~\ref{s7} includes the topics of entropy, Olber's paradox, black holes, astrophysical jets and large-number coincidences that are particularly relevant for CC but are not important for choosing between the two cosmologies.

Although the explanation for the deficiency in observed neutrinos from the sun can be explained by neutrino oscillations it is include here because curvature pressure makes excellent estimates of the expected numbers without any free parameters. The heating of the solar corona is a very old problem and still not fully explained. It is treated here simply to show that curvature redshift offers no help.

Finally it is shown  the {\it Pioneer} 10 anomalous acceleration can be explained by the effects of curvature redshift that is produced by interplanetary dust provided the density of the dust is a little higher than current estimates.

Except for cross-references the sub-sections on observational topics are self contained and can be read independently. Many of the topics use statistical estimation methods and in particular linear regression. A brief summary of the general linear regression and the treatment of uncertainties is provided in the appendix.

\section{Cosmographic Parameters}
\label{s2}
Just like the Doppler shift the cosmological redshift $z$ is independent of the wavelength of the spectral line.   In terms of wavelength, the redshift $z$  is $z= (\lambda_0/\lambda - 1)$ where $\lambda_0$ is the observed wavelength and $\lambda$ is the emitted wavelength. In terms of frequency, $\nu$, and photon energy, $E$, the redshift is $z=\nu/\nu_0 -1= E/E_0-1$.
The basic cosmological equations needed to analyst observations provide the conversion from apparent magnitude to absolute magnitude, the relationship between actual lateral measurement and angle and the volume as a function of redshift.

The conversion from apparent magnitude, $m$, to absolute magnitude, $M$, is given by the equation
\begin{equation}
\label{cpe1}
M = m - \mu(h) -K_z(\lambda_0),
\end{equation}
where $\mu(h)$ is the distance modulus that strongly depends on the assumed cosmology and $K_z(\lambda_0)$ is the K-correction that allows for the difference in the spectrum between the emitted wavelength and the observed wavelength \citep{Rowan-Robertson85, Hogg02} and is independent of the assumed cosmology.
For a small bandwidths and luminosity $L(\lambda)$ it is
\begin{equation}
\label{cpe2}
K_z(\lambda_0) = -2.5\log\left(\frac{L(\lambda)}{(1+z)L(\lambda_0)}\right).
\end{equation}
Note that the bandwidth ratio is included in the definition of the K-correction.
The Hubble constant $H_0$ is the constant of proportionality between the apparent recession speed $v$ and distance $d$. That is $v=Hd$. It is usually written $H_0=100h \mbox{\,km\,s}^{-1}\mbox{\,Mpc}^{-1}$ where $h$ is a dimensionless number. Unless otherwise specified it is assumed to have the value $h=0.7$.

\subsection{Big Bang cosmology (BB)}
\label{s2.1}
The fundamental premise of Big Bang cosmology is that the universe is expanding with a scale factor proportional to ($1+z$). A more detailed account can be found in \citet{Peebles93,Peacock99}. The analysis is simplified  by using comoving coordinates that describe the non-Euclidean geometry without expansion.  Note that in BB the Hubble constant is a function of redshift hence the use of a zero subscript to denote the current value. A problem with BB is that it is only the distances between large objects that are subject to the expansion. It is generally accepted that any objects smaller than clusters of galaxies which are gravitationally bound do not follow the Hubble flow.

The current version of Big-Bang cosmology is the cold dark matter ($\Lambda$CDM) or concordance cosmology that has a complex expression for its parameters that depends on the cosmological energy density $\Omega_\Lambda$. Regardless of the name, the BB model used here is defined by the following equations. Following \citet{Goobar95} (with corrections from \citet{Perlmutter97,Hogg99}),  the function $f(x)$ is defined by
\begin{equation}
\label{cpe3}
f(z)= \int_0^z \frac{dz}{(\sqrt{((1+z)^3 -1)\Omega_M  +1}}.
\end{equation}
where $\Omega_M$ is the cosmological energy-density parameter.
For observations on the transverse size of objects, such as galactic diameters that do not follow the Hubble flow, the linear size  $S_\BB$ is
\begin{equation}
\label{cpe4}
S_\BB=\frac{2.998\times10^9 \theta f(z)}{h (1+z)}\,\mbox{pc/radian}.
\end{equation}
where $\theta$ is its angular size in radians. For $\theta$ in arcseconds the constant is $1.453\times10^4$.
The total comoving volume out to a redshift $z$ is
\begin{equation}
\label{cpe5}
V_\BB=\frac{4 \pi}{3}\left(\frac{2.998 f(z)}{h}\right)^3 .
\end{equation}
Note that the actual volume, which would be relevant for the cosmic gas density, is the comoving volume divided by $(1+z)^3$, which shows that the density of the cosmic gas (i.e. inter-galactic gas outside clusters of galaxies) increases rapidly with increasing redshift.
The distance modulus is
\begin{equation}
\label{cpe6}
\mu_\BB= 5\log \left(\frac{(1+z)f(z)}{h}\right) + 42.384.
\end{equation}

\subsection{Curvature cosmology (CC)}
\label{s2.2}
Curvature cosmology (Section~\ref{s6} \citealt{Crawford06,Crawford09a})  is a complete cosmology that shows excellent agreement with all major cosmological observations without needing dark matter or dark energy and is fully described in section~\ref{s5}. It is compatible with both (slightly modified) general relativity and quantum mechanics and obeys the perfect cosmological principle that the universe is statistically the same at all places and times. This new theory is based on two major hypotheses. The first hypothesis is that the Hubble redshift  is due to an interaction of photons with curved spacetime where they lose energy to other very low energy photons. Thus it is a tired-light model. It assumes a simple universal model of a uniform high temperature plasma (cosmic gas) at a constant density.
The important result of curvature redshift is that the rate of energy loss by a photon (to extremely low energy secondary photons) as a function of distance, $ds$, is given by
\begin{equation}
\label{cpe10}
\frac{1}{E}\frac{dE}{ds} =-\left(\frac{8\pi GNM_{\rm H}}{c^2} \right)^{1/2},
\end{equation}
where $M_{\rm H}$ is the mass of a hydrogen atom and the density in hydrogen atoms per cubic meter is $N=\rho/M_{\rm H}$. Eq.~\ref{cpe10} shows that the energy loss  is proportional to the integral of the square root of the density along the photon's path.
This equation can be integrated to get
\begin{eqnarray*}
\label{cpe11}
\ln(E/E_0) &=&\ln(1+z)\\
&=&\left(\frac{8\pi GM_{\rm H}}{c^2} \right)^{1/2}\int_0^x\sqrt{N(x)}dx.
\end{eqnarray*}
The Hubble constant is predicted to be
\begin{eqnarray*}
\label{cpe12}
H & = &-\frac{c}{E}\frac{{DE}}{{des}} =\left( 8\pi GM_{\rm H} N \right)^{1/2} \nonumber \\
  & = & 51.69 N^{1/2}\,\mbox{kms}^{-1}\,\mbox{Mpc}^{-1} \\
  & = & 64.4\pm0.2\,\mbox{kms}^{-1}\,\mbox{Mpc}^{-1}\;(N=1.55\pm0.01\,\mbox{m}^{-3}),\nonumber
\end{eqnarray*}
where the density $N$ comes from the background X-ray analysis.

The second hypothesis is that there is a pressure, curvature pressure, that acts to stabilize expansion and provides a static stable universe.
This hypothesis leads to modified Friedmann equations which have a simple solution for a unform cosmic gas.
In this model the distance travelled by a photon from a redshift, $z$, to the present is $r=R\chi$, where
\begin{equation}
\label{cpe12a}
 \chi =\ln(1+z)/\sqrt{3}
 \end{equation}
and $R$ is the radius of the universe. Since the velocity of light is a universal constant the time taken is $R\chi/c$. There is a close analogy to motion on the surface of the earth with radius $R$. Light travels along great circles and $\chi$  is the angle subtended along the great circle between two points. The geometry of this curvature cosmology is that of a three-dimensional surface of a four-dimensional hypersphere. It is identical to that for Einstein's static universe. For a static universe, there is no ambiguity in the definition of distances and times. One can use a universal cosmic time and define distances in light travel times or any other convenient measure.

The linear size $S_\CC$, at a redshift $z$ with an angular size of $\theta$ is
\begin{equation}
\label{cpe13}
S_\CC=R\sin(\chi)\theta=\frac{5.193 \sin(\chi) \theta}{h}\,\mbox{kpc}.
\end{equation}
For this geometry the area of a three dimensional sphere with radius, $r=R\chi$, is given by
\begin{equation}
\label{cpe14}
A(r) = 4\pi R^2 \sin ^2 (\chi ).
\end{equation}
The surface is finite and  $\chi$ can vary from 0 to $\pi$.
The total volume $V$, is given by
\begin{eqnarray*}
\label{cpe15}
V(r)& = &2\pi R^3 \left[ {\chi  - \frac{1}{2}\sin (2\chi )}\right] \approx \frac{4\pi}{3}(R\chi)^3\nonumber \\
 & = & \frac{32.648}{h^3}\left[ \chi  - \frac{1}{2}\sin (2\chi )\right]\,\mbox{kpc}^3.
\end{eqnarray*}

The distance modulus is obtained by combining the energy loss rate with the area equation to get
\begin{equation}
\label{cpe16}
\mu_\CC= 5\log \left[{\frac{\sqrt{3} \sin (\chi)}{h}} \right] + 2.5\log(1+z)  \\
+ 42.384.
\end{equation}

\subsection{Numerical comparison of BB and CC}
\label{s2.3}
It turns out that the two cosmologies can be simply related as a function of ($1+z$).  The approximation equations are
\begin{eqnarray*}
S_\CC/S_\BB      & \approx & a_S (1+z)^{b_S}  \nonumber \\
V_\CC/V_\BB      & \approx & a_V (1+z)^{b_V}  \nonumber \\
\mu_\CC - \mu_\BB& \approx & a_M + b_M(1 + z)
\end{eqnarray*}
where the parameters were determined by averaging them from $z=0$  to  the listed value. To avoid any bias the redshifts used were for 15,339 quasars from the Sloan Digital Sky Survey  \citep[Quasar catalogue]{Schneider07,Schneider05}. Table~\ref{cpt1} shows these the relevant parameters for angular size and total volume and Table~\ref{cpt2} shows them for the distance modulus. The uncertainties in the parameters were all less that 0.003 in the exponents and less than 0.002 in the coefficients. The first column (labelled Point) is the actual value of the parameter at that redshift.
Examination of Table~\ref{cpt1} shows that the major difference in angular size comes from the ($1+z$) aberration factor in the denominator of the BB equation. Except for a constant scale factor the CC volume is almost the same as the BB comoving volume. Again note that the actual BB volume has an additional factor of $(1+z)^{-3}$.

\begin{table}
\begin{center}
\caption{Comparison angular size and volume\label{cpt1}}
\begin{tabular}{c|ccc|ccc}\\
\hline
& \multicolumn{3}{|c}{$S_\CC/S_\BB$} & \multicolumn{3}{|c}{$V_\CC/V_\BB$}\\
$z$ & Point & $a_S$ & $b_S$ & Point & $a_V$ & $b_V$   \\
\hline

 1& 1.749 & 0.974 & 0.839 & 0.703 & 0.699 & -0.138  \\
 2& 2.546 & 0.950 & 0.890 & 0.691 & 0.715 & -0.093  \\
 3& 3.350 & 0.942 & 0.903 & 0.717 & 0.718 & -0.072  \\
 4& 4.133 & 0.933 & 0.915 & 0.747 & 0.720 & -0.138  \\
 5& 4.920 & 0.932 & 0.918 & 0.775 & 0.719 & -0.138  \\
 \hline
\end{tabular}
\end{center}
\end{table}

Table~\ref{cpt2} shows that for the same apparent magnitude CC predicts a fainter absolute magnitude. For example for $z=5$ we get $M_\CC - M_\BB=\mu_\BB - \mu_\CC=2.376$ which shows that for the same apparent magnitude, in CC the absolute luminosities are fainter than they are in BB by a factor of 0.112.

\begin{table}
\begin{center}
\caption{Comparison of distance modulus\label{cpt2}}
\begin{tabular}{c|ccc}\\
\hline
&  \multicolumn{3}{|c}{$\mu_\CC - \mu_\BB$}\\
$z$ &  Point & $a_M$ & $b_M$  \\
\hline

 1&  -1.043 & -0.057 & -1.322 \\
 2&  -1.549 & -0.112 & -1.221 \\
 3&  -1.890 & -0.131 & -1.194 \\
 4&  -2.155 & -0.151 & -1.169 \\
 5&  -2.376 & -0.154 & -1.165 \\
 \hline
\end{tabular}
\end{center}
\end{table}

Examination of these tables and in particular $b_S$ and $b_M$ shows that the major difference between BB and CC is the inclusion of the expansion factor $(1+z)$ in the equations for BB.

\section{Theoretical}
\label{s3}
This section briefly examines the ad hoc hypotheses for BB and the basic hypotheses for CC.
Although the basis of BB in general relativity is sound it has acquired some additional hypotheses that are questionable. \citet{Peacock99} has listed some of the basic theoretical problems with BB as being the horizon problem, the flatness problem, the antimatter problem, the structure problem and the expansion problem. Although it does not overcome all these problems the solution that was suggested by \citet{Guth81} and developed by \citet{Linde90,Liddle00} was the concept of inflation.  The idea is that very early in the evolution of the universe there is a brief but an extremely rapid acceleration in the expansion. There is no support for this concept outside cosmology and it must therefore be treated warily. Nevertheless with the adjustment of several parameters it can explain most of the listed problems.

In 1937 \citet{Zwicky37} found  in an analysis of the Coma cluster of galaxies that the ratio of total mass obtained by using the virial theorem to the total luminosity  was 500 whereas the expected ratio was 3. This huge discrepancy was the start of the concept of dark matter. It is surprising that in more than seven decades since that time there is no direct evidence for dark matter. Similarly the concept of dark energy (some prefer quintessence) has been introduced to explain discrepancies in the observations of type 1a supernovae.  The important point is that these three concepts have been introduced in an ad hoc manner to make BB fit the observations. None has any theoretical or experimental support outside the field of cosmology.

As already stated CC is based on the hypotheses of curvature redshift and that of curvature pressure. Both are described in section~\ref{s6} and are supported by strong physical arguments. Curvature redshift is testable in the laboratory  and some support for curvature pressure may come from solar neutrino observations. Nevertheless they are new hypotheses and must be subject to strong scrutiny.

\section{Expansion and Evolution}
\label{s4}
The original models for BB had all galaxies being formed shortly after the beginning of the universe. Then as stars aged the characteristics of the galaxies changed. Consequently these characteristics are expected to show an evolution that is a function of redshift. Since nearly all cosmological objects are associated with galaxies we would also expect their characteristics to show evolution. Clearly in CC there is no evolution in the average characteristics of any object. The individual objects will evolve but the average characteristics of the population remain constant. Any strong evidence of evolution of the average characteristics would constitute  serious  evidence against CC.

The simple BB evolution concept is complicated by galactic collisions and mergers \citep{Struck99,Blanton09}. In many cases these can produce new large star-forming regions. Clearly the influx of a large number of new stars will alter the characteristics of the galaxy. It is likely that a significant number of galaxies have been reformed by collisions and mergers and consequently the average evolution may be very little or at least somewhat reduced over the simple model.

\subsection{Tolman surface brightness}
\label{s4.1}
This test, suggested by \citet{Tolman34}, relies on the observation that the surface brightness of objects does not depend on the geometry of the universe. Although it is obviously true for Euclidean geometry it is also true for non-Euclidean geometries. For a uniform source, the quantity of light received per unit angular area is independent of distance. However, the quantity of light is also sensitive to non-geometric effects, which make it an excellent test to distinguish between cosmologies. For expanding universe cosmologies the surface brightness is predicted to vary as $(1+z)^{-4}$, where one factor of ($1+z$) comes from the decrease in energy of each photon due to the redshift, another factor comes from the decrease in rate of their arrival and two factors come from the apparent decrease in area due to aberration. This aberration is simply the rate of change of area for a fixed solid angle with redshift. In a static, tired-light, cosmology (such as CC) only the first factor is present. Thus an appropriate test for Tolman surface brightness is the value of this exponent.

\subsubsection{Surface brightness in BB}
\label{s4.1.1}
The obvious candidates for surface brightness tests are elliptic and S0 galaxies  which have minimal projection effects compared to spiral galaxies . The major problem is that surface brightness measurements are intrinsically difficult due to the strong intensity gradients across their images. In a series of papers \citet{Sandage01,Lubin01a,Lubin01b,Lubin01c} (hereafter SL01) have investigated the Tolman surface brightness test for elliptical and S0 galaxies.  More recently \citet{Sandage10} has done a more comprehensive analysis but since he came to the same conclusion as the earlier papers and since the earlier papers are better known this analysis will concentrate on them. The observational difficulties are thoroughly discussed by \citet{Sandage01} with the conclusion that the use of Petrosian metric radii helps solve many of the problems. \citet{Petrosian76,Djorgovski81,Sandage90} showed that if the ratio of the average surface brightness within a radius is equal to $\eta$ times the surface brightness at that radius then that defines the Petrosian metric radius, $\eta$. The procedure is to examine an image and to vary the angular radius  until the specified Petrosian radius is achieved.

Thus, the aim is to measure the mean surface brightness for each galaxy at the same value of $\eta$.  The choice of Petrosian radii greatly diminishes the differences in surface brightness due to the luminosity distribution across the galaxies. However, there still is a dependence of the surface brightness on the size of the galaxy which is the  Kormendy relationship \citep{Kormendy77}.
The purpose of the preliminary analysis done by SL01 is not only to determine the low redshift absolute luminosity but also to determine the surface brightness verses linear size  relationship that can be used to correct for effects of size variation in distant galaxies. The data on the nearby galaxies used by SL01 was taken from \citet{Postman95} and consists of extensive data on the brightest cluster galaxies (BCG) from 119 nearby Abell clusters. All magnitudes for these galaxies are in the $R_{\rm \scriptscriptstyle C}$ (Cape/Landolt) system. Since the results for different Petrosian radii are highly correlated the analysis repeated here using similar procedures will use only the Petrosian $\eta=2$ radius.
Although the actual value used for $h$ does not alter any significant results here it is set to $h=0.5$ for numerical consistency. A minor difference is that the angular radius used here is provided by Eq.~\ref{cpe4} whereas they used the older Mattig equation.

The higher $z$ data also comes from SL01. They made Hubble Space Telescope observations of galaxies in three clusters and measured their surface brightness and radii. The names and redshifts of these clusters are given in Table~\ref{sbt1} which also shows
the number of galaxies in each cluster, $N$,  the logarithm of the average metric radius in kpc, $\log(S_\BB)$, and the average absolute magnitude.  Note that the original magnitudes for Cl 1324+3011 and Cl 1604+4304 were observed in the {\it I} band.

\begin{table}
\begin{center}
\caption{Galactic properties for Petrosian radius $\eta=2.0$}
\label{sbt1}
\begin{tabular}{lrccc}
\hline
Cluster   & $N$ &   $\overline{{\log(S_\BB)}}$  & $\overline{SB}$   & $\overline{M_\BB}$  \\
Nearby    & 74  &  4.69 (0.28) & 22.56 (0.84) & -23.84 (0.66) \\
1324+3011 & 11  &  3.99 (0.21) & 22.87 (0.75) & -23.28 (0.65) \\
1604+4304 & 6   &  4.05 (0.17) & 22.34 (0.60) & -23.51 (0.68) \\
1604+4321 & 13  &  4.00 (0.15) & 22.35 (0.78) & -23.33 (0.64) \\
\hline
\end{tabular}
\end{center}
\end{table}

The bracketed numbers are the root-mean-square (rms) values for each variable. In order to get a reference surface brightness at $z=0$ all the surface brightness values, SB, of the nearby galaxies were reduced to absolute surface brightnesses by using Eq.~\ref{sbe1}. Since all the redshifts are small, this reduction is essentially identical for all cosmological models.
However the calculation of the metric radii for the distant galaxies is very dependent on the cosmological model. This procedure of using the BB in analyzing a test of BB is discussed in SL01. Their conclusion is that it reduces the significance of a positive result from being {\it strongly supportive} to being {\it consistent with the model}. Of interest is that Table~\ref{sbt1} shows that on average the distant galaxies are smaller than the nearby galaxies.

Then a linear least squares fit of the absolute surface brightness as a function of $\log(S_\BB)$, the Kormendy relationship, for the nearby galaxies results in the equation
\begin{equation}
\label{sbe1}
SB =9.29\pm0.50 +(2.83\pm0.11)\log(S_\BB)
\end{equation}
whereas SL01 found a slightly different equation
\begin{equation}
\label{sbe2}
SB =8.69\pm0.06 + (2.97\pm0.05)\log(S_\BB).
\end{equation}

Although a small part of the discrepancy is due to slightly different procedures the main reason for the discrepancy is unknown. Of the 74 galaxies used there were 19 that had extrapolated estimates for either the radius or the surface brightness or both. In addition there were only three galaxies that differed from the straight line by more than 2$\sigma$. They were A147 (2.9$\sigma$), A1016 (2.0$\sigma$)and A3565 (-2.4$\sigma$). omission of all or some of these galaxies did not improve the agreement. The importance of this preliminary analysis is that Eq.~\ref{sbe1} contains all the information  that is needed from the nearby galaxies in order to calibrate the distant cluster galaxies.

Next we use the galaxies' radius and Eq.~\ref{sbe1} to correct the apparent surface brightness of the distant galaxies for the Kormendy relation and  then do least squares fit to the difference between the corrected  surface brightness and its absolute surface brightness as a function of $2.5\log(1+z)$ to estimate the exponent, $n$, where $SB\propto(1+z)^n$. If needed the non-linear corrections given by \citet{Sandage10} were applied to the nearby surface brightness values. For the {\it I} band galaxies the absolute surface brightness included the color correction $<R-I>=0.62$ \citet{Lubin01c}. The results for the exponent, $n$, for each cluster are shown in Table~\ref{sbt2} together with the values from SL01  (column 5) where the second column is the band (color) in which the cluster was observed.

\begin{table}
\begin{center}
\caption{Fitted exponents for  distant clusters ($\eta=2.0$)}
\label{sbt2}
\begin{tabular}{lrccc}
\hline
Cluster   & Col & $\overline{z}$ &   $n_\BB$            &$n_{\rm SL01}$  \\
1324+3011 & {\it I} &  0.757         &   1.98$\pm$0.19 & 1.99$\pm$0.15\\
1604+4304 & {\it I} &  0.897         &   2.22$\pm$0.22 & 2.29$\pm$0.21\\
1604+4321 & {\it R} &  0.924         &   2.24$\pm$0.18 & 2.48$\pm$0.25\\
\hline
\end{tabular}
\end{center}
\end{table}

Because the definition of magnitude contains a negative sign the expected value for $n$ in BB is four. Nearly all of the difference between these results and those from SL01 arises from the use of a different Kormendy relationship. If the Kormendy relationship  used by SL01 Eq.~\ref{sbe2} is used instead of Eq.~\ref{sbe1}) the agreement is excellent. If it is assumed that there is no evolutionary or other differences between the three clusters and all the data are combined the resulting exponent is $n_\BB=2.16\pm0.13$.

Clearly there is a highly significant disagreement between the observed exponents and the expected exponent of four. Both SL01 and \citet{Sandage10} claim that the difference is due to the effects of luminosity evolution. Based on a range of theoretical models SL01 show that the amount of luminosity evolution expressed as the exponent, $p=4-n_\BB$,  varies between $p=$0.85--2.36 in the {\it R} band and $p=$0.76--2.07 in the {\it I} band. In conclusion to  their analysis they assert that {\it they have either (1) detected the evolutionary brightening directly from the SB observations on the assumption that the Tolman effect exists or (2) confirmed that the Tolman test for the reality of the expansion is positive, provided that the theoretical luminosity correction for evolution is real.}

SL01 also claim that their results are completely inconsistent with a tired light cosmology. Although this is explored for CC in the next sub-section it is interesting to consider a very simple model. The essential property of a tired light model is that it does not include the time dilation factor of ($1+z$)  in its angular radius equation. Thus assuming  BB but without the ($1+z$) term all values of $\log(S_\BB)$ will be increased  by $\log(1+z)$. Hence the predicted absolute surface brightness will be (numerically) increased by (2.83/2.5)$\log(1+z)$. For example, the exponent for all clusters will be changed to
\[
n_{\rm tired\_light}=2.16\pm0.16 - \frac{2.83}{2.5} = 1.03\pm0.16
\]
This is clearly close to the expected value of unity predicted by a tired-light cosmology and thus disagrees with the conclusion of SL01 that the data are incompatible with a tired light cosmology.

There are two major criticisms of this work. The first is that relying on theoretical models to cover a large gap between the expected index and the measured index makes the argument very weak. Although SL01 indirectly consider the effects of relatively common galaxy interactions and mergers in the very wide estimates they provide for the evolution,  the fact that there is such a wide spread makes the argument that Tolman surface brightness for this data is consistent with BB possible but weak. Ideally there would be an independent estimate of $p$ based on other observations. The second criticism is that the nearby galaxies are not the same as the distant cluster galaxies. The nearby galaxies are all brightest cluster galaxies (BCG) whereas the distant cluster galaxies are normal cluster galaxies. It is well known that BCG \citep{Blanton09} are in general much brighter and larger than would be expected for the largest member of a normal cluster of galaxies. Whether or not this amounts to a significant variation is unknown but it does violate the basic rule that like should be compared with like.

\subsubsection{Surface brightness in CC}
Unsurprisingly  it is found that using CC the relationship between absolute surface brightness and radius is identical to that shown in Table~\ref{sbt1}. What is different is the average radii, the absolute magnitudes and the observed exponent $n_\CC$.  These are shown in Table~\ref{sbt3} where as before the bracketed terms are the rms spread in the radii.

\begin{table}[hb]
\begin{center}
\caption{Radii and fitted exponents for distant clusters ($\eta=2.0$)}
\label{sbt3}
\begin{tabular}{lcccc}
\hline
Cluster   &  $N$  &  $\bar{\log(S_\CC)}$  & $\bar{M_\CC}$     & $n_\CC$  \\
nearby    &   74  &   4.70 (0.28)     & -23.78 (0.66) & \\
1324+3011 &   11  &   4.18 (0.21)     & -22.41 (0.66) & 1.19$\pm$0.19 \\
1604+4304 &    6  &   4.27 (0.17)     & -22.54 (0.65) & 1.45$\pm$0.21 \\
1604+4321 &   13  &   4.23 (0.15)     & -22.33 (0.68) & 1.48$\pm$0.17 \\

\hline
\end{tabular}
\end{center}
\end{table}

The result for all clusters is $n_\CC=1.38\pm0.13$ which is in agreement with unity. Note that the critical difference from BB is in the size of the radii. They are not only much closer to the nearby galaxy radii but because they are larger they do not require the non-linear corrections for the Kormendy relation. As before we note that the nearby galaxies are BCG which may have a brighter SB than the normal field galaxies. If this is true it would bias the exponent to a larger value. If we assume that CC is correct then this data shows that on average the BCG galaxies are $-0.64\pm0.08$ mag (which is a factor of 1.8 in luminosity) brighter than the general cluster galaxies.
\subsubsection{Conclusion for surface brightness}
The SL01 data for the surface brightness of elliptic galaxies is consistent with BB but only if a large unknown effect of luminosity evolution is included. The data do not support expansion and are in complete agreement with CC.
\subsection{Angular size}
\label{s4.2}
Closely related to surface brightness is relationship between the observed angular size of a distant object and its actual linear transverse size. The variation of angular size as a function of redshift is one of the tests that should clearly distinguish between BB and CC. The major distinction is that CC like all tired-light cosmologies does not include the ($1+z$) aberration factor. Its relationship (Eq.~\ref{cpe13}) between the observed angular size and the linear size is very close (for small redshifts) to the Euclidean equation. \citet*{Gurvits99} provide a comprehensive history of studies for a wide range of objects that generally show a $1/z$ or Euclidean dependence. Most observers suggest that the probable cause is some form of size evolution. Recently \citet{Lopez10} used 393 galaxies with redshift range of $0.2<z<3.2$ in order to test many cosmologies.  Briefly his conclusions are
{\em
\begin{description}
\item The average angular size of galaxies is approximately proportional to $z^{-\alpha}$ with $\alpha$ between 0.7 and 1.2.
\item Any model of an expanding universe without evolution is totally unable to fit the angular size data \ldots
\item Static Euclidean models with a linear Hubble law or simple tired-light fit the shape of the angular size vs $z$ dependence very well: there is a difference in amplitude of 20\%--30\%, which is within the possible systematic errors.
\item  It is also remarkable that the explanation of the test results with an expanding model require four coincidences:
\begin{enumerate}
\item The combination of expansion and (very strong evolution) size evolution gives nearly the same result as a static Euclidean universe with a linear Hubble law: $\theta\propto z^{-1}$.
\item This hypothetical evolution in size for galaxies is the same in normal galaxies as in quasars, as in radio galaxies, as in first ranked cluster galaxies, as the separation among bright galaxies in cluster
\item The concordance ($\Lambda$CDM) model gives approximately the same (differences of less than 0.2 mag within $z<4.5)$ distance modulus in a Hubble diagram as the static Euclidean universe with a linear law.
\item The combination of expansion, (very strong) size evolution, and dark matter ratio variation gives the same result for the velocity dispersion in elliptical galaxies (the result is that it is nearly constant with $z$) as for a simple static model with no evolution in size and no dark matter ratio variation.
\end{enumerate}
\end{description}
}

With a redshift range of $z<3$ the value of $S_\CC$ is approximately proportional to $ z^{0.68}$ which shows that it is consistent with these results. A full analysis requires a fairly complicated procedure to correct the observed sizes for variations in the absolute luminosity.

A simple example of the angular size test can be done using double-lobed quasars. Using quasar catalogues, \citet{Buchalter98} carefully selected 103 edge-brightened, double-lobed sources from the VLA FIRST survey and measured their angular sizes directly from the FIRST radio maps. These are Faranoff-Riley type II objects \citep{Faranoff74} and exhibit radio-bright hot-spots near the outer edges of the lobes. Since \citet{Buchalter98} claim that three different Friedmann (BB) models fit the data well but that a Euclidean model had a relatively poor fit a re-analysis is warranted. The angular sizes were converted to linear sizes for each cosmology and were divided into six bins so  that there were 17 quasars in each bin. Because these double-lobed sources are essentially one dimensional a major part of their variation in size is due to projection effects. For the moment assume that in each bin they have the same size, $\hat{S}$, and the only variation is due to projection then the observed size is $\hat{S}\sin(\theta)$ where $\theta$ is the projection angle. Clearly we do not know the projection angle but we can assume that all angles are equally likely so that if the $N$ sources, in each bin, are sorted into increasing size the i'th source in this list should have, on average, an angle  $\theta_i=\pi(2i -1)/4N$. Thus the maximum likelihood estimate of $\hat{S}$ is
\[
\hat{S}_{\rm est}=\frac{\sum_{i=1}^{N} \sin(\theta_i)S_{i}}{\sum_{i=1}^{N}\sin^2(\theta_i)}.
\]
Note that the sum in the denominator is a constant and that the common procedure of using median values is the same as using only the central term in the sum.
Next a regression (Appendix~\ref{appendix}) was done between logarithm  of the estimated linear size in each bin and $\log(1+z)$ where $z$ is the mean redshift. Then the significance of the test was how close was the exponent, $b$, to zero. For BB the exponent was $b=-0.79\pm0.44$ and for CC it was $b=0.16\pm0.44$. Although the large uncertainties  show that this is not a decisive discrimination between the two  cosmologies the slope for BB suggests that no expansion is likely. The overall conclusion is strongly in favor of no expansion.

\subsection{Type 1a supernovae}
\label{s4.3}
Type 1a supernovae make ideal cosmological probes. Nearby observations show that they have an essentially constant peak absolute magnitude and the widths of the light curves provide an ideal probe to investigate the dependence of time delay  as a function of redshift.
The current model for type 1a supernovae \citep{Hillebrandt00} is that of  a white dwarf  steadily acquiring matter from a close companion until the mass exceeds the Chandrasekhar limit at which point  it explodes.  The light curve has a rise time of about 20 days followed by a fall of about 20 days and then a long tail that is most likely due to the decay of $^{56}$Ni. The widths are measured in the light coming from the expanding shells before the radioactive decay dominates. Thus the widths are a function of the structure and opacity of the initial explosion and have little dependence on the radioactive decay. The type 1a supernovae are distinguished from other types of supernovae by the absence of hydrogen lines and the occurrence of strong silicon lines in their spectra near the time of maximum luminosity. Although the theoretical modeling is poor, there is much empirical evidence, from nearby supernovae, that they all have remarkably similar light curves, both in absolute magnitude and in their time scales. This has led to a considerable effort to use them as cosmological probes. Since they have been observed out to redshifts with $z$ greater than one they have been used to test the cosmological time dilation that is predicted by expanding cosmologies.

Several major projects have used both the Hubble space telescope and large earth-bound telescopes to obtain a large number of type 1a supernova observations, especially to large redshifts. They include the Supernova Cosmology Project \citep{Perlmutter99,Goldhaber01,Knop03}, the Supernova Legacy Survey \citep{Astier05}, the Hubble Higher z supernovae Search \citep{Strolger04} and the ESSENCE supernova survey \citep{Wood-Vasey08,Davis07}. Recently \citet{Kowalski08} have provided a re-analysis of these survey data and  all other relevant supernovae in the literature and have provided new observations of some nearby supernovae. Because these {\em Union} data are comprehensive, uniformly analyzed and include nearly all previous observations, the following analysis will be confined to this data. The data provided (Bessel) {\it B}-band magnitudes,  stretch factors, and $B-V$ colors for supernovae in the range $0.015\le z\le 1.551$. Since there is a very small but significant dependence of the absolute magnitude on the $B-V$ color , in effect the K-correction, following \citet{Kowalski08} the magnitudes were reduced by a term $\beta(B-V)$ where $\beta$ was determined by minimizing the $\chi^2$ of the residuals after fitting $M_\BB$ verses $2.5\log(1+z)$. This gave $\beta=1.54$ which can be compared with $\beta=2.28$ given by \citet{Kowalski08}.

The widths (relative to the standard width), $w$, of the supernova light curves are derived from the stretch factors, $s$, provided by \citet{Kowalski08} by the equation $w=(1+z)s$.  The uncertainty in each width was taken to be ($1+z$) times the quoted uncertainty in the stretch value. For convenience in determining power law exponents a new variable $W$ is defined by $W=2.5\log(w)$. Since the width is relative to a standard template  the reference value for $W$ is $W_0=0$. Fig.~/ref{snf1} shows a plot of these widths as a function of redshift.
An preliminary fit for $W$ as a function of $2.5\log(1+z)$ showed an offset $-0.095\pm0.014$. This offset was removed from the supernova widths, $W$, before further analysis was done. The same color correction and width offset will be used for both cosmologies.

What is relevant for both cosmologies is the selection procedure.  The current technique for the supernova observations is a two-stage process \citep{Perlmutter03,Strolger04,Riess04}. The first stage is to conduct repeated observations of many target areas to look for the occurrence of supernovae. Having found a possible candidate the second stage is to conduct extensive observations of magnitude and spectra to identify the type of supernova and to measure its light curve. This second stage is extremely expensive of resources and it is essential to be able to determine quickly the type of the supernova so that the maximum yield of type 1a supernovae is achieved. Since current investigators assume that the type 1a supernovae have essentially a fixed absolute BB magnitude (with possible corrections for the stretch factor), one of the criteria they used is to reject any candidate whose predicted absolute peak magnitude is outside a rather narrow range. The essential point is that the absolute magnitudes are calculated using BB and hence the selection of candidates is dependent on the BB luminosity-distance modulus.
In a comprehensive description of the selection procedure for a major survey \citet{Strolger04} state: {\em Best fits required consistency in the light curve shape and peak color (to within magnitude limits) and in peak luminosity in that the derived absolute magnitude in the rest-frame B band had to be consistent with observed distribution of absolute B-band magnitudes shown in \citet{Richardson02}}.

\subsubsection{Supernovae in BB}
Fundamental to any cosmology that explains the Hubble redshift as being due to an expanding universe is the requirement that exactly the same dependence must apply to time dilation. The raw data of the widths of the type 1a supernovae light curves as a function of redshift is shown in Fig.~/ref{snf1} for the {\em Union}  data  provided by \citet{Kowalski08}. The fitted straight line shows that the exponent of ($1+z$) is $1.097\pm0.032$ which is in good agreement with the expected value  of unity.  These results, which show excellent quantitative agreement with the predicted time dilation, have been hailed as one of the strongest pieces of evidence for an expanding cosmology. However the regression of $M_\BB$ on $2.5\log(1+z)$ shows that the luminosity is proportional to $(1+z)^{0.230\pm0.070}$ which shows that the absolute luminosity is slowly decreasing as the universe evolves. The standard explanation for this change is the ad hoc introduction of  dark energy \citep{Turner99} or quintessence \citep{Steinhardt98}.

\begin{figure}[!htb]
\includegraphics[width=\figurewidth,trim=35 27 70 55,clip=true]{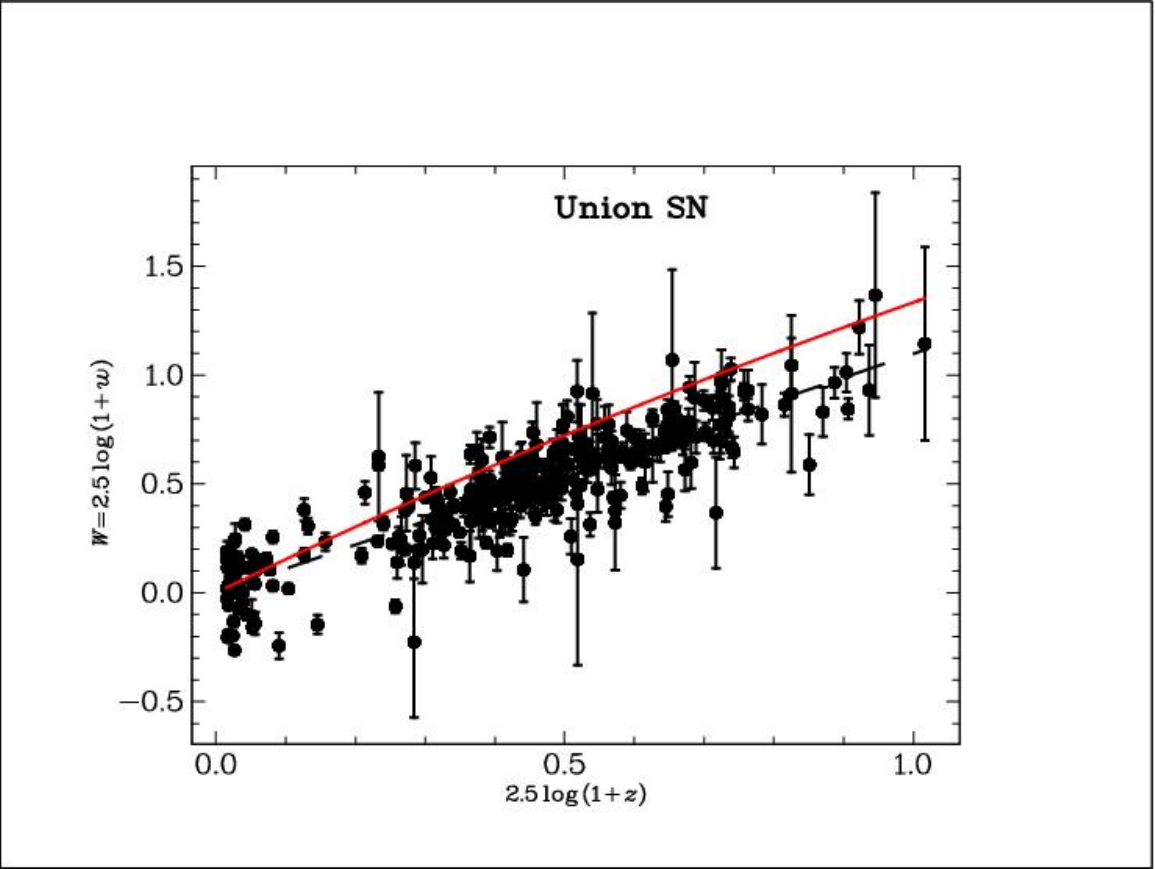}
\caption{Width of supernovae type 1a verses redshift for Union data. The dashed line (black) is expected time dilation in an expanding cosmology.  The solid line (red) is the function $\mu_\BB(z)-\mu_\CC(z)$.  \label{snf1}}
\end{figure}

A further problem with BB is a shortfall in the number of high redshift supernovae that are found.  Since to the first order the discovery of a supernova depends only on apparent magnitude this search procedure (at high $z$) should find all the candidates out to a redshift where the apparent magnitude is too faint for the telescope. Then the expected distribution of supernovae as a function of redshift should be proportional to the comoving volume.
To check the redshift distribution the 300 acceptable supernovae  were put in six bins in ascending order of redshift so that there were 50 supernovae in each bin.  Then the cutoff apparent magnitude (24.95 mag) was chosen so that all supernovae in the first five bins would be included. The results are shown in Table~\ref{snt1} where the columns are: the bin number, the redshift range, the included number in the bin, the ratio of BB volume in that bin to the BB volume in bin 2 and an estimate based on CC (see below).  The results for bin one are not unexpected. It simply shows that there have been many more searches done for supernovae at nearby redshifts. The results for bin six show that 33 out of 50 supernovae had an apparent magnitude brighter than the cut-off. To compensate the volume for bin 6 was computed for a redshift limit of $z=1.139$ which was the highest redshift for an included supernova. The problem with the BB results is that there is a dramatic shortage of supernovae in the high redshift bins. The usual explanation is that this shortage  is due to evolutionary effects. However this explanation must be able to show why there is a dramatic decrease in the rate of occurrence of supernovae at redshifts for $z$ near one when there is no obvious change in the stellar contents of galaxies with these redshifts.

\begin{table}
\begin{center}
\caption{Density distribution}\label{snt1}
\begin{tabular}{ccccc}
\hline
bin & $z$ range & $N_{\rm bin}$ & V$_\BB$\tablenotemark{a} & $p_sV_\CC$\tablenotemark{b}\\
1 & 0.014--0.123 & 50~&  0.04  & 0.10 \\
2 & 0.123--0.387 & 50~ &  1.00  & 1.00 \\
3 & 0.387--0.495 & 50~ &  0.89  & 0.49 \\
4 & 0.495--0.612 & 50~ &  1.29  & 0.43  \\
5 & 0.612--0.821 & 50~ &  3.09  & 0.51 \\
6 & 0.821--1.560\tablenotemark{c} & 33 \tablenotemark{d}&  6.40 & 0.37  \\
\hline
\end{tabular}
\end{center}
\tablenotetext{a}{The ratio of volumes in BB: $V_\BB(n)/V_\BB(2)$}
\tablenotetext{b}{The ratio of selection probability times CC volume}
\tablenotetext{c}{Redshift range used columns 4 and 5 is 0.821--1.139}
\tablenotetext{d}{The number brighter than the cutoff of 24.95 mag}
\end{table}

\subsubsection{Supernovae in CC}
Since the redshift in CC arises from an interaction with the intervening gas, it is not always a good measurement of distance. In particular the halo around our galaxy and that around any target galaxy will produce an extra redshift that results in an overall redshift that is larger than would be expected from the distance and a constant inter-galactic gas density. Since this is an additive effect it is important only for nearby objects. In fact the nearby supernovae (defined as those with $z<0.15$) show an average absolute magnitude that is brighter than the extrapolated magnitude from more distant supernovae. In order to make a partial correction for this bias all redshifts were reduced by subtracting 0.006 from each redshift, $z$. This correction brought the near and more distant magnitudes into agreement.
A plot of absolute magnitude, $M_\CC$, verses width, $W$, is shown in Fig.~/ref{snf2}. For later analysis the data are divided into the same 6 redshift bins used in Table~\ref{snt1}. The best-fit straight line to all the supernova has a (global) slope of $0.695\pm0.072$ (for $M_\BB$ it is $-0.391\pm0.056$).  This implies that supernovae that are brighter have narrower widths, or the weaker are wider. Table~\ref{snt2} gives the rms of the reduced magnitudes and the slope of the reduced magnitudes verses the width, $W$, for each bin. However \citet{Phillips93,Hamuy96,Guy05} argue for a local dependence of magnitude on stretch that has the opposite sign to the fitted straight line.

In the first bin the stretch and magnitude are essentially identical so that we can compare the local result of $-1.19\pm0.21$ with $-1.56\pm0.25$ reported by \citet{Guy05} to show good agreement. The slope for the other bins shows that although this local slope is not so well defined it is clearly present at higher redshifts. The challenge is to explain why the slope in a small redshift range is opposite to that for the global redshift range. If $\Delta M$ is a variation in magnitude and $\Delta W$ is a variation in width then the
variations can be summarized:
\begin{enumerate}
\renewcommand{\theenumi}{(\arabic{enumi})}
\item Local: $\Delta M\propto -(1.56\pm0.25)\Delta W$ \citep{Guy05}.
\item Global: $\Delta M \propto (0.70\pm0.07)\Delta W$.
\item Constant energy: $\Delta M \propto  \Delta W$.
\end{enumerate}
where the last item (constant energy) assumes that the shape of the supernova light curve is the same for all supernova  with a height proportional to the peak luminosity. Hence the total energy is proportional to the peak luminosity multiplied by the width.

\begin{figure}[!htb]
\includegraphics[width=\figurewidth,trim=35 27 70 55,clip=true]{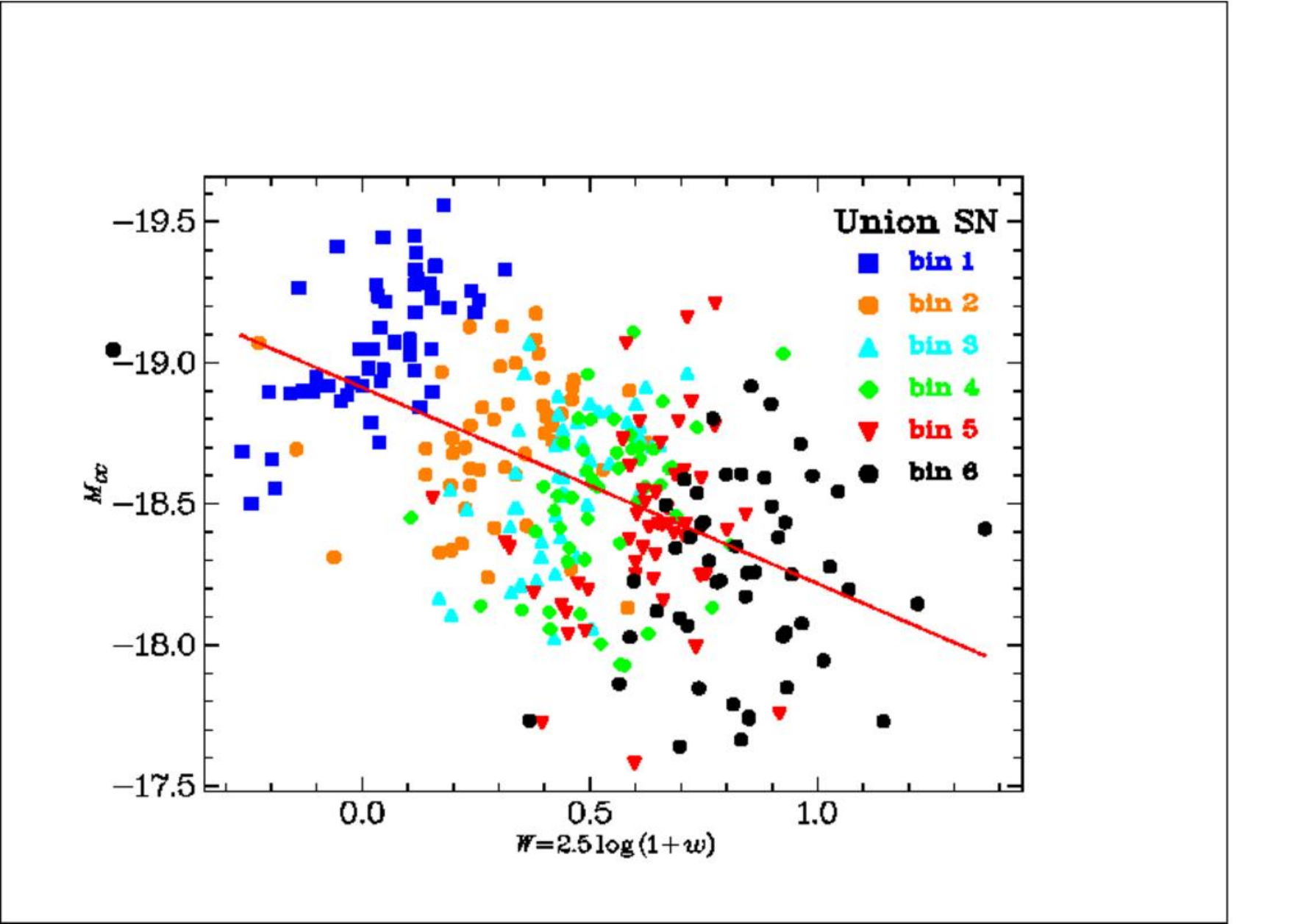}
\caption{Absolute magnitudes, $M_\CC$, (Bessel B-band) of type 1a supernovae verses widths of their light curves for Union data. The solid line (red) is the line of best fit. The data are divided into 6 redshift bins identical to the bins shown in Table~\ref{snt1}. Note that the brightest supernovae are at the top.\label{snf2}}
\end{figure}

\begin{table}
\begin{center}
{\caption{rms and local dependence of $M_\CC$ verses width, $W$.}\label{snt2}}
\begin{tabular}{cccc}
\hline
Bin & No. &  $M_{\rm rms}$ & slope\tablenotemark{a}  \\
1 & 50 & 0.239 & $-1.19\pm0.21$ \\
2 & 50 & 0.250 & $-0.46\pm0.30$ \\
3 & 50 & 0.258 & $-1.34\pm0.30$ \\
4 & 50 & 0.280 & $-0.99\pm0.40$ \\
5 & 50 & 0.323 & $-1.74\pm0.35$ \\
6 & 50 & 0.323 & $-0.72\pm0.41$ \\
\hline
\end{tabular}
\end{center}
\tablenotetext{a}{The slope of $M_\CC$ verses width $W$}
\end{table}

For type 1a supernovae it will be shown that the choice of BB magnitudes does have an important effect on the selection of supernovae and the use of BB leads to a biased sample. Because the Chandrasekhar limit places  a well-defined limit on the supernova mass and hence its energy, it is expected that each supernova has the same total energy output.
Since the total energy of a supernova is proportional to the area under its light curve it is proportional to the product of the maximum luminosity and the width of the light curve. The proposed model is based on the principle that the most unchanging characteristic of type 1a supernovae is their total energy and not their peak magnitude. Due to local effects the total energy will have small variations about a constant value.

Since the distance modulus for  BB is always larger than that for CC  a selection based on absolute magnitude that uses BB will select at greater redshifts  supernovae that are fainter and hence supernovae with wider light curves. Thus finding that a selection that has width increasing with redshift could mimic time dilation.

Define the magnitude of the total energy, $E$, in the same way as the magnitude of the luminosity, that is $E=-2.5\log({\cal E/E_0})$ where $\cal E_0$ is a reference energy.  Then the first assumption is that although individual supernovae will show variation in total energy, magnitude and width we expect that the averages over many supernovae will satisfy the equation $E=M-W$. Since, by definition, the reference width, $W_0$, is zero the expected value of the energy is $E_0=M_0$ where $M_0$ is the reference magnitude. Since its expected value is constant $E$ will be a better {\em standard candle} than~$M$.

To summarize, by relying on BB distance modulus the (distant) supernovae search method consistently selects supernova that are weaker than expected. This pushes the selection towards the limits of the natural variation and also selects supernovae with wider light curves. If the error in the BB magnitudes is due to the inclusion of the time dilation term the correct absolute luminosity is smaller than the BB luminosity by a factor of $(1+z)$. Thus with constant energy, the width is larger by a factor of $(1+z)$ which agrees with the results shown in  Fig.~\ref{snf1}.

Using CC and a constant energy model the dependence is  $\Delta M= M_\BB - M_\CC=\mu_\CC -  \mu_\BB$ we get $\Delta W=\mu_\BB -\mu_\CC$ which is shown as the solid (red) line in Fig.~/ref{snf1}. Considering that the selected supernovae are a biased sample the agreement with the widths is reasonable. There is still a problem of explaining the local slope. In BB the average of the local slope over the six bins is $-1.24\pm0.15$ compared with the global slope of $-0.391\pm0.056$ and for CC the average local slope is $-1.07\pm0.20$. Consider a supernovae with above average energy. If the only change in the shape of the light curve is a larger scaling factor the peak luminosity will be proportional the width which agrees with the average local slope.

Finally  this model can be used to get an approximate estimate of the expected number of supernovae that would be selected. The nearby supernovae come from a wide range of heterogenous surveys and serendipitous observations. Consequently their selection probability is essentially unknown. However the more recent, distant supernovae come from deliberate surveys that scan a small area of the sky looking for sudden outbursts. The crucial point is that provided the apparent magnitude of the supernova is within the observational constraints the probability of detection is independent of redshift. Thus as a first approximation all the surveys may be combined in order to determine the expected number of supernovae that should be detected.

Assuming that the intrinsic  distribution in magnitude is normal (with a standard deviation of $\sigma$) and that a supernova is selected if $M_\BB$ falls with a narrow range about the reference value, then the probability of selection is proportional to $\exp(-(\mu_\BB(z)-\mu_\CC(z))^2/2\sigma^2)$. The test is whether the number of expected supernovae as a function of redshift is similar to the observed number. The expected numbers (probability times volume, $V_\CC$), are shown for $\sigma=0.37$ in the last column of Table~\ref{snt1}. Note that the redshift ranges for each bin are determined by the observed supernovae which  are heavily biased in bin 1 and probably also in bin 2 due to the inclusion of supernovae from many local surveys that do not fit the selection model.  Consequently the value of $\sigma=0.35$ was chosen to provide roughly equal values for bins 3, 4 and 5. The average of the rms of $M_\CC$ shown in Table~\ref{snt2}  is 0.32 mag which is reasonably consistent with $\sigma=0.37$.

It has been argued that the total energy of type 1a supernovae makes a good {\em standard candle}.  Now we investigate whether the energies of type 1a supernovae are independent of redshift.  Fig.~/ref{snf3} shows the expected energy, defined here for each supernova as  $E_\CC=M_\CC-W$, of the Union supernovae as a function of redshift.

\begin{figure}[!htb]
\includegraphics[width=\figurewidth,trim=35 27 70 55,clip=true]{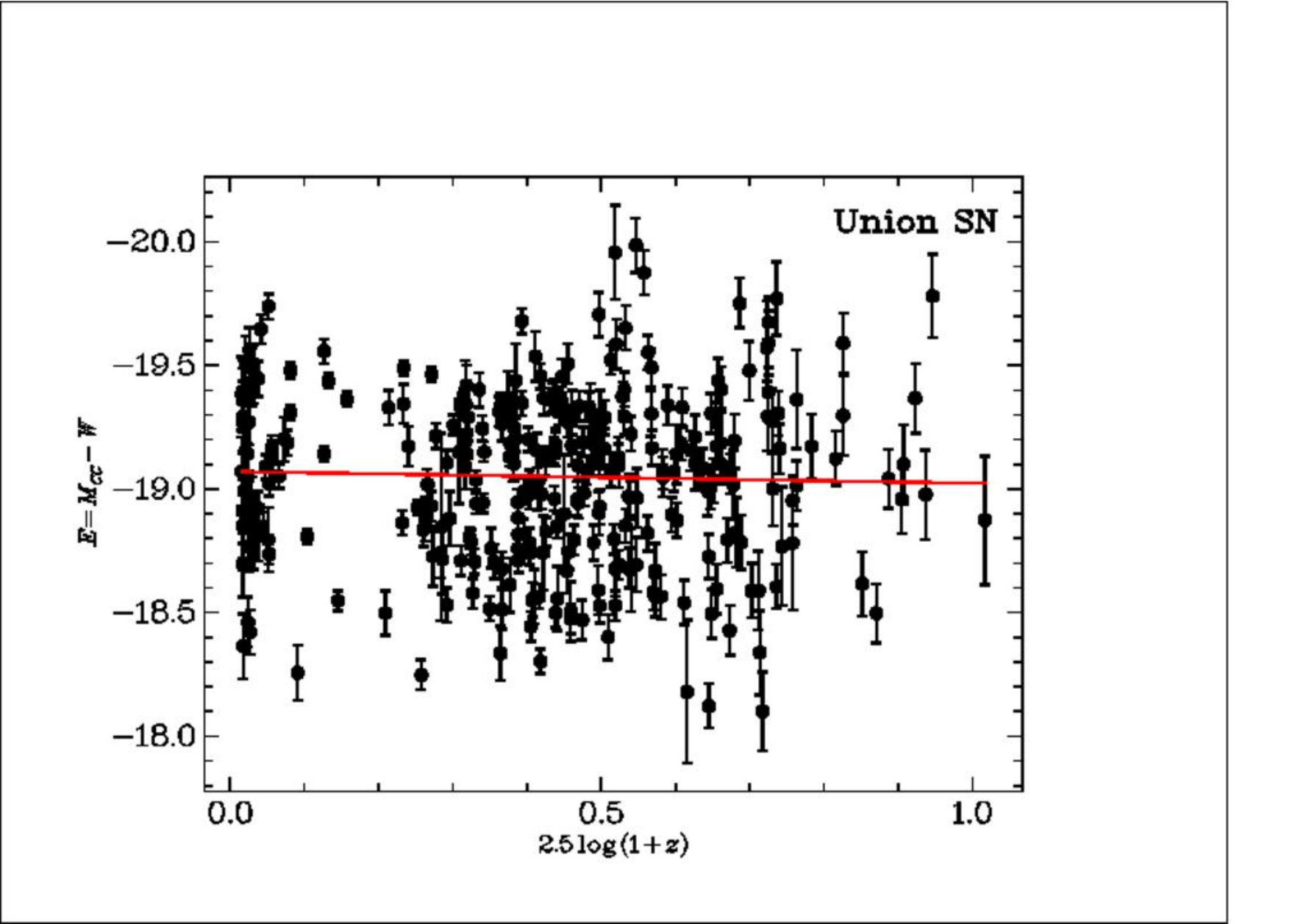}
\caption{Absolute energy, $M_\CC-W$,  verses redshift, $2.5\log(1+z)$, for Union data. Solid line (red) is the line of best fit with the equation $(-19.070\pm0.042) + (0.047\pm0.089)2.5\log(1+z)$ \label{snf3}}
\end{figure}

The slope of the fitted line is $0.047\pm0.089$ which is in excellent agreement with zero. Thus showing that CC can provide an good fit to the supernova data without the fitting of any free parameters and that $E$ is an good {\em standard candle}.  Since there is no deviation from the straight line at large redshifts there is no need for dark energy \citep{Turner99} or quintessence \citep{Steinhardt98} (both of which are meaningless in CC). The estimate of the intercept, that is $M_0$, is $-19.070\pm0.042$. \citet{Riess05} have measured accurate distances to two galaxies containing nearby supernovae. Together with two earlier measurements, they derive an absolute magnitude of type 1a supernovae of $-19.17\pm0.07$.  Hence the reduced Hubble constant, $h$, can be estimated from $-19.070 -5\log(0.7)=-19.17 -5\log(h)$ to get $h=0.638\pm0.05$. Thus  the measured Hubble constant is $63.8\pm0.5\,$kms$^{-1}$ Mpc$^{-1}$.

\subsubsection{Supernova conclusion}
It has been shown  that there is very strong support for the proposition that the most invariant property of type 1a supernovae is their total energy and not their peak magnitude. Given an essentially constant energy there is an inverse relationship between the peak luminosity and the width of the light curve.  Since the prime characteristic used for selecting these supernovae is the peak magnitude which is computed using BB there is a strong bias that results in intrinsically weaker supernovae being selected at higher redshifts. For constant energy these weaker supernovae must have wider light curves. Using a simple model for the selection process it was shown that it predicts the observed  dependence for the light curve widths on redshift (Fig.~\ref{snf1}). It is also consistent with the observed local variation of magnitudes on widths. When the observed magnitudes are corrected for the supernova width, they are independent of redshift (Fig.~/ref{snf3}). The conclusion is that with a simple selection model these supernovae observations are fully compatible with CC and there is no need for dark energy or quintessence. This is strong support for the premise that there is no time dilation and hence no universal expansion.

\subsection{Gamma ray bursts (GRB)}
\label{s4.4}
Gamma-Ray bursts (GRB) are transient events with time scales of the order of seconds and with energies in the X-ray or gamma-ray region. \citet{Piran04} provides (a mainly theoretical) review and \citet*{Bloom03} give a review of observations and analysis. Although the reviews by \citet{Meszaros06} and \citet{Zhang07} cover more recent research and provide extensive references they are mainly concerned with GRB models. This paper considers only the direct GRB observations and makes no assumptions about GRB models.

The search for the time dilation signature in data from GRB has a long history and before redshifts were available \citet{Norris94, Fenimore95, Davis94} claimed evidence for a time dilation effect by comparing dim and bright bursts. However \citet{Mitrofanov96} found no evidence for time dilation. \citet*{Lee00} found rather inconclusive results from a comparison between brightness measures and timescale measures. They also provide a brief summary of earlier results.  Since redshifts have become available \citet{Chang01} and \citet*{Chang02} using a Fourier energy spectrum method and \citet{Borgonovo04} using an autocorrelation method claim evidence of time dilation. The standard understanding, starting with \citet{Norris02} and \citet{Bloom03}, is that time dilation is present but because of an inverse relationship between luminosity and time measures it cannot be seen in the raw data. Their argument is that because a strong luminosity-dependent selection produces an average luminosity that increases with redshift there will be a simultaneous selection for time measures that decrease with redshift which can cancel the effects of time dilation.

\citet{Crawford09b} has argued that within the paradigm of BB that there is no evidence for strong luminosity selection or luminosity evolution.  Consequently those time measures that show a strong relationship with luminosity must have evolved in a similar manner. Although it is possible that a combination of luminosity selection, selection of GRB by other characteristics and evolution may be sufficient to cancel time dilation it does require a fortuitous coincidence of these effects to completely cancel the effects of time dilation in the raw data. Another explanation is that the universe is not expanding and thus there is no time dilation.

\subsection{Galaxy distribution}
\label{s4.5}
Recently, large telescopes with wide fields and the use of many filters have enabled a new type of galactic survey. The light-collecting capability of the large telescopes enables deep surveys to apparent magnitudes of 24 mag or better and the wide field provides a fast survey over large areas. A major innovation is the use of many filters whose response can be used to classify the objects with great accuracy. Thus, galaxies can be separated from quasars without needing morphological analysis. This photometric method of analysis works because photometric templates are available for a wide range of types of galaxies and other types of objects. In addition, accurate redshifts are obtained from fitting the templates without the tedious procedure of measuring the spectrum of each object.

A typical example of this photometric method is the COMBO-17  survey (Classifying Objects by Medium-Band Observations in 17 filters) provided by \citet{Wolf04}. The goal of this survey was to provide a sample of ~50,000 galaxies and ~1000 quasars with rather precise photometric redshifts based on 17 colors. In practice, such a filter set provides a redshift accuracy of  ~0.03 for galaxies and ~0.1 for quasars. The central wavelength of the 17 filters varied from 364 nm to 914 nm and consisted of 5 broadband filters ($U, B, V, R, and I$) and 12 narrower-band filters. \citet{Wolf03} have analyzed this  data and claim that there is strong evolution for $0.2 <z <1.2$. Instead of using generic K-corrections, the restframe luminosity of all galaxies are individually measured from their 17-filter spectrum. For each galaxy, three restframe passbands are considered, (i) the SDSS $r$-band, (ii) the Johnston $B$-band and (iii) a synthetic UV continuum band centered at $\lambda_{\rm rest}$= 280 nm with 40 nm FWHM and rectangular transmission function.  A spectral energy distribution, SED, was determined for each galaxy by template matching. For the evolution analysis they were assigned to one of four types. The only type that showed a well defined peak in their luminosity distribution was Type 1 which covers the E-S$_{\rm a}$ galactic types. The characteristics of the luminosity distribution were obtained by fitting a Schechter function which is
\[
\phi(L)dL\phi^*(L/L^*)^\alpha e^{L/L^*} dL
\]
where the luminosity $L^*$ (and its magnitude $M^*$) is a measure of location and $\alpha$ is a measure of shape. They found that a fixed value for $\alpha$ works quite well for the luminosity functions of individual SED types. Examination of their estimate of $M^*$ for Type 1 galaxies showed that if they were converted to CC magnitudes they were independent of redshift. This is shown in Table~\ref{galt1} where the data are taken from the appendix to \citet{Wolf03}. The second column is the difference, $\mu_\CC - \mu_\BB$, between BB and CC distance moduli. The remaining columns show the absolute magnitude for the three restframe passbands.
The last row shows the $\chi^2$ for the five magnitudes relative to their mean using the given uncertainties (all in the range 0.14-0.23).

\begin{table}
\begin{center}
\caption{$M^*_\CC$ for SED Type 1 galaxy luminosity distributions.\label{galt1}}
\begin{tabular}{lrrrr}
\hline
$z$ &  $\Delta \mu$ &  $M^*_r$\tablenotemark{a} & $M^*_B$ & $M^*_{280}$  \\
\hline
0.3 & 0.426 & -20.49 & -19.06 & -17.38 \\
0.5 & 0.642 & -20.49 & -19.15 & -17.84 \\
0.7 & 0.822 & -20.77 & -19.37 & -17.62 \\
0.9 & 0.975 & -20.54 & -19.09 & -17.79 \\
1.1 & 1.107 & -20.87 & -19.23 & -18.23 \\
$\chi^2$& & 3.70 & 2.32 & 12.81 \\
\hline
\end{tabular}
\end{center}
\tablenotetext{a}{Absolute magnitude for the SDSS $r$-band}
\end{table}

With four degrees of freedom the first two bands show excellent agreement with a constant value. The values  for $M^*_{280}$ have less than  a  2.5\% chance of being constant. However since most of the discrepancy comes from the $z=0.3$ value of -17.38 mag and most of this band at small redshifts is outside the range of the 17 filters this discrepancy can be ignored. If this value is ignored the  $\chi^2$ is reduced from 12.81 to 6.12 (with 3 D0F) which is consistent with being constant. Since $\alpha$ is independent of redshift the result is that if the data had been analyzed using CC the magnitude for these Type 1 galaxies does not vary with redshift.
Thus we have the surprising result that using BB a class of galaxies has a well defined luminosity evolution that is predicted by CC. In other words there is no expansion.

\subsection{Quasar distribution}
\label{s4.6}
A major difference between BB and CC is that at a redshift just greater than $z=5$ the absolute luminosity of a quasar is a factor of ten smaller for CC than for BB.
\citet{Richards06} have made a comprehensive study of quasars from the Sloan Digital Sky Survey (SDSS) and provide tables of absolute (BB) magnitudes and selection probabilities for 15,343 quasars. The sample extends from $i=$ 15 to 19.1 at $z\le3$ and $i=$ 20.2 for $z>3$. There was an additional requirement that the absolute magnitude  was $M_\BB < -22$ mag. Only some low redshift quasars failed this test. The final selection criterion was that each had a full width at half-maximum of lines  from the broad-line region greater than 1000 km\,s$^{-1}$.
\citet{Richards06} provided the redshift, apparent magnitude, selection probability and the K-correction for each quasar. The K-correction had two parts. The first part was a function only of the redshift and therefore it was independent of the nature of each quasar. However the second part was very important since it depended on the color difference $g-i$. They computed the quasar luminosity function in eleven redshift bins and in each case it was close to a power law in luminosity or an exponential function in magnitude. Effectively this meant that distributions were scale free and that there was no way the magnitudes could be directly used to compare cosmologies.

Let us assume that the magnitude distribution is exponential and can be written as
\[
\phi(M)\,dM= V \rho\beta \exp(\beta M)\,dM
\]
where $\beta$ is the basic parameter of the exponential distribution, $V$ is the accessible volume and $\rho$ is the quasar density.
Then using Eq.~\ref{cpe1} we get
\begin{equation}
\label{quae1}
\phi(M)\,dM=V  \rho\beta \exp{\beta (m-\mu(z)-K(z))}\,dM
\end{equation}
Now consider a small range of redshifts centered on $z_k$, then because $m$ will also has an exponential distribution the expected number in this redshift range is
\begin{eqnarray}
\label{quae2}
\phi_k(M)\,dM & = & V_k \rho \exp(\beta_k (\widehat{m}_k-\mu_k-\overline{K}_k))\times \nonumber \\
& &   \beta_k  \exp(\beta_k (m-\widehat{m}_k-K(z)+\overline{K}_k))\,dm
\end{eqnarray}
where the $\widehat{m}_k$ is the cutoff for the apparent magnitude  and $\overline{K}_z$ is the average K-correction for this $z$ range.
This is necessary because there are color-dependent corrections that are a property of the individual quasars. Another change is to change the increment in the independent variable from $dM$ to $dm$. The reason for the separation of the two exponents in Eq.~\ref{quae2} is the first line is  the same for all quasars in the range and the second line contains all the details of the quasar distribution. All variables that are common to the all the quasars in the redshift range have a suffix of $k$.  The result of integrating with respect to $M$ on the left and with respect to $m$ on the right is the expected number of quasars in this redshift range, $N_k$.  Thus rearranging Eq.~\ref{quae2} we get
\begin{equation}
\label{quae3}
\mu_k = \log\left( \frac{V_k \rho}{N_k}\right) /\beta +\widehat{m}_k -\overline{K}_k
\end{equation}
The essence of this method is that because the luminosity distribution is a power law we can easily change the independent variable from absolute magnitude to apparent magnitude. Thus Eq.~\ref{quae3} provides an estimate of the distance modulus where  cosmology enters only through the volume, $V_k$. The overall density $\rho$ is common to all redshift ranges and can be estimated by a least squares fit to all the ranges.

The next step is to estimate the exponential parameter $\beta_k$. A small complication is that the apparent magnitudes have a measurement uncertainty so that assuming a Gaussian error distribution the expected distribution is the convolution of a Gaussian with the exponential distribution to get
\[
p(m)\,dm=\beta \exp(-\beta(\widehat{m}-m) + \frac{1}{2} \beta^2\sigma^2)\,dm
\]
where $\sigma$ is the standard deviation of the magnitude uncertainty. Note that the second term shows that there is an excess of quasars moved from fainter magnitudes compared to those moved to fainter magnitudes. The maximum likelihood estimate for $\beta$ is the solution to the quadratic equation $\beta^2\sigma^2-(\widehat{m}-\overline{m})\beta +1$ where $\overline{m}$ is the mean magnitude  and its variance is
\[
\mbox{var}(\beta)= \frac{\beta^2}{N(1-\beta^2\sigma^2)}
\]

\begin{figure}[!htb]
\includegraphics[width=\figurewidth,trim=35 27 70 55,clip=true]{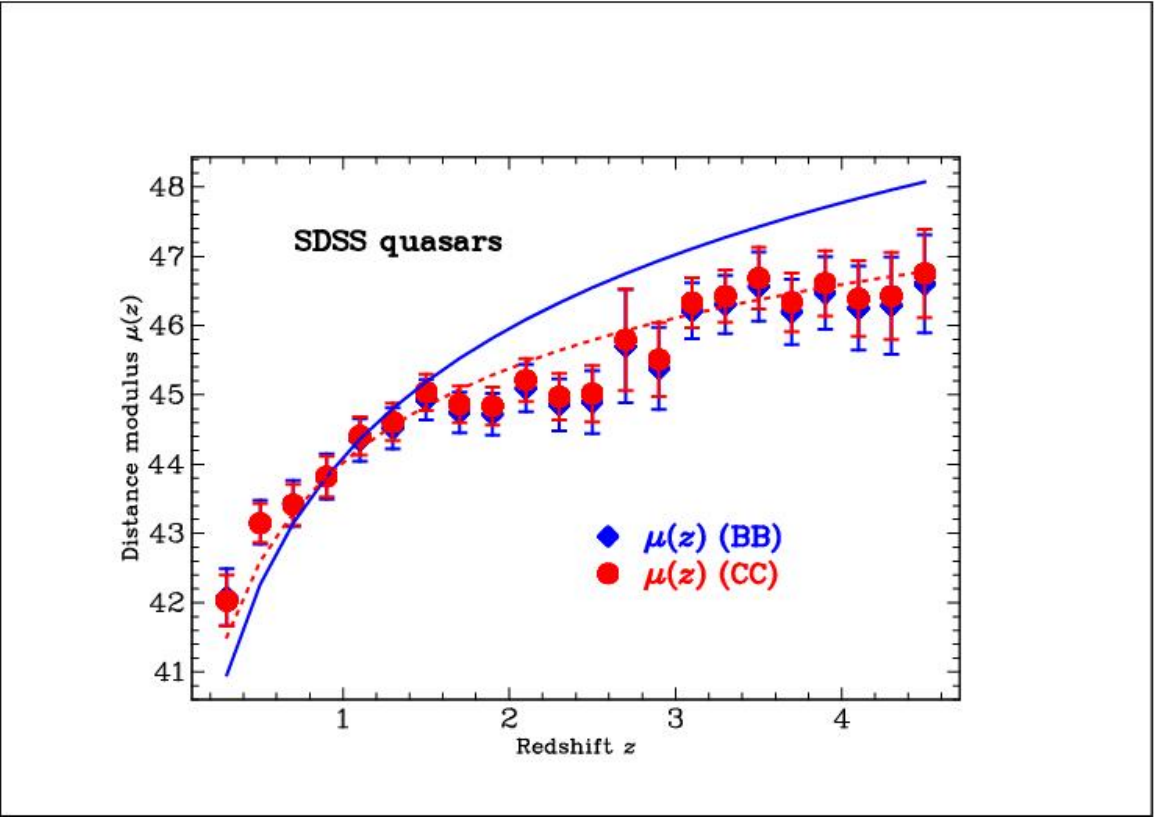}
\caption{The distance modulus of SDSS quasars as a function of redshift. The diamonds (blue) are for BB and the circles (red) are for CC. The full (blue) line is the theoretical distance modulus for BB and the dotted (red) line is for CC. All have been normalized to be the same at $z=0.9$.\label{quaf1}}
\end{figure}

This analysis has been done for the SDSS data \citep{Richards06} and the results are shown in Fig.~\ref{quaf1} where filled circles (red) are for CC volumes and the filled diamonds (blue) are for BB volumes. Both sets of points have been normalized to be the same for $z=0.9$. The selection probability for each quasar is allowed for by including only those quasars that have a selection probability greater than 0.3 and by dividing each quasars contribution to the distribution by the selection probability. As expected the different volumes between the cosmologies have only a small effect. The full (blue) line shows the BB distance modulus and the dashed (red) line is for the CC modulus.  The uncertainties were estimated from the deviations  from a smooth curve: in this case $\mu_\CC(z)$. The plotted lines are the theoretical distance moduli normalized in the same way as the data. The apparently lower values for observations near z=2.8 are probably due to confusion between the spectra of stars and that for quasars in this region which not only produces lower selection probabilities but also makes their estimates more uncertain. The very clear result is that the quasars are consistent with CC but they are not consistent with BB without evolution.

The conclusion is that CC clearly fits the data whereas BB would require evolution that cancels the expansion term in its distance modulus.

\subsection{Radio Source Counts}
\label{s4.7}
The count of the number of radio sources as a function of their flux density is one of the earliest cosmological tests that arose from the development of radio astronomy after World War II. Indeed, this test played a pivotal role in the rejection of the steady state cosmology of Bondi, Gold, and Hoyle in favor of the Big-Bang evolutionary model. In recent years, the study of radio source counts has declined for several reasons both theoretical and experimental. An important experimental problem is that many radio sources are double or complex in structure. Whether or not they are resolved depends on their angular size  and the resolution of the telescope. Since their distance is unknown, the counts are distorted in a way that cannot be readily determined. The main theoretical problem in Big-Bang  cosmology is that the counts are of a collection of quite different objects such as galaxies and quasars that can have different types of evolution. Thus, the radio source counts are not very useful in the study of these objects. However, in CC, the source number density must be the same at all places and at all times. Curvature cosmology demands that radio source counts are consistent with a reasonable luminosity number density distribution that is independent of redshift.  Thus it provides a critical test of CC.

In order to clarify the nature of the radio source count distribution let us start with a simple Euclidean model. Let the observed flux density of a source at an observed frequency of $\nu_0$ be $S(\nu_0)$ in units of W\,m$^{-2}$\,Hz$^{-1}$ and its luminosity at the emitted frequency $\nu$ be $L(\nu)$ in units of W\,Hz$^{-1}$. For simplicity, let us assume that all the sources have the same luminosity and that they have a volume density of $N$ sources per unit volume. Then using the inverse square law, the observed number of sources with a flux density greater than $S$ is
\[
n(>S)=\frac{4\pi N}{3}\left( \frac{L}{4\pi S} \right)^{3/2}.
\]
Thus the number density of observed sources is ${dn/ds}=S^{-5/2} $. The importance of this result is that it is customary to multiply the observed densities by $S^{5/2}$ so that if the universe had Euclidean geometry the distribution as a function of $S$ would be constant. It has the further advantage in that it greatly reduces the range of numbers involved. However this practice is not used in this analysis.

For CC the area at a distance $r$ is $A(r)=4\pi R^2\sin^2(\chi)$
where $r=\sin(\chi)$ and $\chi=\ln(1+z)/\sqrt{3}$. Note that the actual light travel distance is $R\chi$ . Thus
\[
S(\nu_0)d\nu_0=\frac{L(\nu )d\nu}{4\pi R^2 \sin ^2 (\chi )(1 + z)},
\]
where $\nu=(1+z)\nu_0$ and the ($1+z$) in the divisor allows for the energy loss due to curvature redshift. Since the ratio of the differentials (the bandwidth factor) contributes a factor that cancels the energy loss, the result is
\begin{equation}
\label{rade1}
S(\nu_0)=\frac{L((1+z)\nu_0)}{4\pi R^2 \sin^2 (\chi)}.
\end{equation}
It is convenient to replace the distance variable by the redshift parameter $z$. Then the differential volume is
\[
dV(r)=\frac{4\pi R^3 \sin^2 (\chi)}{\sqrt{3}(1+z)}\,dz.
\]

If the luminosity number density is $N(L,(1+z)\nu_0)$ the expected radio-source-count distribution (allowing for the $dS/dL$ term needed to match the differentials) is
\[
n(S,\nu_0)=\frac{16\pi ^2 R^5}{\sqrt{3}}
\int_0^{z_m}{\frac{\sin ^4 \left(\chi\right)}{1+z}
N\left( {L,(1+z)\nu_0} \right)\,dz},
\]
where $z_m$ is the limiting redshift.

A major problem with the observations is the difficulty in knowing the selection criteria. Typically, all sources greater than a chosen flux density are counted in a defined area. Since the flux density measurements are uncertain and the number of sources is a strong function of the flux density, it is difficult to assess a statistically valid cut-off for the survey. In a static cosmology, the change in the distribution due to the change in emitting frequency as a function of $z$ is an added complication. Thus, an essential test of CC is to show that there is an intrinsic distribution of radio sources that is identical at all redshifts. Unfortunately, it is not feasible to obtain a definitive distribution. What will be done is to show that the observations are consistent with a possible distribution. Thus the aim of this section is to show that there is a distribution $N(L,\nu_e)$ that provides a reasonable fit to the observations at all frequencies.

A simple distribution has been found that provides a good first approximation to the intrinsic distribution. Define the variable $x$ by
\begin{equation}
\label{rade2}
x=\left( \frac{L}{L^*} \right)\left( {\frac{\nu_0(1+z)}{1{\mbox{\,GHz}}}} \right)^\beta,
\end{equation}
where $L^*$ and  $\beta$ are constants. The first term is the ratio of the absolute flux density to a reference value, $L^*$, and the absolute flux density is obtained from the measured flux density by using Eq.~\ref{rade1}. The second term in Eq.~\ref{rade2} contains the frequency of emission, $\nu_0(1+z)$, and is the only frequency contribution in this simple model. Then the model for the intrinsic radio-source distribution is
\begin{equation}
\label{rade3}
N(L,\nu )=Ax^{-\alpha } \exp (-\gamma x).
\end{equation}
where $\alpha$, $\beta$, $\gamma$, and $A$ are constants that are found by fitting the model to the data listed in Table~\ref{radt1}. In order to provide realistic values a value of $h$=0.7 has been adopted.  The results of fitting this distribution to the data are shown in Table~\ref{radt2}. The  $\chi^2$ goodness of fit was 4360 for 252 DoF. Because of the poor fit, the estimate of statistical uncertainties has been omitted.

\begin{table}
\begin{center}
\caption{Origin of radio source-count data.\label{radt1}}
\begin{tabular}{llrl}
\hline
Survey name& Telescope & MHz & Reference\\
\hline
7C& CLST&191&   \citet{McGilchrist90} \\
5C6&   One-Mile&408& \citet{Pearson78} \\
B2& Bologna&408&    \citet{Colla75} \\
All Sky&&   408&    \citet{Robertson73} \\
Molonglo &  Cross&408&\citet{Robertson77a, Robertson77b} \\
Molonglo&   MOST&843&  \parbox{2.5 in}{\citet{Subrahmanya87}} \\
FIRST&  VLA&1400&   \citet{White97} \\
Virmos& VLA&1400&   \citet{Bondi03} \\
Phoenix&    ATCA&1400&  \citet{Hopkins98} \\
ATESP&  ATCA&1400&  \citet{Prandoni01} \\
ELIAS&  ATCA&1400&  \citet{Gruppioni99} \\
Parkes& Parkes&2700&    \citet{Wall85} \\
RATAN&  RATAN-600&3945& \citet{Parijskij91} \\
100m& (MPIfR)&4850& \citet{Maslowski84a, Maslowski84b} \\
VLA&    &5000&  \citet{Bennett83} \\
VLA& &5000& \citet{Partridge86} \\
MG II&  NRAO 300&5000&  \citet{Langston90} \\
\hline
\end{tabular}
\end{center}
\end{table}

\begin{table}
\begin{center}
\caption{Parameters for radio source distribution.\label{radt2}}
\begin{tabular}{lcc}
\hline
Parameter&  Value&  Unit \\
\hline
$\alpha$&   1.652&  \\
$\beta$&    0.370&  \\
$\gamma$&   0.0141&  \\
$L^*$&  $0.0237$&   W\,Hz$^{-1}$ \\
$A$&    $63.5$&   Gpc$^{-3}$ \\
\hline
\end{tabular}
\end{center}
\end{table}

\begin{figure}[!htb]
\includegraphics[width=\figurewidth,trim=35 27 70 55,clip=true]{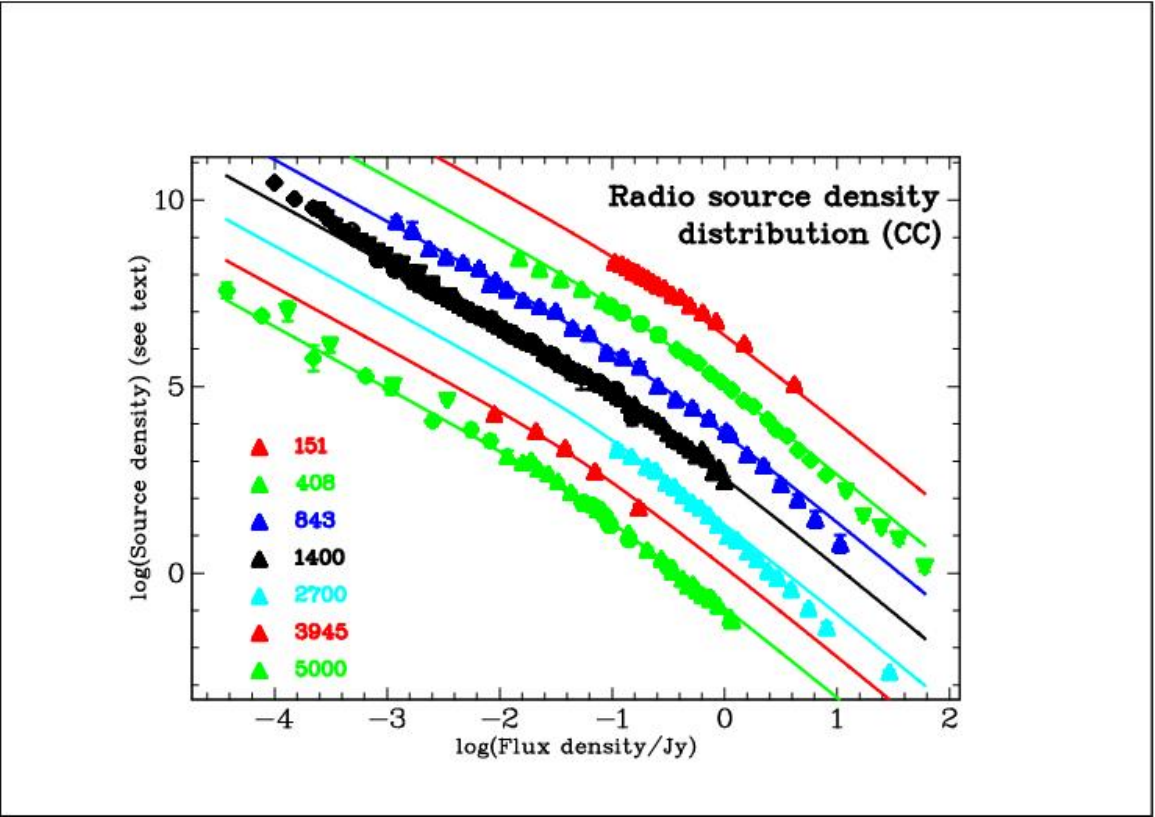}
\caption{Logarithm of the radio source volume density ($Gpc^{-3}$) distribution as a function of flux density. The legend shows the frequency (in Mhz) for each group. Each group is displaced vertically from the adjacent group by a factor of 10. The vertical scale is exact for the 1400 MHz group.\label{radf1}}
\end{figure}

A plot of the data (references in Table~\ref{radt1}) with the flux densities in Jy, and the results of this model is shown in Fig.~\ref{radf1}. For clarity the source density (number per Gpc$^3$) for each set of points for a given frequency has been multiplied by a factor of 10 relative to the adjacent group. The density scale is correct for the 1400 MHz data.

Clearly, the model satisfies the basic structure of the distributions but there is a poor fit at low flux densities, which is probably due to the limitations of the simple distribution (Eq.~\ref{rade2}). It should be emphasized that the only free parameters are those shown in Table~\ref{radt2}. Since the observed densities at low flux densities are larger than those predicted by the model it suggests that there may be a second population that is intrinsically fainter that the flux densities fitted by the model. This analysis shows that a simple distribution of radio source flux densities can be found that is consistent with the observations. In Big-Bang cosmology even with the inclusion of evolution there is no model of radio sources that is as simple as the one described here. Thus, the observations of the number distribution of radio sources as a function of flux density and frequency shows that they have no need for luminosity evolution.

\subsection{Quasar variability in time}
\label{s4.8}
One of the major differences between a tired-light cosmology and an expanding universe cosmology is that any expanding universe cosmology predicts that time variations and clocks have the same dependence on redshift as does the frequency of radiation. \citet{Hawkins10, Hawkins01, Hawkins03} has analyzed the variability of 800 quasars covering time scales from 50 days to 28 years. His data permitted the straightforward use of Fourier analysis to measure the time scale of the variability. He showed that there was no significant change in the time scale of the variability with increasing redshift.  He considered and rejected various explanations including that the time scales of variations were shorter in bluer pass bands or that the variations were not intrinsic but were due to intervening processes such as gravitational micro-lensing. His conclusion was either that the quasars are not at cosmological distances or that the expanding universe cosmologies are incorrect in this prediction. Curvature cosmology on the other hand would predict just such results.
To summarize, the sparse data on quasar variability strongly supports no time dilation.

\subsection{The Butcher--Oemler effect}
\label{s4.9}
If there were evidence of significant change in the universe as a function of redshift,  it would be a detrimental to any static cosmology. Probably the most important evidence for this cosmic evolution that appears to be independent of any cosmological model is the \citet{Butcher78} effect. Although they had discussed the effect in earlier papers, the definitive paper is \citet{Butcher84}. They observed that the fraction of blue galaxies in galactic clusters  appears to increase with redshift. Clusters allow the study of large numbers of galaxies at a common distance and out to large redshifts, which makes them ideal for studies in evolution. The core regions in a cluster are dominated by early-type (elliptical and lenticular) galaxies, which have a tight correlation between their colors and magnitudes. We can calculate $R_{30}$, the projected cluster-centric radius that contains 30\% of the total galaxy population. The blue fraction, $f_B$, is defined to be the fraction of galaxies within $R_{30}$ which are bluer than the color-magnitude relationship for that cluster. At first sight, this may appear to be a simple test that could be done with apparent magnitudes. However to compare the ratio for distant clusters with that for nearby ones the colors must be measured in the rest frame of each cluster, hence the need to use K-corrections. The major advantage of the Butcher--Oemler effect is that it is independent of the luminosity-distance relationship that is used. Therefore, to be more precise $f_B$ is the fraction that has an absolute magnitude $M_V$, whose rest frame (B-V) color is at least 0.2 magnitudes bluer than expected. A review by \citet{Pimbblet03} summaries the important observations.

In its original form the Butcher--Oemler effect is dependent on the apparent magnitude cut-off limits. It is essential that selection effects are the same in the rest frame for each cluster. There are further complications in that the percentage of blue galaxies may or may not depend on the richness of the cluster and the effect of contamination from background galaxies. Although \citet{Pimbblet03} concluded there was a definite effect, his Fig.~1 shows that this conclusion is open to debate. Since then there have been several attempts to measure an unambiguous effect. Even though they attempted to duplicate the original methodology of Butcher \& Oemler, \citet{dePropris03a} found essentially no effect for K-selected galaxies. \citet*{Andreon04} examined three clusters around $z$=0.7 and did not find clear-cut evidence for the effect. To quote one of their conclusions: {\em Twenty years after the original intuition by Butcher \& Oemler, we are still in the process of ascertaining the reality of the Butcher--Oemler effect}.
The Butcher--Oemler effect remains uncertain, and therefore does not provide evidence to refute a static cosmology.

\section{Non-expansion Observations}
\label{s5}
\subsection{The Hubble redshift}
\label{s5.1}
The increase in the redshift of an object with distance is well documented and both cosmologies provide a solution. In BB it is due to the expansion of the universe which is predicted by  general relativity. The basic instability of the  Einstein model is well known \citep{Tolman34,Ellis84}. Although the inclusion of a cosmological constant provides a static solution it is still unstable and after a small perturbation will either expand or contract. Thus the BB is in full agreement with general relativity. On the other hand in CC the Hubble redshift is due to a gravitational interaction between the photons and the cosmic gas whose density produces a curved space-time. Both cosmologies predict that for nearby distances the Hubble redshift is a linear function of distance. However in CC it depends on the integral of the square root of the density along the path length and can because of density variations in a particular situation it may vary from the average  value.
Another important difference is that CC predicts the actual value for the Hubble constant. For the measured density of $N=1.55\pm0.01\,\mbox{m}^{-3}$ the calculated value of the Hubble constant is  $H=64.4\pm0.2\,\mbox{km\,s}^{-1}\,\mbox{Mpc}^{-1}$ whereas the value estimated from the type 1a supernova data  is $63.8\pm0.5\,$kms$^{-1}$ Mpc$^{-1}$ and the result from the Coma cluster (Section~\ref{s5.14}) is $65.7\,$kms$^{-1}$ Mpc$^{-1}$.

\subsection{X-ray background radiation}
\label{s5.2}
Since \citet{Giacconi62} observed the X-ray background there have been many suggestions made to explain its characteristics. Although much of the unresolved X-ray emission comes from active galaxies, there is a part of the spectrum between about 10 keV and 1 MeV that is not adequately explained by emission from discrete sources.
The very high energy range is most likely due to external point sources. It is the intermediate range that is examined here.

\subsubsection{X-rays in BB}
In Big-Bang  cosmology for the intermediate X-ray range of about 10--300 keV, the production of X-rays in hot cosmic plasma through the process of bremsstrahlung has been suggested by \citet{Hoyle62,Gould63,Field64,Cowsik72}. In a review of the spectrum of the X-ray background radiation \citet{Holt92} concluded that the measured spectra of discrete sources are not consistent with the observations in the intermediate energy range but there is a remarkable fit to a 40 keV ($4.6 \times 10^8\,$K) bremsstrahlung spectrum from a diffuse hot gas. However, in an expanding universe most of the X-rays are produced at redshifts of $z\ge 1$ where the density is large enough to scatter the CMBR. This scattering known as the Sunyaev--Zel'dovich  effect (see Section~\ref{s5.5}) makes a distinctive change in the spectrum of the CMBR. This predicted change in the spectrum  has not been observed and this is the major reason why the bremsstrahlung model is rejected in Big-Bang cosmology.

\subsubsection{X-rays in CC}
In CC, the basic component of the universe is plasma with a very high temperature, and with low enough density to avoid the Sunyaev--Zel'dovich effect (Section~\ref{s5.5}). The background X-ray emission is produced in this plasma by the process of free-free emission (bremsstrahlung). The observations of the background X-ray emission are analyzed in order to measure the density and temperature of the plasma. In CC, this density is the major free parameter and it determines the size of the universe and the value of the Hubble constant. In addition, the temperature of the plasma determined from the X-ray measurements can be compared with the predicted value from CC of $2.56 \times 10^9\,$K.

The first step is to calculate the expected X-ray emission from high temperature plasma in thermal equilibrium. Here the dominant mechanism is bremsstrahlung radiation from electron-ion and electron-electron collisions. With a temperature $T$ and emission into the frequency range  $\nu$ to $\nu +d\nu$  the volume emissivity per steradian can be written as
\begin{eqnarray}
\label{xre1}
j_\nu(\nu )d\nu & = &\left(\frac{16}{3} \right)\left( \frac{\pi}{6} \right)^{1/2} r_0^3 m_e c^2
 \left( \frac{m_e c^2}{kT} \right)^{1/2} \nonumber \\
 & & \times g(\nu ,T) \exp \left( {-\frac{h\nu}{kT}} \right) N_e
\sum {Z_i^2 N_i}\,d\nu,
\end{eqnarray}
where $g(\nu,T)$ is the Gaunt factor, $N_e$ is the electron number density, $N_i$ is the ion number density and $r_0$ is the classical electron radius and the other symbols have their usual significance \citep*{Nozawa98}. In SI $j_\nu (\nu)$ has the units of W\,m$^{-3}$\,Hz$^{-1}$.
As it stands this equation does not include the electron-electron contribution. \citet{Nozawa98} and \citet{Itoh00} have done accurate calculations for many light elements. Based on their calculations Professor Naoki Itoh (http://www.ph.sophia.ac.jp/) provides a subroutine to calculate the Gaunt factor that is accurate for temperatures greater than $3 \times 10^8\,$K. It is used here. Let the average density be expressed as the number of hydrogen atoms per unit volume ($N= \rho/M_H$ m$^{-3}$). Then it is convenient to define $n_e=N_e/N$ and
\[
n_i =\sum {N_i Z_i^2} / N.
\]
where the sum is over all species present. Because of the very high temperature, we can assume that all atoms are completely ionized. Thus, Eq.~\ref{xre1} including the Gaunt factor provides the production rate of X-ray photons as a function of the plasma temperature and density.
\begin{table}
\begin{center}
\caption{List of background X-ray data used.\label{xrbt1}}
\begin{tabular}{lll}
\hline
Name     &  Instrument    & Reference\\
\hline
Gruber   &  HEAO 1 A-4    & \citet{Gruber99} \\
Kinzer   &  HEAO 1 MED    & \citet{Kinzer97} \\
Dennis   &  OSO-5         & \citet{Dennis73} \\
Mazets   &  Kosmos 541    & \citet{Mazets75} \\
Mandrou  &  Balloon       & \citet{Mandrou79} \\
Trombka  &  Apollo 16, 17 & \citet{Trombka77} \\
Horstman &  Rocket        & \citet{Horstman-Morretti74} \\
Fukada   &  Rocket        & \citet{Fukada75} \\
\hline
\end{tabular}
\end{center}
\end{table}
The next step is to compute the expected intensity at an X-ray detector. Consider an X-ray photon that is produced at a distance $R\chi$  from the detector. During its travel to the detector, it will have many curvature-redshift interactions. Although the photon is destroyed in each interaction, there is a secondary photon produced that has the same direction but with a slightly reduced energy. It is convenient to consider this sequence of photons as a single particle and to refer to it as a primary photon. The important result is that the number of these primary photons is conserved. Therefore, we need the production distribution of the number of photons per unit energy interval. The number of photons emitted per unit volume per unit time in the energy interval $\varepsilon$ to $\varepsilon + d\varepsilon$ is given by
\begin{equation}
\label{xre2}
j_n (\varepsilon )\,d\varepsilon= \frac{j_\nu(\nu)}{\varepsilon }h\,d\nu,
\end{equation}
where  $\varepsilon=h\nu$, $h$ is Plank's constant and $j_\nu(\nu)$ is the energy distribution per unit frequency interval. Now consider the contribution to the number of X-rays observed by a detector with unit area. Because the universe is static, the area at a distance $R$  from the source is the same as the area at a distance $R$  from the detector. Since there is conservation of these photons, the number coming from a shell at radius $R$  per unit time and per steradian within the energy interval $\varepsilon$ to $\varepsilon +d\varepsilon$  is
\[
\frac{{dn(r)}}{{dt}}\,d\varepsilon  = j_n (\varepsilon )d\varepsilon R\,d\chi.
\]
Next, we integrate the photon rate per unit area and per steradian from each shell where the emission energy is $\varepsilon$  and the received energy is $\varepsilon_0$  to get
\[
I_n (\varepsilon _0 )\,d\varepsilon _0  = R\int_0^{\chi_m } {j_n (\varepsilon )}\,d\varepsilon\,d\chi,
\]
where $\varepsilon=(1+z)\varepsilon_0$ and it is assumed that the flux is uniform over the 4$\pi$ steradians. Furthermore, it is useful to change the independent coordinate to the redshift parameter $z$. Then using Eq.~\ref{xre2} we get
\[
I_\nu  (\nu _0 )\,d\nu _0 = \frac{c}{H}\int_0^{z_m}{\frac{{j_\nu(\nu)}}
{{1 + z}}}\,dz\,d\nu _0,
\]
where $H$ is the Hubble constant  and the change in bandwidth factor $d\nu/d\nu_o$,  cancels the ($1+z$) factor that comes from the change in variable from $d\chi$ to $dz$ but there is another divisor of ($1+z$) that accounts for the energy lost by each photon. Thus the energy flux per unit area, per unit energy interval, per unit frequency and per solid angle is given by Eq.~\ref{xre3} where Plank's constant is included to change the differential from frequency to energy. The $z_m$ limit of 8.2 comes from the limit of  $\chi \le \pi$.

\begin{eqnarray}
\label{xre3}
I_\nu \left({\nu_0}\right) &=& \left( \frac{16}{3} \right)
\left(\frac{\pi}{6}\right)^{1/2}
\frac{r_0^3 m_e c^3}{h}
\left(8\pi GM_H\right)^{-1/2}
\left(\frac{mc^2}{kT}\right)^{1/2}\nonumber \\
& & \times n_e n_i N^{3/2}
\int\limits_0^{z_m } {\frac{g\left((1+z)\nu_0 ,T \right)}{(1+z)}
\exp \left(-\frac{h(1+z)\nu _0}{kT} \right)}\,dz \nonumber \\
 &=& \frac{1.9094 \times 10^3\,\mbox{keV}}{\mbox{keV\,m}^2\mbox{\,s\,sr}}
\left({\frac{mc^2}{kT}}\right)^{1/2}n_e n_i N^{3/2} \nonumber \\
& &
\times \varepsilon _0 \int\limits_0^{z_m}{\frac{g\left((1+z)\nu_0,T \right)}{(1+z)}\exp {\left({-\frac{h(1+z)\nu_0}{kT}} \right)}}\, dz.
\end{eqnarray}

The density $N$ is obtained by fitting Eq.~\ref{xre3} to the observed data as a function of the temperature $T$, and then extracting $N$ from the normalization constant. The X-ray data used is tabulated in Table~\ref{xrbt1}. It consists of the background X-ray data cited in the literature and assessed as being the latest or more accurate results. Preliminary analysis showed that there were some discrepant data points that are listed in Table~\ref{xrbt2} in order of exclusion.

\begin{table}
\begin{center}
\caption{Background X-ray data: rejected points.\label{xrbt2}}
\begin{tabular}{lrcr}
\hline
Source &  Energy  & Flux density       & $\chi^2$ \\
       &  keV     & keV/(keV\,cm$^2$\,s\,sr) & (1 DoF) \\
\hline
Gruber &    98.8  & 0.230$\pm$0.012 &   108.6\\
Gruber &    119.6 & 0.216$\pm$0.022 &    65.2\\
Fukada &    110.5 & 0.219$\pm$0.011 &    66.6\\
Gruber &    152.6 & 0.140$\pm$0.022 &    50.9\\
Fukada &    179.8 & 0.110$\pm$0.005 &    41.5\\
Gruber &    63.9  & 0.484$\pm$0.034 &    25.1\\
\hline
\end{tabular}
\end{center}
\end{table}

Very hard X-rays cannot be produced even by this hot plasma and are presumably due to discrete sources \citep{Holt92}. Since bremsstrahlung is very sensitive to the presence of heavy elements, results are presented for four different abundances of hydrogen, helium, and `metals'. The `metals', which is a descriptor for all the other elements, were simulated by an element with $<Z>$=8.4, $<Z^2>$=75.3 and $<A>$=17.25.

\begin{table}
\begin{center}
\caption{Abundances for four models.\label{xrbt3}}
\begin{tabular}{crrrcc}
\hline
Model & \%H & \%He & \%metals & $N_e/N$ &$\sum {N_i Z_i^2/N}$ \\
\hline
A & 100.0 & 0.0  &  0.0  &  1.000 & 1.000\\
B & 92.17 & 8.5  &  0.0  &  0.875 & 1.002\\
C & 92.06 & 7.82 &  0.12 &  0.868 & 1.061\\
D & 91.91 & 7.82 &  0.28 &  0.860 & 1.135\\
\hline
\end{tabular}
\end{center}
\end{table}

These values were derived from the abundances given by \citet{Allen76}. The details of the four different abundances are shown in Table~\ref{xrbt3} where the percentages are by number and the last two columns show the relative number of electrons and average value of $Z^2$ per unit hydrogen mass. Thus the models are A: pure hydrogen, B: hydrogen with 8.5\% helium, C: normal abundance and D: similar to C but with enhanced `metals'.

\begin{table}
\begin{center}
\caption{Fitted parameters for four abundance models.\label{xrbt4}}
\begin{tabular}{cccccc}
\hline
Model   &  $N$ &   $T_9$\tablenotemark{a} & $\chi^2$\tablenotemark{b}& $N_e$\tablenotemark{c} \\
\hline
A    & 1.93$\pm$0.02 & 2.62$\pm$0.04 & 167.4& 1.93\\
B    & 1.55$\pm$0.01 & 2.62$\pm$0.04 & 167.6& 1.35\\
C    & 1.16$\pm$0.01 & 2.61$\pm$0.04 & 168.5& 1.01\\
D    & 0.88$\pm$0.01 & 2.61$\pm$0.04 & 169.0& 0.75\\
\hline
\end{tabular}
\end{center}
\tablenotetext{a}{Temperature in units of $10^9$K}
\tablenotetext{b}{all for 74 DoF}
\tablenotetext{c}{the electron number density (m$^{-3}$)}
\end{table}

\begin{figure}[!htb]
\includegraphics[width=\figurewidth,trim=35 27 70 55,clip=true]{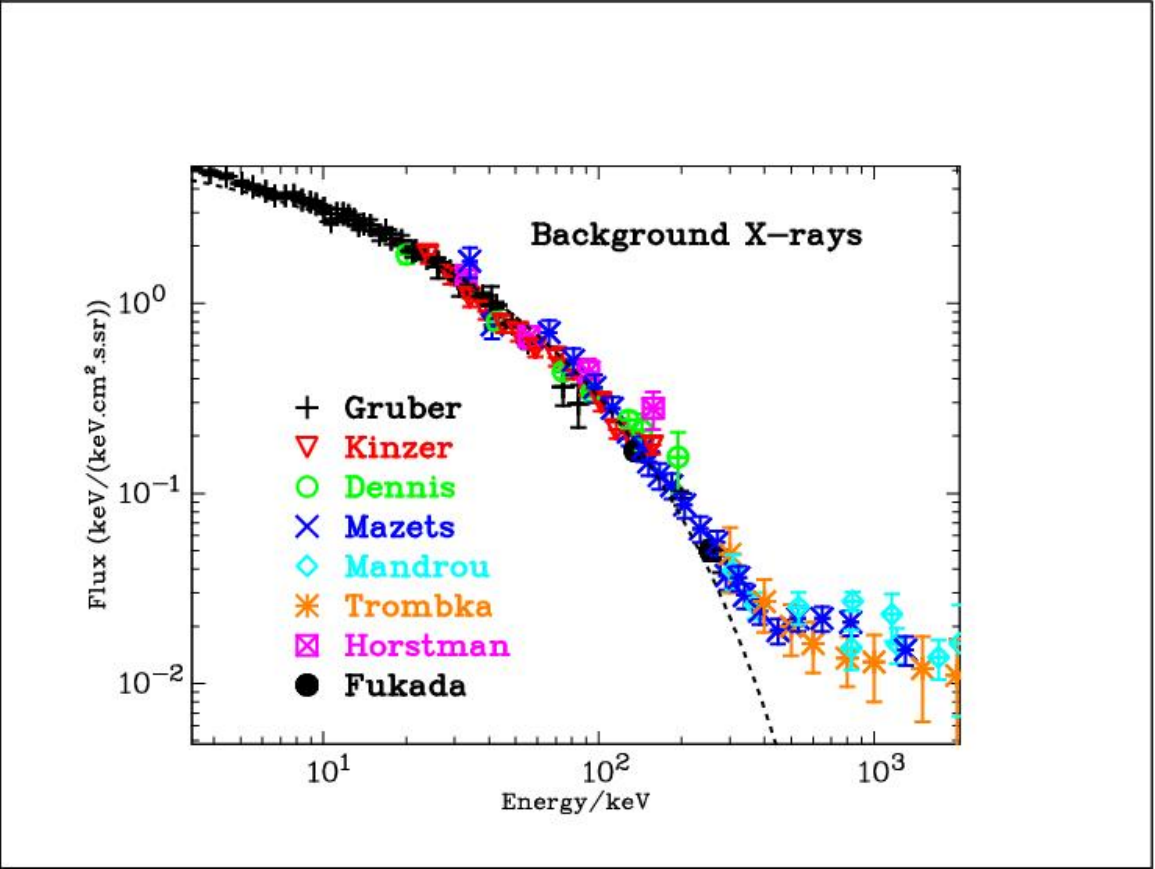}
\caption{Background X-ray spectrum. See Table~\ref{xrbt1} for list of observations. The dashed (black) line is best fit from 10 keV to 300 keV.\label{xrbf1}}
\end{figure}

The results of the fit of the data to these models is given in Table~\ref{xrbt4} where the errors are the fitted uncertainties ($1\sigma$). Fig.~\ref{xrbf1} shows the flux density for the fitted curve for model B and for the observations as a function of energy.

Most of the X-ray flux below 10 keV and part of the flux just above 10 keV is emission from discrete sources. The deviation from the curve at energies above about 300 keV arises from X-rays coming from discrete sources. In the intermediate region where bremsstrahlung should dominate, there are clear signs of some minor systematic errors. In addition, there appears to be some variation between the data sets. It is not clear whether the discrepancy between the observed points and the predicted flux densities is due to an inadequate theory, inadequate X-ray emission model, or systematic errors in the observations. After all the measurements are very difficult and come from a wide range of rocket, balloon and satellite experiments. In particular, the recent HEAO results \citet{Kinzer97} differ from earlier results reported by \citet{Marshall80}.

It is apparent from Table~\ref{xrbt4} that although the measured temperature is relatively insensitive to the assumed abundance the density estimate is strongly dependent. This is because bremsstrahlung depends mainly on the number and to a lesser extent the type of charged particles whereas the density also depends on the number of neutrons in each nucleus.

\begin{table}
\begin{center}
\caption{Fitted parameters for model B.\label{xrbt5}}
\begin{tabular}{lcl}
\hline
Quantity &  Symbol &    Value \\
\hline
Mean density              &$N$  &$1.55 \pm 0.01\,\mbox{m}^{-3}$ \\
Electron density          &$N_e$&$1.35 \pm 0.01 \,\mbox{m}^{-3}$ \\
Electron temperature      &$T_e$&$(2.62 \pm 0.04) \times 10^9\,\mbox{K}$ \\
Power law value at 1 Mev  &a    &$0.019 \pm 0.001$ \\
Exponent                  &b    &$-0.49 \pm 0.19$ \\
\hline
\end{tabular}
\end{center}
\end{table}

The power law fit parameters are the same for all the models and are shown in Table~\ref{xrbt5} for model B. This model was chosen because it uses the standard abundances that might be expected in the plasma and it has a relatively good fit to the observations. The quoted errors are the formal uncertainties of the fit. There are certainly larger, unknown systematic errors.

For the measured density of $N=1.55\pm 0.01\,\mbox{m}^{-3}$ the calculated value of the Hubble constant is $H=64.4\pm 0.2\mbox{\,km\,s}^{-1}\mbox{\,Mpc}^{-1}$. Further properties of the universe based on this density are shown Table~\ref{lnt1}.
That said, this bremsstrahlung model for the background X-ray emission within CC provides a good fit to the relevant observations. A crucial test of CC is that it predicts a temperature of $2.56\times 10^9\,$K for the cosmic plasma. The temperature estimated from fitting the X-ray data is $(2.62\pm 0.04)\times10^9\,$K. There is remarkable agreement between these values. It should be emphasized that the predicted temperature is a pure prediction from the theory without any dependence on observations. This agreement and the good fit to the observations gives strong support to CC.

In CC the argument against  bremsstrahlung based on the Sunyaev--Zel'dovich effect is not valid because  the density of the gas is much less and the CMBR has a different source. It has been shown that the X-ray data in the range from about 10 Kev to about 300 kev can be explained by  bremsstrahlung from the cosmic gas. The fitted temperature was $2.62\pm0.04\times 10^9\,$K and the fitted density was  $N=1.55\pm0.01$ hydrogen atoms per cubic meter (2.57$\times10^{-27}\,$kg\,m$^{-3}$).

At present BB does not have a good explanation for background X-ray radiation in the intermediate range of energies. Curvature cosmology can completely explain these observations as coming from  bremsstrahlung in a hot cosmic plasma.

\subsection{Cosmic microwave background radiation}
\label{s5.3}
The cosmic microwave background radiation (CMBR) is one of the major success stories for BB. The observed radiation has a spectrum that is extremely (by normal cosmological standards) close to a black body spectrum which means that it can be described by a single parameter, its temperature. Observations of the CMBR spectrum were obtained from the FIRAS  instrument on the {\it Cobe} satellite by \citet{Mather90}. They measured the temperature of the CMBR to be 2.725 K. This temperature is in agreement with the observations of \citet{Roth95} who measured a temperature of $2.729 (+0.023, -0.031)\,$K using cyanogen excitation in diffuse interstellar clouds.
It must be remembered that there is nothing special about a black body spectrum. If the radiation is quantized and all energy levels are freely available the black body (Plank function) is the thermal equilibrium spectrum. It is the maximum entropy solution. The black body is only required to permit the number of photons in each energy level come into equilibrium.  Thus in a general sense the black body spectrum is the default spectrum.

\subsubsection{CMBR in BB}
In BB the CMBR is the relic radiation that has been redshifted from the high temperature radiation that was in equilibrium with matter at the time when the ions combined with electrons to produce neutral atoms which are transparent. This decoupling of the radiation from matter occurred at a redshift of about $z=1000$. The exact temperature of the subsequent CMBR depends on the baryonic density parameter. Over time the redshift of the photons results in a decrease of their energy corresponding to an identical decrease in temperature without changing the shape of the spectrum. Thus BB predicts a black body spectrum with only a  poor estimate of its temperature.

\subsubsection{CMBR in CC}
In CC, the CMBR comes from the curvature-redshift process acting on the high-energy electrons and ions in the cosmic plasma. Examination of  Eq.~\ref{ce31} shows that even for very high temperature plasma the emission from electrons will dominate that from other ions. The energy loss occurs when an electron that has been excited by the passage through curved spacetime interacts with a photon or charged particle and loses its excitation energy. Ideally, the theoretical model would provide the number distribution of secondary photons as a function of their energy. This distribution would then be combined with the distribution of electron energies to obtain the production spectrum of low-energy secondary photons  from the plasma. The final step would be to integrate this production spectrum over all distances allowing for the geometry and curvature redshift. The result would be the spectrum of photons  that would be observed at any point in space.

We assume that the production spectrum for the photons is peaked at much larger energies than the cosmic microwave background photons. Then the cosmic microwave photons are seen to have had many curvature-redshift interactions. At each of these interactions the photons lose a small fraction of energy to very low energy photons that have frequencies less than the plasma frequency. Thus these low energy secondaries only exist as evanescent waves with their energy heating the plasma. Now since the radiation field is quantized the choice of the precise frequency of the large secondary photon is controlled by the number of available modes of oscillation. Being evanescent the very low energy photons are not relevant and the number of modes of oscillation is determined by the wavelength of the large secondary photon. Thus although the average energy loss rate is determined by the average density the selection of the number and occurrence of individual interactions depends on the quantization of the radiation field. All modes are available and in equilibrium the rate of photons entering a mode will equal the rate of photons leaving a mode. Because of the very large number of curvature-redshift interactions that have occurred since the original photon was produced the distribution of number of photons in each mode is essentially determined by the curvature-redshift interactions and not by the original spectrum. Thus the observed  spectrum will be the maximum entropy spectrum determined by the allowed modes of the radiation field. In equilibrium there is a constant energy density for these photons and as originally shown by Einstein \citep{Longair91} the maximum entropy  solution is that for a black body with a well defined temperature. The assumption of equilibrium enables us to equate the energy loss by the electrons to the energy loss by the CMBR photons and then to use Stefan's law to determine the temperature of the CMBR.

This brings up the problem of how the excited electrons produce the CMBR photons. Since conservation of energy and momentum prevents an excited free electron from emitting a photon, there must be an interaction with a third particle. A quick calculation shows that Thompson (Compton) scattering with the existing CMBR photons is too infrequent. The only other suitable interaction is Rutherford scattering off other electrons and ions. Since its last gravitational interaction, the electron will have become excited and have an excess of energy due to its passage through curved spacetime. At the Rutherford scattering this excitation  energy is transferred to secondary photons which become the CMBR after many curvature-redshift interactions.

The balance between the energy loss by the X-ray  electrons and the energy loss by the CMBR photons implies that there is an overall conservation of energy with the photon energy loss being returned to the electrons. Since the secondary photons produced by curvature-redshift interaction of the CMBR photons have frequencies well below the plasma frequency (of about 10 Hz), their energy must go into plasma waves which are dissipated with their energy going to heat the plasma. These processes are not driven by temperature differences so that there is no change in entropy.

For equilibrium, the energy gained by these secondary photons  must equal the energy lost by the electrons. Since the dominant energy loss by photons in the cosmic space is via curvature redshift, we can equate the two loss rates to determine the average energy of these photons. For electrons, or indeed any non-zero rest mass particle, the energy loss rate is given by Eq.~\ref{ce31}. Thus the energy loss rate for an electron is
\begin{equation}
\label{cme1}
\frac{{d\varepsilon }}
{{dt}}=H\left[{\beta^3 (\gamma ^2-\frac{1}{2})^{1/2} (\gamma-1)}\right]m_e c^2,
\end{equation}
where to prevent confusion with the symbol for temperature the electron's kinetic energy is denoted by  $\varepsilon=(\gamma -1)m_ec^2$ and the an extra factor of $\beta$ comes from conversion of distance rate to time rate. The next step is to average this energy loss over the distribution of electron energies. Since the electrons are relativistic, the appropriate distribution is J\"{u}ttner distribution, which is \citep*{deGroot80}
\begin{equation}
n(p)d^3 p = \frac{d^3p}{h^3}\exp \left({-\frac{\gamma mc}{kT_e}} \right).
\end{equation}
With a change of variable to  $\gamma$  it becomes
\begin{equation}
n(\gamma)\,d\gamma  \propto \gamma (\gamma^2-1)^{1/2} \exp \left({-\frac{\gamma m_e c^2}{kT_e}} \right)d\gamma.
\end{equation}
Then integrating Eq.~\ref{cme1} over all the electron energies we get
\begin{equation}
\label{cme2}
\overline {\frac{d\varepsilon}{dt}}  = HN_e m_e c^2 f(T_e ),
\end{equation}
where $N_e$ is the density of the electrons and $f(T_e)$ is average of the $\gamma$ terms.
Where we have
\begin{equation}
f(T_e) = \frac{\int_1^\infty  {\left[ {\left({\gamma^2-\frac{1}{2}} \right)^{1/2} \beta^3(\gamma-1)} \right]  n(\gamma)\,d\gamma}}
{\int_1^\infty  {n(\gamma)\,d\gamma}}.
\end{equation}
Although the J\"{u}ttner distribution can be integrated analytically in terms of modified Bessel functions, it is just as easy to evaluate both integrals numerically. Table~\ref{cmt1} shows some values for the function $f(T_e)$ as a function of the electron temperature $T_e$.

\begin{table}
\begin{center}
\caption{Some values for function $f(T)$.\label{cmt1}}
\begin{tabular}{cccccc}
\hline
$T_e/10^9$ &$f(T_e)$ &  $T_e/10^9$ & $f(T_e)$ & $T_e/10^9$ & $f(T_e)$\\
\hline
1.2 & 0.138 &   1.8 & 0.443 &   2.4 & 0.967\\
1.3 & 0.175 &   1.9 & 0.515 &   2.5 & 1.076\\
1.4 & 0.217 &   2.0 & 0.592 &   2.6 & 1.193\\
1.5 & 0.265 &   2.1 & 0.676 &   2.7 & 1.316\\
1.6 & 0.318 &   2.2 & 0.767 &   2.8 & 1.445\\
1.7 & 0.378 &   2.3 & 0.863 &   2.9 & 1.582\\
\hline
\end{tabular}
\end{center}
\end{table}

Consider the CMBR photons at one point in space. All of these photons will have been produced in one of the shells surrounding that point. In a static cosmology the contribution from each shell depends only on the thickness of the shell and is independent of the radius of the shell. However in CC there is an energy loss due to curvature redshift which means that the average energy that comes from a shell at redshift $z$ is reduced by the factor $(1+z)^{-1}$. Thus the average energy at the select point is less than the average energy of production by the integration of this factor with respect to distance. By using  Eq.~\ref{cpe10} to convert from distance to redshift the ratio is
\begin{equation}
\frac{1}{\sqrt{3}}\int_0^\infty \frac{1}{(1+z)^2}dz=\frac{1}{\sqrt{3}}.
\end{equation}
Thus the electron energy loss rate must be $\sqrt 3$ times larger than the energy loss rate by the CMBR photons. This is because the CMBR photons have already lost a major part of their energy since production during which time their spectrum is transformed in that for a black body.

The next step is to calculate the energy loss rate for the CMBR photons. If the CMBR photons are the result of curvature redshift acting on the cosmic electrons and the system is in equilibrium these two loss rates should be equal. For a black body spectrum then the energy density of the CMBR photons near us must be the same as that for a uniform black body radiation with the same temperature. However, because the universe is homogeneous, the energy density must be the same everywhere.  Then using Eq.~\ref{cme2} and Stefan's equation we get
\begin{eqnarray*}
\frac{4\sigma}{c}T_p^4 H &=&\frac{ N_e m_e c^2}{\sqrt{3}} f(T_e)H{\mbox{,   hence}}\nonumber \\
  T_p^4  &=& 62.4786 N_e f(T_e ).\nonumber
\end{eqnarray*}
where $\sigma$ is the Stefan-Boltzmann constant and, not surprisingly, the Hubble constant cancels. Then from Table~\ref{cmt1} we get $N_e=1.35$ and for a temperature of $(2.62\pm0.04)\times 10^9$\,K the calculated value of the function $f(T_e)$ is 1.215. These numbers result in a predicted temperature for the CMBR of 3.18 K. Probably the largest error in these temperature estimates comes from the uncertainty in the nuclear abundances. For the four abundance models (section~\ref{s5.2}) the predicted temperatures of the CMBR are 3.48 K, 3.18 K, 2.95 K and 2.74 K for the models 1, 2, 3 and 4 respectively. The main dependence is due to the differences in the electron density. Another important factor is the assumption that the universe has uniform density when it is apparent that it has large density variations.

\subsubsection{CMBR conclusions}
Both cosmologies offer argument to support the black body spectrum. Those for BB are well founded those for CC are less well founded. Against that BB does not have a good prediction for the temperature while CC predicts a narrow range of temperatures that is in excellent agreement with the observed temperature.
\subsection{CMBR at large redshifts}
\label{s5.4}
The temperature of the CMBR has been measured at large redshifts using two different methods. The first method measures the column density ratio of the fine structure absorption lines originating from the fundamental and first excited states of carbon \citep*{Ge97,Lima00,Srianand00,Srianand08}. These lines are seen in the Lyman-$\alpha$  forest that is observed in the spectra from a bright quasar. The temperature estimate is based on the relative strengths of these spectral lines. For these measurements to be valid, it is essential that the line widths and column densities are well understood. In CC the width of a spectral line is increased by the differential redshift as the photons traverse the absorbing gas. This change in the widths of spectral lines makes the very complex interpretation of the spectra required to estimate the temperature of the radiation suspect. Thus, until this interpretation is fully understood in the context of CC, CMBR temperature results from this method cannot be trusted.

The second method uses the Sunyaev--Zel'dovich  effect acting on the CMBR by the gas in clusters  of galaxies \citet{Battistelli02}. By using multiple frequencies, it is possible to minimize the effects due to properties of the clusters on the result. However the method is flawed in CC because the CMBR has a different cause from that in Big-Bang cosmology. Thus, these results cannot be taken as showing a dependence of the temperature of the CMBR on redshift until the complete mechanism is understood in the context of CC.
\subsection{Fluctuations in the CMBR}
\label{s5.4a}
One of the arguments for the interpretation of the CMBR in BB is that there are extensive models that can explain the density and polarization of spatial fluctuations in the observed radiation. In the model proposed for curvature radiation these fluctuations will also occur but in this case they are due to variations in the density of the cosmic plasma.  The CMBR seen through the denser gas within a galactic cluster will have lower than average temperature. \citet{Cabre06} show some support for this model in that they have correlated data from the Wilkinson Microwave Anisotropy Probe (WMAP) with galaxy samples from the SDSS DR4 galaxy survey  and found a significant correlation for the intensity fluctuations with galaxy density.

\subsection{Dark matter}
\label{s5.5}
The theoretical problems with dark matter have already been canvassed. Here we  concentrate on the observational evidence. All observational evidence for dark matter comes from the application of Newtonian gravitational physics to either clusters of objects or the rotation of galaxies. Galaxy rotation will be dealt with in Section~\ref{s5.11}. The original concept for dark matter comes from applying the virial theorem to the Coma cluster of galaxies \citep{Zwicky37}. The virial theorem \citep{Goldstein80} is a statistical theorem that states that for an inverse square law the average kinetic energy of a bound system is equal to half the potential energy (i.e. $2T+V=0$). Then with knowing the linear size of the cluster and measuring the mean square spread of velocities we can estimate the total mass of the cluster. There is no doubt that applying the virial theorem to the Coma and other clusters of galaxies provides mass estimates that can be several hundred times the mass expected from the total luminosity. Even the mass of inter-galactic gas  is not enough to overcome this imbalance. In BB cosmology dark matter has been introduced to make up for the shortfall of mass.

However if CC is valid then it is possible that the observed redshifts  are not due to kinematic velocities but are curvature redshifts produced by the inter-galactic gas. The purpose of this section is to show that curvature redshift can explain the galactic velocities without requiring dark matter. For simplicity, we will use the Coma cluster as a test bed. Not only is it very well studied, but it also has a high degree of symmetry and the presence of an inter-galactic gas cloud is known from X-ray  observations. \citet{Watt92} and \citet{Hughes89} have fitted the density of the gas cloud to an isothermal-model with the form
\begin{equation}
\label{dme1}
\rho =\rho_0\left( {1 + \left(\frac{r}{r_e} \right)}\right)^{-\alpha},
\end{equation}
with a center at 12$^h$59$^m$10$^s$, 27$\arcdeg$59$\arcmin$56$\arcsec$  (J2000) and with $r_e=8.8\arcmin\pm0.7\arcmin$,  $\alpha=1.37\pm0.09$,  $\rho_0=(2.67\pm 0.22)\times 10^3  h_{50}^2 \,\mbox{m}_H\,\mbox{m}^{-3}$. The central density is obtained from the X-ray luminosity and has a strong dependence on the distance. \citet{Watt92} assumed a Hubble constant  of 50 km\,s$^{-1}$\,Mpc$^{-1}$. With a mean velocity of 6,853 km\,s$^{-1}$ \citep{Colless96} and with this Hubble constant, the distance to the Coma cluster is 137 Mpc. Recently \citet{Rood88} using the Tully--Fisher relation to measure the distance modulus  to the galaxies in the Coma cluster, to observe a value of 34.4$\pm$0.2 mag whereas \citet{Liu01} using infrared surface brightness fluctuations get 34.99$\pm$0.21 mag. The average is 34.7 mag that corresponds to a distance of 87.1 Mpc. This is consistent with the distance of 85.6 Mpc given by \citet{Freedman01}. Thus putting $h=0.7$ gives a corrected central gas density of  $\rho_0=(6.61\pm0.54) \times 10^{-3}\,\mbox{m}_H\,\mbox{m}^{-3}$.

The galactic velocity data are taken from \citet{Beijersbergen04} who provide information for 583 galaxies. The velocity centroid of the Coma cluster is 12$^h$59$^m$19$^s$, 27$\arcdeg52\arcmin 2\arcsec$  (J2000). They find that early-type galaxies (E+S0+E/S0) have a mean velocity of 9,926 km\,s$^{-1}$ and a rms (root-mean-square) velocity, dispersion velocity, of 893 km\,s$^{-1}$. Let us assume that all the galactic velocities are due to curvature redshift. That is we assume that the actual velocities, the peculiar velocities, are negligible. Then the redshifts for the galaxies are calculated (in velocity units) by
\begin{equation}
\label{dme2}
v = v_0+\int_0^Z {51.691\sqrt{N\left(Z\right)}}\,dZ\,\mbox{km\,s}^{-1},
\end{equation}
where $Z$ is the distance from the central plane of the Coma cluster to the galaxy measured in Mpc, $N(Z)$ is the density of the inter-galactic gas cloud and $v_0$ is the average velocity of the galaxies in the cluster. The problem here is that we do not know $Z$ distances. Nevertheless, we can still get a good estimate by assuming that the distribution in $Z$ is statistically identical to that in $X$ and in $Y$. In a Monte Carlo simulation, each galaxy was given a $Z$ distance that was the same as the $X$ (or $Y$) distance of one of the other galaxies in the sample chosen at random. For 50 trials, the computed dispersion was 554 km\,s$^{-1}$ which can be compared with the measured dispersion of 893 km\,s$^{-1}$. Curvature cosmology has predicted the observed dispersion of galactic velocities in the Coma cluster to within a factor of two. Considering that this is a prediction of the cosmological model without fitting any parameters and ignoring all the complications of the structure both in the gas and galactic distributions the agreement is remarkable.

Since the distance to the Coma cluster is an important variable, the computed velocity dispersion from the Monte Carlo simulation for some different distances (all the other parameters are the same) is shown in Table~\ref{t14}. Thus, the redshift dispersion (in velocity units) is approximately a linear function of the Coma distance. This is not surprising since in this context the distance is mainly a scale factor.

\begin{table}
\begin{center}
\caption{Coma velocity dispersions for some distances.\label{t14}}
\begin{tabular}{lrrrr}
\hline
Distance/Mpc            &50  &   87 &   100 &   150\\
Dispersion /km\,s$^{-1}$ &318 &  554 &   636 &   955\\
\hline
\end{tabular}
\end{center}
\end{table}

\citet{Beijersbergen04} note that a better fit to the velocity distribution is provided by the sum of two Gaussian curves. Their best fit parameters for these two Gaussians are $v_1=7,501\pm 187\,\mbox{km\,s}^{-1}$, with  $\sigma_1=650\pm 216\, \mbox{km\,s}^{-1}$ and $v_2=6641\pm 470\,\mbox{km\,s}^{-1}$, with  $\sigma_2=1,004\pm 120\,\mbox{km\,s}^{-1}$. This double structure is supported by \citet{Colless96} who argue for an ongoing merger between two sub clusters centered in projection on the dominant galaxies NGC 4874 and NGC 4889. In addition, \citet*{Briel92} found evidence for substructure in the X-ray  emission and \citet{Finoguenov04} and \citet*{White93} have measured the X-ray luminosity of individual galaxies in the Coma  cluster showing that the model for the gas used above is too simple. The net effect of this substructure is that the observed velocity dispersion would be different from that predicted by a simple symmetric model. Thus, it appears that substructure makes it very difficult to achieve a more accurate test of CC using the Coma cluster.

There is an important difference between curvature redshift and models that assume that the redshifts of the galaxies within a cluster are due to their velocities. Since the laws of celestial mechanics are symmetric in time, any galaxy could equally likely be going in the opposite direction. Thus a galaxy with a high relative (Z) velocity could be in the near side of the cluster or equally likely on the far side of the cluster. However, if the redshifts are determined by curvature redshift  then there will be a strong correlation in that the higher redshifts will come from galaxies on the far side of the cluster. A possible test is to see if the apparent magnitudes are a function of relative redshift. With a distance of 87.1 Mpc the required change in magnitude is about 0.025 mag\,Mpc$^{-1}$. A simple regression between magnitude of Coma  galaxies (each relative to its type average) and velocity did not show any significant dependence. Although this was disappointing, several factors can explain the null result. The first is the presence of substructure; the second is that the magnitudes for a given galactic type have a standard deviation of about one magnitude, which in itself is sufficient to wash out the predicted effect; and thirdly mistyping will produce erroneous magnitudes due to the different average velocities of different types. In support of the second factor we note that for 335 galaxies with known types and magnitudes, the standard deviation of the magnitude is 1.08 mag and if we assume that the variance of the $Z$ distribution is equal to the average of the variances for the $X$ and $Y$ distributions then the expected standard deviation of the slope is 0.076 mag\,Mpc$^{-1}$. Clearly, this is such larger than the expected result of 0.025 mag\,Mpc$^{-1}$. It is expected that better measurements or new techniques of measuring differential distances will in the future make this a very important cosmological test.

In BB observations of the velocity dispersion of clusters of galaxies cannot be explained without invoking an ad hoc premise such as dark matter. However CC not only explains the observations but also makes a good prediction, without any free parameters, of its numerical value.

\subsection{The Sunyaev--Zel'dovich effect}
\label{s5.6}
The Sunyaev--Zel'dovich effect \citep{Sunyaev70,Peebles93} is the effect of Thompson scattering of background radiation by free electrons in the intervening medium. The technique depends on knowing the spectrum of the background source and then measuring the changes in the spectrum due to the intervening plasma.  In particular, it is the scattering in both angle and frequency of the cosmic microwave background radiation (CMBR) by electrons in the cosmic plasma. Because of the rapidly increasing density (like $(1+z)^3$) with redshift this is an important effect in BB.

The effect is often characterized by the dimensionless Compton y-parameter, which for a distance $x$ through non-relativistic thermal plasma with an electron density of $N_e$ has the value
\begin{equation}
\label{sze1}
y = \frac{kT_e}{m_e c^2}\sigma_T N_e x = 3.46 \times 10^{-16} N_e T_e x\,{\mbox{Mpc}},
\end{equation}
where  $\sigma_T$ is the Thompson cross-section. An object at redshift $z$ is at the distance $x = R\chi = 5.80\times 10^3 N_e^{1/2}\log(1+z)\,{\mbox{Mpc}}$.
Hence, using $T_e =2.62\times10^9\,$K, $N_e =1.35\mbox{\,m}^{-3}$  we get $y = 9.2\times10^{-6} \log (1+z)$.

Using the CMBR as a source the Sunyaev--Zel'dovich effect has been observed and \citet{Mather90} report an observed upper limit of $y = 0.001$,
and more recently \citet{Fixsen96} report $y=1.5\times10^{-5}$. Using this limit with Eq.~\ref{sze1} shows that there is no effect in CC if $z < 4.1$. Although in CC the CMBR has a more local origin it is of interest to note that this analysis assumes that each photon has many Compton interactions. For this electron density the Compton mean free path is 575 Gpc whereas the distance to $z=4.1$ is about 3.7 Gpc which means that a negligible number of the photons will have an interaction with the high temperature electrons. Furthermore the photon energy distribution for a single interaction has a different spectrum for that for the normal Sunyaev--Zel'dovich effect \citep{Longair91,Sunyaev80}. \citet{Bielby07} extend the results of \citet*{Lieu06} to show that not only was the Sunyaev--Zel'dovich effect less than what was expected but that it tendered to disappear as the redshift went from 0.1 to 0.3.  The conclusion is that CC is completely consistent with the experimental observations of the Sunyaev--Zel'dovich effect on the CMBR.
The conclusion is that although the Sunyaev--Zel'dovich effect is important in BB it is not important in CC.

\subsection{Gravitational lensing}
\label{s5.7}
There are more than 50 known gravitational lens where a quasar  or distant galaxy  has one or more images produced by a nearer lensing galaxy or cluster of galaxies. A set of these lensing systems has been examined in the context of CC to see if it offers a consistent and possibly simpler explanation. The two important measures are the prediction of the mass of the lensing galaxy and the determination of the Hubble  constant from the time delays between variations in the luminosity of different images. Since the delay measurement is easily done all that is needed is to measure the different path lengths. This path difference involves both geometric and general relativistic corrections.

One of the remarkable properties of gravitational lenses is that the geometry is completely determined by a two-dimensional lensing potential which can be expressed in terms off a surface density at the  position of the lensing galaxy. For thin lenses, any two systems with the same surface density distribution have the same lens effect. Now the usual way to determine the surface density is to measure the widths of spectral lines, assume that the width is due to velocity and then use the virial theorem to obtain the surface density. However in CC the widths of spectral lines are likely to have a large component due to the effects of curvature redshift from dust and gas in the lensing object. Thus the widths are not a reliable measure of area density and this method cannot be used.
Instead some double image gravitational lens were  investigated using a very simple point sources lens in order to see if the observations could be consistent with CC. However because of the paucity of examples and the wide range of characteristics there was no test that showed a significant difference between BB and CC. The data was consistent with both cosmologies. Currently the modeling and the data are not sufficient to choose between the cosmologies. However a more thorough analysis within the paradigm of CC may be more definite.

\subsection{Lyman\_alpha forest}
\label{s5.8}
The Lyman-$\alpha$ (Ly$\alpha$) forest is the large number of absorption lines seen in the spectra of quasars. Most of the lines are due to absorption by clouds of neutral hydrogen in the line of sight to the quasar. Some of the lines are due to other elements or due to Lyman-$\beta$ absorption. Because of the redshift  between the absorbing cloud and us, the lines are spread out over a range of wavelengths. Usually the analysis is confined to lines between the Ly$\alpha$ (at a wavelength of 121.6 nm) and Ly$\beta$ (at 102.5 nm). Thus, each quasar provides a relatively narrow spectrum of Ly-$\alpha$ lines at a redshift just less than that for the quasar. Since the advent of spacecraft telescopes, in which can observe the ultraviolet lines, and by using many quasars the complete redshift range up to the most distant quasar has been covered. The large redshift range makes the Lyman $\alpha$ spectra potentially a powerful cosmological tool.

The obvious cosmological observation is the density of lines as a function of redshift but as discussed by \citet{Rauch98} in an excellent review, there are many important observational problems. The first, which has now been overcome, is that the spectra must have sufficient resolution to resolve every line. The second is that most lines are very weak and the number of resolved lines can depend greatly on the signal to noise ratio. This is accentuated because the steep spectrum for the density of lines as a function of their strength means that a small decrease in the acceptance level can drastically increase the number of observed lines. The third problem is that each quasar only provides a set of lines in a narrow range of redshift and there are considerable difficulties in getting uniform cross-calibrations. In addition to these problems, it will be shown that curvature redshift can have a profound effect on the interpretation of the line widths and column densities.

Since in CC the distribution of clouds is independent of time or distance the expected density of lines  as a function of redshift is
\begin{equation}
\label{ce123}
\frac{dn}{dz} = \frac{AcN_0}{H(1 + z)},
\end{equation}
where $N_0$ is the volume density and $A$ is the average area of a cloud. Most observers have fitted a power law with the form $(1+z)^\gamma$  to the observed line densities with a wide range of results. They vary from  $\gamma=1.89$ to $\gamma =5.5$ \citep{Rauch98}. All of which are inconsistent with the CC prediction of  $\gamma=-1$.
In CC there is the additional effect that much of the line broadening may be due to curvature redshift. Curvature redshift will be operating within the clouds so that the observed line width will be a combination of the usual Voigt  profile and the change in the effective central frequency as the photons pass through the cloud. If the cloud has a density  $\rho(x)$ at the point $x$, measured along the photon trajectory then the change in frequency from the entering frequency due to curvature redshift is
\[
\frac{\Delta \nu}{\nu} = \frac{1}{c}\int {\sqrt{8\pi G\rho (x)}}dx.
\]
In units of $N(x)= \rho(x)/m_H$ this is (with $N$ in m$^{-3}$ and $dx$ in kpc)
\[
\frac{\Delta \nu}{\nu} = -\frac{\Delta \lambda}{\lambda} = \int{1.724 \times 10^{-7} \sqrt{N(x)} } dx.
\]
Then the final profile will be the combination of the natural line width, the Doppler width due to temperature, any width due to bulk motions and the curvature redshift width. Now assuming pure hydrogen, the hydrogen column density is given by $N_H  = \int {N(x)} dx$.
Although it is unlikely that the line of sight goes through the center of the cloud, it is reasonable to expect a roughly symmetric distribution of gas with a shape similar to a Gaussian. We can define an effective density width by
\[
x_w^2 = \int {\left({x-\overline x}\right)^2 N(x)}dx\left / \right . \int{N(x)}dx.
\]
Also define $N_0=N_H/x_w$ and an effective velocity width $\Delta v =51.68\eta x_w \sqrt{N_0 }$ and where $\eta$ is a small numeric constant that depends on the exact shape of the density distribution. Eliminating the central density, we get (with $x_w$ in kpc)
\begin{equation}
\label{ce128}
\Delta v^2 =8.656 \times 10^{-17} \eta^2 N_H x_w.
\end{equation}
For values $N_H=10^{19} \mbox{\,m}^{-2}$, $x_w$=1 kpc and with $\eta$=1 we get $\Delta v$=29 km\,s$^{-1}$. Since there is a wide variation in column densities and the effective widths are poorly known it is clear that curvature redshift could completely dominate many of the Lyman-$\alpha$  line widths and the others would require a convolution of the Doppler profile with the curvature  redshift density effect. What is also apparent is that the very broad absorption lines may be due to curvature redshift acting in very dense clouds. Although there is uncertainty about the observed relationship between the line width and the column density, we note that for a fixed effective density width, Eq.~\ref{ce128} predicts a square relationship that may be compared with the exponent of $2.1\pm 0.3$ found by \citet{Pettini90}. Clearly, there needs to be a complete re-evaluation of profile shapes, column-densities, and cloud statistics that allows for the effects of CC. We must await this analysis to see whether the Lyman-$\alpha$ forest can provide a critical test of CC.

\subsection{The Gunn--Peterson trough in high redshift quasars}
\label{s5.9}
The Gunn--Peterson trough is a feature of the spectra of quasars probably due to the presence of neutral hydrogen in the intergalactic medium. The trough is characterized by suppression of electromagnetic emission from the quasar at wavelengths less than that of the Lyman-$\alpha$  line at the redshift  of the emitted light. This effect was originally predicted by \citet{Gunn65}. Although the Gunn--Peterson-trough has now been seen in several high redshift quasars \citet{Becker01}, \citet{Peng05}, \citet{White05} it is not seen in all quasars and appears to be strongly dependent on redshift. In BB the explanation  is that it is only seen in very high redshift quasars where the intergalactic medium is still neutral but it is not seen in closer quasars because the medium has been re-ionized. In CC it has a more prosaic explanation.

We assume that the quasar is surrounded by a large halo or it lies within a cluster of galaxies which has, like many clusters, an internal gas cloud. The hypothesis is that the halo or gas cloud is cool enough to have a small but important density of neutral hydrogen that absorbs much of the quasar emission at shorter wavelengths than the Lyman-$\alpha$ emission. The important point is that as the radiation traverses the cloud it is redshifted by curvature redshift  due to the density of the cloud. The Lyman-$\alpha$  radiation from the quasar is redshifted by the curvature redshift due to the density of the whole cloud. The absorption lines are only shifted by part of the cloud and appear at a shorter wavelength than the quasar Lyman-$\alpha$ emission. Consider the scattering probability of a photon with wavelength $\lambda$ by the neutral hydrogen. The differential probability is
\[
dp = \pi c f_H r_0 f\int_{\nu_\alpha}^\nu {N\left(r\right)}
g\left({\nu-\nu \alpha} \right)dr,
\]
where $r_0$ is the classical electron radius, $f_H$ is the fraction of neutral hydrogen atoms, $f$ is the oscillator strength (here $f$ = 0.416), ${N(r)}$ is the gas density and $g$ is the profile function that is strongly peaked about $\nu=\nu_\alpha$. At a distance $r$ inside the cloud $\nu =\nu _q \exp \left({- ar}\right)$  where $\nu_q)$ is its frequency at the beginning of the cloud and
\[
a = \sqrt{\frac{8\pi GM_H n(r)}{c^2}}=5.588\times 10^{-27}
\sqrt{n \left( r\right)} \mbox{\,m}^{-1}.
\]
A change of variable from $r$ to $\nu$ and noting that the width of $g(\nu)$ is small compared to the value of $\nu_\alpha$ and assuming a constant density $N=N(r)$ results in the approximation
\[
p \approx \frac{\pi r_0 \lambda _\alpha  f_H \sqrt{N}}{a}
= 8.01 \times 10^4 f_H \sqrt{N}.
\]
Clearly all the photons are scattered for quite low densities and a very small density of hydrogen atoms.

Having shown that total scattering is feasible the next step is to see why the effect is more pronounced for high redshift quasars. The basic equation for curvature redshift is $1+z=\exp(cr/H)$ where $r$ is the distance, $H$ is the Hubble constant and $dz=a(1+z)dr$. Thus the wavelength range that is observed over which the Gunn--Peterson effect is seen should scale as ($1+z$). Thus the probability of detection which depends on the wavelength range will be lower at smaller redshifts.  Naturally the effect may not be seen or could be partially obscured depending on the individual characteristics of the cloud around each quasar.

\citet{Becker01} report details of the quasar SDSS 1030+0524 which has an emission redshift of 6.29 and a Gunn--Peterson depression that has a width $dz=0.21$. With this value we find that it can be explained by a cloud that has the constraint $N^{1/2}dr = 167$ where $dr$ is in Mpc. With $dr=2$ Mpc the required density is $7 \times 10^3 \mbox{\,m}^{-3} $ which is a feasible size and density for a large cluster cloud. Note that if the gas has high enough temperature $f_H$ will be too small and no effect will be seen.

\subsection{Nuclear abundances}
\label{s5.10}
One of the successes of BB is in its explanation of the primordial abundances of the light elements. Since the proposed CC is static, there must be another method of getting the `primordial' abundances of light elements. In CC, the primordial abundance refers to the abundance in the cosmic gas from which the galaxies are formed.
The first point to note is that in CC the predicted temperature of the cosmic gas is $2.56 \times 10^9 {\mbox{ K}}$
at which temperature nuclear reactions can proceed. The major difference with the production of helium and deuterium in the BB early universe is that the densities were incredibly higher in BB than they are in CC. It is postulated that in CC there is a continuous recycling of material from the cosmic gas to galaxies and stars and then back to the gas. Because of the high temperature, nuclear reactions will take place whereby the more complex nuclei are broken down to hydrogen, deuterium, and helium. Although this cycling can take many billions of years the very low density of the gas means that the cycle time may not be long enough for the nuclei densities to achieve statistical equilibrium. In addition, the major reactions required are the breaking down of heavier nuclei to lighter ones and not those that construct nuclei. It is through the interactions of cosmic gas in CC that the light nuclei abundances are produced.

\subsection{Galactic rotation curves}
\label{s5.11}
One of the most puzzling questions in astronomy is: why does the observed velocity of rotation in spiral galaxies not go to zero towards the edge of the galaxy. Simple Keplerian mechanics suggests that there should be a rapid rise to a maximum and then a decrease in velocity that is inversely proportional to the square root of the radius once nearly all the mass has been passed. Although the details vary between galaxies, the observations typically show a rapid rise and then an essentially constant tangential velocity as a function of radius out to distances where the velocity cannot be measured due to lack of material. The BB explanation is that this is due to the gravitational attraction of a halo of dark matter  that extends well beyond the galaxy. We examine whether this rotation curve can be explained by curvature redshift.

Observations show that our own Galaxy and other spiral galaxies have a gas halo that is larger than the main concentration of stars. It is clear that if the observed redshifts are due to curvature redshift acting within this halo, the halo must be asymmetric; otherwise, it could not produce the asymmetric rotation curve. Now the observed velocities in the flat part of the curves are typically 100 to 200 km\,s$^{-1}$. The first step is to see if curvature redshift provides the right magnitude for the velocity. For a gas with an average density of $N_H$  the predicted redshift (in velocity units) is $5.17 \times10^{-2} d\sqrt{N}\mbox{\,km\,s}^{-1}$ where $d$ is the distance in kpc. For realistic values of $d=10\,$kpc and $N=1.0\times 10^5$ m$^{-3}$ the velocity is 163 km\,s$^{-1}$. Thus, the magnitude is feasible.
Although there could be a natural asymmetry in a particular galaxy, the fact that the flattened rotation curve is seen for most spiral galaxies suggests that there is a common cause for the asymmetry. One possibility is that the asymmetry could arise from ram pressure. Since most galaxies are moving relative to the cosmic medium, it is expected that there will be an enhanced density towards the leading point of the galaxy. This asymmetric density could produce an apparent velocity gradient across the galaxy that could explain the apparent rotation curve. Naturally, there would be range of orientations and the apparent velocity gradient must be added to any intrinsic rotation curve to produce a wide diversity of results. Thus, curvature redshift could explain the galactic rotation curves if there is an asymmetric distribution of material in the galactic halo.
Both cosmologies have problems with galactic rotation curves. BB not only requires dark matter but does not have any definite models for its distribution. Curvature cosmology has the problem of achieving sufficient asymmetry to mimic a rotation curve.

\subsection{Redshifts in our Galaxy}
\label{s5.12}
In our Galaxy, the Milky Way, there is an interesting prediction. The density of the inter-stellar ionized gas is high enough to inhibit curvature redshift for radio frequencies. From  Eq.~\ref{ce25} it was shown that for wavelengths longer than about 20.6$N^{-1/2}$ m the effect of refractive index in fully ionized plasma will inhibit curvature redshift. The refractive index of neutral hydrogen is too low to inhibit curvature redshift. However, any fully ionized plasma with $N>10^4$m$^{-3}$ will inhibit curvature redshift for the 21 cm hydrogen line. Since the local interstellar medium has an electron density of about $10^5$ m$^{-3}$ \citep{Redfield06}, curvature redshift will be inhibited for the 21 cm hydrogen in regions of the galaxy near the sun. Thus for sight lines close to the Galactic plane we can assume a similar density and thus a similar inhibition with the result that the observed radio redshifts can be correctly interpreted as velocities. Thus, there is little change needed to the current picture of Galactic structure and rotation derived from 21 cm redshifts. However, there may be some curvature redshift present in sight lines away from the plane and especially in the Galactic halo.

Since optical redshifts have the full effects of curvature redshift, it should be possible to find objects with discrepant redshifts where the optical redshift is greater than the radio redshift. The difficulty is that the two types of radiation are produced in radically different environments: the optical in compact high temperature objects, such as stars, and the radio in very low-density cold clouds. In addition, there is the complication that within the galactic plane, optical extinction due to dust limits the optical range to about one kpc.

Curvature redshift may help to explain an old stellar mystery. There is a long history provided by \citet{Arp92} of observations of anomalous redshifts in bright hot stars, which is called the K-term  or K-effect. \citet{Allen76} states that B$_0$ stars typically show an excess redshift of 5.1 m\,s$^{-1}$, A$_0$ have 1.4 km\,s$^{-1}$ and F$_0$ have 0.3 km\,s$^{-1}$. This can be explained if these stars have a large corona that produces a curvature redshift.  It is probably no coincidence that such stars have large stellar winds and mass outflows.  In order to see if it is feasible let us consider a simple model for the outflow in which the material has a constant velocity $v_0$, and conservation of matter (Gauss's Law) then requires that the density has inverse square law dependence. Although this is incorrect at small stellar radii, it is a reasonable approximation further from the star. Then if $\rho_1$ is the density at some inner radius $r_1$, then integration of  Eq.~\ref{cpe10} out to a radius $r_2$, the expected redshift in velocity units is
\[
v = \sqrt{\frac{2G\dot M}{v_o}} \log \left( \frac{r_2}{r_1} \right),
\]
where $\dot M$ is the observed stellar mass-loss-rate. Then with $\dot M$
in solar masses per year, with $v$ and $v_0$ in km\,s$^{-1}$, the redshift is
\[
v =91.7\sqrt\frac{\dot M}{v_o} \log \left( \frac{r_2}{r_1}\right)
\mbox{\,km\,s}^{-1},
\]
With $\dot M = 10^{-5} M_\odot\mbox{\,yr}^{-1}$ \citet{Cassinelli79}, $v_0= 1\,$km\,s$^{-1}$ and $r2/r1=10^3$ the predicted redshift (in velocity units) is 2 km\,s$^{-1}$ which is in reasonable agreement with the observed K-effects mentioned above.
\subsection{Anomalous redshifts}
\label{s5.13}
\citet{Arp87, Ratcliffe10} have argued that there is strong observational evidence for anomalous redshifts between quasars and galaxies. Typically there is a quasar very close to a galaxy with a material bridge or other evidence that suggests that they are associated. \citet{Chu98} report on five X-ray emitting blue stellar objects located less than 12\arcmin~  from the X-ray Seyfert galaxy NGC 3516. In this case the association is that the objects lie close to a straight line on either side of the galaxy and that their redshifts are proportional to $\log(\theta)$ where $\theta$ is the angular distance from the central galaxy. Furthermore the line of objects is within a few degrees of the minor axis of NGC 3516. The measured redshifts are 0.33, 0.69, 0.93, 1.4 and 2.1. NGC 3516 is a barred spiral galaxy and it has a redshift of 0.00884.

Can CC explain this redshift anomaly? If the objects are seen through  a large dense cloud, such as a galactic halo, then curvature redshift will produce an extra redshift due to the photons passage through the cloud.  the extra redshift, $\delta$, is
\[
\delta = 1.72\times10^{-10}\int\sqrt{N(x)}\,dx,
\]
where $N(x)$ is the number density and distances are measured in pc. If $z$ is the cosmological redshift then the extra-observed redshift is $\Delta z = (1+z)(e^\delta -1)$.
In order to achieve an extra redshift $\delta \approx 1$ with a distance of $10^4$ pc the gas number density must be about $3\times10^{11}$ m$^{-3}$. Now although cold interstellar molecular clouds can have densities reaching this value it is still a very high density. But if the size is increased by a factor of two the required density is decreased by a factor of four. Moreover the objects with the largest redshifts are the furthermost away from the galaxy. These redshifts could be explained by curvature  redshift in a very large, very dense galactic halo with a hole in the middle. Since NGC 3516 has a very low redshift and is seen nearly face on the implication is that this gas cloud is probably shaped like a torus and it lies in the galactic plane of NGC3516. A further test is to compare an estimate of the mass of this torus with that for a typical galaxy. Since a torus formed by the rotation of a circle with radius  $r$ about a axis in the plane of the circle where the radius of rotation is $R$, its volume is $V=2\pi^2Rr^2$. With $R$ and $r$ in kpc and an average density of $N$ its mass is $M=0.484Rr^2N\,M_{\sun}$. Then with $R=15\,$kpc, $r=10\,$kpc and $N=3\times10^{11}$ the mass is
$2\times10^{14}M_{\sun}$ which considerably larger than a normal galaxy.
Since these anomalous redshifts are completely outside any BB model the only reason that these observations are not fatal to BB is their the controversial nature.

\subsection{Voids}
\label{s5.14}
If CC is valid then the redshift of the galaxies in the Coma cluster (Section~\ref{s5.5}) will have been increased, on average, by the additional redshift due to the inter-galactic gas. Thus, they will have, on average, a larger redshift than an isolated galaxy at the same distance. Table~\ref{vt1} shows the predicted (effective) velocity for a galaxy in the center plane of the Coma cluster as a function of the projected radius. The second column is the velocity at that exact radius and the third column shows the average velocity of galaxies (uniformly spread in area) within that radius. This simulation also showed that the average velocity offset for the galaxies in the Coma cluster is 1206 kms$^{-1}$ which means that the redshift of the center of the Coma cluster is 6926-1206=5720 kms$^{-1}$. This offset is important for calculating the Hubble constant  which from these figures is 5270/87.1=65.7 kms$^{-1}$\,Mpc$^{-1}$.

\begin{table}
\begin{center}
\caption{Velocity at, and average velocity within various projected radii in the Coma cluster (distance = 87.1 Mpc).\label{vt1}}
\begin{tabular}{lrr}
\hline
Radius\tablenotemark{a} &  Velocity    & Mean velocity\\
/Mpc & /km\,s$^{-1}$ &  /km\,s$^{-1}$\\
\hline
0.0 &   2327.7 &    2327.7\\
0.5 &   1477.7 &    1764.8\\
1.0 &   1033.4 &    1342.5\\
1.5 &   803.3  &    1096.9\\
2.0 &   658.6  &    933.2\\
2.5 &   557.0  &    814.4\\
3.0 &   481.0  &    723.3\\
3.5 &   421.7  &    650.7\\
4.0 &   374.0  &    541.2\\
4.5 &   334.8  &    541.2\\
5.0 &   302.0  &    498.7\\
\hline
\end{tabular}
\end{center}
\tablenotetext{a}{projected radius}
\end{table}

In addition, the redshift of objects seen through a cluster will be increased by curvature-redshift from the inter-galactic gas. \citet*{Karoji76} claim to have seen this effect. They examined radio galaxies  and classified them into region A if their light does not pass through a cluster and region B if their light passes through a cluster. They found no significant differences in magnitudes between the two regions but they did find a significant difference in the average redshifts that was consistent over the complete range. Their result is that radio galaxies seen through a cluster had an average extra redshift (in velocity units) of 2412$\pm$1327 km\,s$^{-1}$. Overall the difference in the distance modulus was  $\mu=0.16\pm0.04$, which is just significant. Since the density and distribution of the gas in the clusters is unknown and the limiting radius of the cluster is not stated it is impossible to get an accurate prediction. Nevertheless, we note that for the Coma cluster with a radius of 2 Mpc the average extra redshift (from Table~\ref{vt1} with a factor of two) corresponds to 1866 km\,s$^{-1}$ showing that curvature-cosmology could explain the effect. In a different study, \citet{Nottale76} and \citet{Nottale77} compared the magnitude of the brightest galaxy in a cluster with that in another cluster with similar redshift. They found that there was no significant difference in magnitudes between clusters  but that the clusters with the largest number of galaxies had the higher redshift difference between the pairs. On average the redshift difference (in velocity units) was 292$\pm$85 km\,s$^{-1}$. This can be explained by the expected correlation between number of galaxies and size and density of the inter-galactic gas. However it should be noted that these observations have been disputed by \citet{Rood82}.

In his review of voids in the distribution of galaxies, \citet{Rood88} quotes \citet{Mayall60} who observed a large void in the distribution of galaxies in front of the Coma cluster.  This void has a magnitude of about 3000 kms$^{-1}$, which although somewhat larger, is not inconsistent with the expected value of about 1200 km\,s$^{-1}$. In other words, the Coma cluster galaxies have an extra curvature-redshift due to the inter-galactic gas. However, the galaxies just outside the cluster nearer to us do not have this extra redshift and would appear to be closer to us. Hence, we see an apparent void in the redshift distribution in front of the Coma cluster.

A consequence of gas clouds and curvature-redshift is that the distribution of redshifts is similar to but not identical to the distribution of $z$ distances. Galaxies that are behind a cloud will have a higher redshift than would be expected from a simple redshift distance relationship. Thus, we would expect to see anomalous voids and enhancements in the redshift distribution. This will be accentuated if the gas clouds have a higher than average density of galaxies. \citet*{deLapparent86} show a redshift plot for a region of the sky that includes the Coma cluster. Their data are from the Center for Astrophysics redshift survey and their plot clearly shows several voids. They suggest that the galaxies are distributed on the surfaces of shells. However, this distribution could also arise from the effects of curvature-redshift in clouds of gas.

\section{Curvature Cosmology Theory}
\label{s6}
Curvature cosmology (CC) is a static tired-light cosmology  where the Hubble redshift (and many other redshifts) is produced by an interaction of photons with curved spacetime called  curvature redshift. It is a static solution to the equation of general relativity that is described by the Friedmann equations with an additional term that stabilizes the  solution. This term called curvature pressure is a reaction of high speed particles back on the material producing the curved spacetime. This sense of this reaction is to try and reduce the curvature.
The basic cosmological model is one in which the cosmic gas dominates the mass distribution and hence the curvature of spacetime. In this first order model, the gravitational effects of galaxies are neglected. The geometry of this CC is that of a three-dimensional surface of a four-dimensional hypersphere. It is almost identical to that for Einstein's static universe. For a static universe, there is no ambiguity in the definition of distances and times. One can use a universal cosmic time and define distances in light travel times or any other convenient measure.  In a statistical sense CC obeys the perfect cosmological principle of being the same at all places and at all times.

CC makes quite specific predictions that can be refuted. Thus, any observations that unambiguously show changes in the universe with redshift would invalidate CC. In CC, there is a continuous process in which some of the cosmic gas will aggregate to form galaxies and then stars. The galaxies and stars will evolve and eventually all their material will be returned to the cosmic gas. Thus, a characteristic of CC is that although individual galaxies will be born, live and die, the overall population will be statistically the same for any observable characteristic.

This paper is the culmination of many years of work and is a complete re-synthesis of many approaches that I have already published. Because hypotheses and notations have changed and evolved, direct references to these earlier versions of the theory would be misleading. Table~\ref{tcc1} (all with author D. F. Crawford) is provided briefly stating each reference and the major topic in each paper. In nearly all cases, the data analyzed in the papers has been superseded by the more recent data that are analyzed in this paper.

\begin{table}
\begin{center}
\caption{Published papers\label{tcc1}}
\begin{tabular}{lll}
\hline
Year & Reference          & Major topic \\ \hline
1975 &  Nature, 254, 313  & First mention of photon extent and gravity     \\
1979 &  \nat, 277, 633    & Photon decay near the sun: limb effect\tablenotemark{a}   \\
1987 & \ajp 40, 440       & First mention of curvature redshift\tablenotemark{b}  \\
1987 & \ajp, 40, 459      & Application to background X-rays      \\
1991 & \apj, 377, 1       & More on curvature redshift and applications \\
1993 & \apj, 410, 488     & A static stable universe: Newtonian cosmology     \\
1995 & \apj, 440, 466     & Angular size of radio sources   \\
1995 & \apj, 441, 488     & Quasar distribution  \\
1999 & \ajp, 52, 753      & Curvature pressure and many other topics \\
2006 & Book \citep{Crawford06} & {"Curvature Cosmology"}  \\
2008 & Web site \tablenotemark{c} & Major update\tablenotemark{d} of the book  \\
\hline
\end{tabular}
\end{center}
\tablenotetext{a}{Not only is the theory discredited but also the observations
have not stood the test of time.}
\tablenotetext{b}{This gives the equation for photons but not for non-zero rest  mass particles.}
\tablenotetext{c}{http://www.davidcrawford.bigpondhosting.com}
\tablenotetext{d}{Superseded by this paper.}
\end{table}

\subsection{Derivation of curvature redshift}
\label{s6.1}
The derivation of curvature redshift is based on the fundamental hypothesis of Einstein's general theory of relativity that spacetime is curved. As a consequence, the trajectories of initially-parallel point particles, geodesics, will move closer to each other as time increases. Consequently in space with a positive curvature, the cross sectional area of a bundle of geodesics  will slowly decrease. In applying this idea to photons, we assume that a photon is described in quantum mechanics as a localized wave where the geodesics correspond to the rays of the wave. Note that this wave is quite separate from an electromagnetic wave that corresponds to the effects of many photons. It is fundamental to the hypothesis that we can consider the motion in spacetime of individual photons. Because the curvature of spacetime causes the focussing of a bundle of geodesics, this focussing also applies to a wave. As the photon progresses, the cross sectional area of the wave associated with it will decrease. However, in quantum mechanics properties such as angular momentum are computed by an integration of a radial coordinate over the volume of the wave. If the cross sectional area of the wave decreases, then the angular momentum will also decrease. However, angular momentum is a quantized parameter that has a fixed value. The solution to this dilemma is that the photon  splits into two very low-energy photons  and a third that has the same direction as the original photon and nearly all the energy. It is convenient to consider the interaction as a primary photon losing a small amount of energy into two secondary photons. Averaged over many photons this energy loss will be perceived as a small decrease in frequency.
Since in quantum mechanics electrons and other particles are considered as waves, a similar process will also apply. It is argued that electrons will interact with curved spacetime to lose energy by the emission of very low-energy photons.

\subsubsection{Photons in Curved Spacetime}
\label{s6.1.1}
Einstein's general theory of relativity requires that the metric of spacetime be determined by the distribution of mass (and energy). In general this spacetime will be curved such that in a space of positive curvature nearby geodesics that are initially parallel will come closer together as the reference position moves along them. This is directly analogous to the fact that on the earth lines of longitude come closer together as they go from the equator to either pole. In flat spacetime, the separation remains constant. For simplicity, let us consider geodesics in a plane. Then the {\em equation for geodesic deviation} can be written \citet*{Misner73}, p 30 as
\[
\frac{{d^2 \xi }}{{ds^2 }} = -\frac{\xi}{{a^2}},
\]
where $\xi$ is normal to the trajectory and $s$ is measured along the trajectory. The quantity $1/a^2$ is the Gaussian curvature at the point of consideration. For a surface with constant curvature, that is the surface of a sphere, the equation is easily integrated to get (ignoring a linear term) $\xi  = \xi _0 \cos(s/a)$.
Note that this equation also describes the separation of lines of longitude as we move from the equator to either pole. Now geodesics describe the trajectories of point particles. Null-geodesics are associated with mass-less particles. However, photons are not point particles. The experiment of using single photons in a two-slit interferometer shows that individual photons must have a finite size.
Quantum mechanics requires that all particles are described by wave functions and therefore we must consider the propagation of a wave in spacetime. Because photons are bosons, the usual quantum mechanical approach is to describe the properties of photons by creation and destruction operators. The emphasis of this approach is on the production and absorption of photons with little regard to their properties as free particles. Indeed because photons travel at the speed of light, their lifetime in their own reference frame between creation and destruction is zero. However, in any other reference frames they behave like normal particles with definite trajectories and lifetimes. \citet{Havas66} has pointed out that the concept of a single photon is rather tenuous. There is no way we can tell the difference between a single photon and a bundle of photons with the same energy, momentum, and spin. However, it is an essential part of this derivation that a single photon  has an actual existence.

Assume that a photon can be described by a localized wave packet  that has finite extent both along and normal to its trajectory. This economic description is sufficient for the following derivation. We define the frequency of a photon as $\nu  = E/h$  and its wavelength as $\lambda = hc/E$ where $E$ is its energy. These definitions are for convenience and do not imply that we can ascribe a frequency or a wavelength to an individual photon; they are properties of groups of photons. The derivation requires that the wavelength is short compared to the size of the wave packet and that this is short compared to variations in the curvature of spacetime. Furthermore, we assume that the rays of any wave follow null geodesics and therefore any deviations from flat spacetime produce change in shape of the wave packet. In other words, since the scale length of deviations from flat space are large compared to the size of the wave packet they act as a very small perturbation to the propagation of the wave packet.

Consider a wave packet moving through a spacetime of constant positive curvature. Because of geodesic deviation, the rays come closer together as the wave packet moves forward. They are focussed. In particular the direction $\theta$, of a ray (geodesic) with initial separation $\xi _0 $ after a distance $s$ is (assuming small angles)
\[
\label{ce3}\theta = -\frac{s\xi_0}{a^2},
\]
where $a$ is the local radius of curvature. Since the central geodesic is the direction of energy flow, we can integrate the wave-energy-function times the component of $\theta$ normal to the trajectory, over the dimensions of the wave packet in order to calculate the amount of energy that is now travelling normal to the trajectory. The result is a finite energy that depends on the average lateral extension of the wave packet, the local radius of curvature, and the original photon energy. The actual value is not important but rather the fact that there is a finite fraction of the energy that is moving away from the trajectory of the original wave packet. This suggests a photon interaction in which the photon interacts with curved spacetime with the hypothesis that the energy flow normal to the trajectory goes into the emission of secondary photons normal to its trajectory. From a quantum-mechanical point of view, there is a strong argument that some interaction must take place. If the spin of the photon is directly related to the angular momentum of the wave packet about its trajectory then the computation of the angular momentum is a similar integral. Then because of {\em focussing} the angular momentum clearly changes along the trajectory, which disagrees with the quantum requirement that the angular momentum, that is the spin, of the photon is constant. The Heisenberg uncertainty principle requires that an incorrect value of spin can only be tolerated for a finite time before something happens to restore the correct value. We now consider the consequences.

Consider motion on the surface of a three dimensional sphere with radius $r$. As described above, two adjacent geodesics will move closer together due to focussing. Simple kinematics tells us that a body with velocity $v$  associated with these geodesics has acceleration  $v^2/r$, where $r$ is the radius of curvature. This acceleration is directly experienced by the body. In addition, it experiences a tidal acceleration within itself. This tidal acceleration is equivalent to the focussing of the geodesics. Although the focussing and acceleration are closely linked, we need to consider whether the occurrence of one implies the occurrence of the other. Does the observation of focussing (tidal acceleration) imply acceleration in the orthogonal direction? It is true in two and three dimensions, but it needs to be demonstrated for four dimensions.

The geometry of a three dimensional surface with curvature in the fourth dimension is essentially the same as motion in three dimensions except that the focussing now applies to the cross-sectional area and not to the separation. Does this acceleration have the same physical significance? Assuming it does, a wave packet that is subject to focussing has acceleration in an orthogonal dimension. For instance if we could constrain a wave packet (with velocity $c$) to travel on the surface of a sphere in three dimensions it would not only show a focussing effect but also experience an acceleration of $c^2/r$ normal to the surface of the sphere. Then a wave packet (and hence a photon) that has its cross-sectional area focussed by curvature in the fourth dimension with radius $r$ would have an energy loss rate proportional to this acceleration. The essence of the curvature-redshift hypothesis is that the tidal distortion causes the photon to interact and that the energy loss rate is proportional to $c^2/r$. For a photon with energy $E$ the loss rate per unit time is $cE/r$, and per unit distance it is $E/r$.

In general relativity the crucial equation for the focussing of a bundle of geodesics was derived by \citet{Raychaudhuri55}, also see \citet{Misner73} and  \citet{Ellis84} and for the current context we can assume that the bundle has zero shear and zero vorticity. Since any change in geodesic deviation along the trajectory will not alter the direction of the geodesics we need consider only the cross-sectional area $A$ of the geodesic bundle to get the equation
\[
\frac{1}{A}\frac{d^2 A}{ds^2}=-{\bf R }_{\alpha\beta}{\bf U }^{\alpha}{\bf U }^{\beta} =-\frac{1}{a^2},
\]
where $\bf R$  is the Ricci tensor (it is the contraction of the Riemann-Christoffel tensor), $\bf U$  is the 4-velocity of the reference geodesic and $a$ is the local radius of curvature. This focussing can be interpreted as the second order rate of change of cross-sectional area of a geodesic bundle that is on the three-dimensional surface in four-dimensional space. Then if we consider that a photon is a wave packet we find that the rate at which the photon  loses energy per unit distance is $E/a$ or more explicitly
\[
\frac{1}{E}\frac{dE}{ds}=-\frac{1}{a} =-\left({{\bf R }_{\alpha\beta} {\bf U }^\alpha  {\bf U }^\beta}\right)^{1/2},
\]
What is interesting about this equation is that, for the Schwarzschild (and Kerr) solutions for the external field for a mass, the Ricci tensor is zero; hence, there is no focussing and no energy loss. A geodesic bundle passing a mass such as the sun experiences a distortion but the wave packet has not changed in area. Hence, this model predicts that photons  passing near the limb of the sun will not suffer any energy loss due to curvature redshift.

The field equation for Einstein's general theory of gravitation is
\[
{\bf R}_{\alpha \beta}=8\pi G\left({{\bf T}_{\alpha \beta}-\frac{1}{2}  T{\bf g}_{\alpha \beta}} \right)+\Lambda {\bf g}_{\alpha \beta },
\]
where $ T$ is the contracted form of ${\bf T}_{\alpha \beta}$   the stress-energy-momentum tensor, ${\bf g}$   is the metric tensor, $G$ is the Newtonian gravitational constant and $\Lambda$ is the cosmological constant. It states that the Ricci tensor describing the curvature of spacetime is determined by the distribution of mass (and energy). Direct application of the field equations (without the cosmological constant) in terms of the stress-energy-momentum tensor ${\bf T}_{\alpha \beta}$, the metric tensor ${\bf g}$ and with the material having a 4-velocity ${\bf V}$  gives
\begin{equation}
\label{ce5}
\frac{1}{a^2}=8\pi G\left({{\bf T}_{\alpha \beta}{\bf U}^\alpha {\bf U}^\beta - \frac{1}{2}  T {\bf g}_{\alpha\beta} {\bf V}^\alpha {\bf V}^\beta} \right).
\end{equation}

For null geodesics ${\bf g}_{\alpha\beta} {\bf V}^\alpha {\bf V}^\beta$  is zero which leaves only the first term. For a perfect fluid the stress-energy-momentum tensor is
\begin{equation}
\label{ce6}
{\bf T}_{\alpha\beta}=\frac{p}
{{c^2}}{\bf g}_{\alpha \beta}+\left({\rho+\frac{p}{c^2}}\right){\bf U}_\alpha {\bf U}_\beta,
\end{equation}
where $p$ is the proper pressure and $\rho$ is the density. Combining Eq.~\ref{ce5} with Eq.~\ref{ce6} gives for null geodesics
\[
\frac{1}{a^2} = \frac{8\pi G}{c^2}\left({\rho+\frac{p}{c^2}}\right).
\]

For cases where the proper pressure is negligible compared to the density we can ignore the pressure and get
\begin{eqnarray*}
\label{ce7}
\frac{1}{E}\frac{dE}{ds}&=&-\frac{1}{a}=-\left(\frac{8\pi G\rho}{c^2} \right)^{1/2}\nonumber\\
&=&-1.366\times 10^{-13}\sqrt{\rho}\,\mbox{m}^{-1}.
\end{eqnarray*}

For many astrophysical types of plasma, it is useful to measure density by the equivalent number of hydrogen atoms per cubic metrae: that is we can put  $\rho=Nm_H$ and get
\begin{equation}
\label{ce8}
\frac{1}{E}\frac{dE}{ds} =-\left(\frac{8\pi GNM_H}{c^2}\right)=-5.588\times10^{-27}\sqrt{N}\, \mbox{m}^{-1}.
\end{equation}
The rate of energy loss per distance travelled depends only on the square root of the density of the material, which may consist of gas, plasma, or gas and dust.
This equation can be integrated to get
\begin{equation}
\label{ce9}
\ln(E/E_0) =\left(\frac{8\pi GM_{\rm H}}{c^2} \right)^{1/2}\int_0^x\sqrt{N(x)}dx.
\end{equation}

\subsubsection{Curvature redshift secondary photons}
\label{s6.1.2}
The above derivation does not define the form of energy loss. The most realistic model is that the photon decays into three secondary photons, one of which takes nearly all the energy and momentum and two very low-energy secondary photons. It is convenient (although not strictly correct) to think of the high-energy secondary as a continuation of the primary but with slightly reduced energy. Two secondary photons are required to preserve spin and, by symmetry, they are emitted in opposite directions with the same energy {This assumption that the two secondary photons have the same energy is made without proper justification. What can be said is that if they are not, they will still have nearly equal energies because the probability of having one with a much longer relative wavelength is very low}.  From symmetry they are ejected at right angles to the original trajectory. Thus, the primary photon is not deflected. We can get an estimate of how often these interactions occur and hence what the secondary energies are by using the Heisenberg uncertainty principle applied to the primary. For linear momentum and distance it is $\Delta p\Delta x \cong h /4\pi$,  and putting $X = \Delta x$ we get $\Delta E = hc /4 \pi X $. Now after the photon with energy $E_0$ has travelled a distance $X$ the energy-loss is $\Delta E=E_0 X/a$, and hence
\begin{equation}
\label{ce12}
X^2 = \frac{{ahc}}{4\pi E_0 } = \frac{a\lambda_0}{4\pi}=
\frac{c\lambda _0 }
{4\pi \sqrt{8\pi G\rho}}.
\end{equation}
If each secondary photon takes half the energy-loss, we find
\begin{equation}
\label{ce13}
\Delta E = \frac{1}
{2}\frac{{E_0 X}}{a}.
\end{equation}
Therefore the secondary photons  have a wavelength of
\begin{equation}
\label{ce14}
\lambda  = \frac{2 a \lambda_0}{X}
= 8\pi X{\mbox{  }}=4\sqrt{\pi a\lambda_0 }.
\end{equation}
For example consider a visible photon with wavelength 600 nm travelling in gas with density $N$, then $X = 2.93 \times 10^9 N^{-1/4} \mbox{\,m}$  and the wavelength is $\lambda  = 7.36 \times 10^{10} N^{-1/4}\,$m
which corresponds to a frequency of $\nu  = 4.07N^{1/4} \mbox{\,mHz}$
Now for fully ionized plasma the plasma frequency is
\[
\nu_p = \left(\frac{Ne^2}{\pi m_e}\right)^{1/2}=8.975N^{1/2}\mbox{\,Hz},
\]
and the ratio is
\[
\frac{\nu}{\nu_p} = 4.55 \times 10^{-4}N^{-1/4}.
\]
Thus, for optical photons and all plasmas with densities greater than $N=0.14$\,m$^{-3}$  the secondary photons have frequencies well below the plasma frequency  and therefore cannot propagate but will be quickly absorbed by the plasma. The energy lost by the primary photon is dissipated into heating the plasma.

\subsubsection{Inhibition of curvature redshift}
\label{s6.1.3}
From the discussion above it is clear that the process of curvature redshift requires a gradual focussing to a critical limit, followed by the emission of secondary photons. It is as if the photon gets slowly excited by the focussing until the probability of secondary emission becomes large enough for it to occur. If there is any other interaction the excitation due to focussing will be nullified. That is, roughly speaking, curvature-redshift interaction requires an undisturbed path length of at least $X$ (Eq.~\ref{ce12}) for significant energy loss to occur. A suitable criterion for inhibition to occur is that the competing interaction has an interaction length less than $X$. Although Compton or Thompson scattering are possible inhibitors there is another interaction that has a much larger cross-section. This is the coherent multiple scattering that produces refractive index.

In classical electro-magnetic theory, the refractive index of a medium is the ratio of the velocity of light in vacuum to the group velocity in the medium. However, in quantum mechanics photons always travel at the velocity of light in vacuum. In a medium, a group of photons   appears to have a slower velocity because the individual photons interact with the electrons in the medium and each interaction produces a time delay. Because the interaction is with many electrons spread over a finite volume, the only possible result of each interaction is the emission of another photon with the same energy and momentum. Now consider the absorption of a wave. In order to cancel the incoming wave a new wave with the same frequency and amplitude but with opposite phase must be produced. Thus, the outgoing wave will be delayed by half a period with respect to the incoming wave. For example if the phase difference was not exactly half a period for an electro-magnetic wave incident on many electrons, the principle of conservation of energy would be violated. This simple observation enables us to compute the interaction length for refractive index $n$. If $L$ is this interaction length then it is
\[
L = \frac{\lambda _0}{2\left| {n - 1} \right|},
\]
where $n$ is the refractive index and the modulus allows for plasma and other materials where the refractive index is less than unity. Note that $L$ is closely related to the extinction length  derived by Ewald and Oseen (see \citep{Jackson75} or \citet{Born99}) which is a measure of the distance needed for an incident electromagnetic wave with velocity $c$ to be replaced by a new wave. For plasmas the refractive index is
\[
n \cong 1 - \frac{{N_e \lambda _0^2 }}
{{2\pi r_0 }},
\]
where $N_e$ is the electron density and $r_0$ is the classical electron radius. We can combine these two equations to get (for a plasma)
\begin{equation}
\label{ce24}
L = (N_e r_0 \lambda_0 )^{-1}.
\end{equation}
Thus, we would expect the energy loss to be inhibited if the average curvature-redshift interaction distance is greater than that for refractive-index interactions, i.e. if $X>L$. Therefore, we can compute the ratio (assuming a plasma with $N\cong N_e$) and using Eq.~\ref{ce12} to get
\begin{equation}
\label{ce25}
{X/L} = 0.0106N^{3/4}\lambda_0^{3/2}
\end{equation}
This result shows that curvature redshift will be inhibited if this ratio is greater than one, which is equivalent to $\lambda_0\ge20.7N^{-1/2}$ m. For example, curvature redshift for the 21 cm hydrogen line will be inhibited if the electron density is greater than about $10^4 \mbox{\,m}^{-3}$.

\subsubsection{Possible laboratory tests}
\label{s6.1.4}
It is apparent from the above analysis that to observe the redshift in the laboratory we need to have sufficient density of gas (or plasma) to achieve a measurable effect but not enough for there to be inhibition by the refractive index. The obvious experiment is to use the M\"{o}ssbauer effect for  $\gamma$-rays that enables very precise measurement of their frequency. Simply put, the   rays are emitted by nuclei in solids where there is minimal recoil or thermal broadening of the emitted   ray. Since the recoil-momentum of the nucleus is large compared to the atomic thermal energies and since the nucleus is locked into the solid so that the recoil momentum is precisely defined, then the $\gamma$-ray energy is also precisely defined. The absorption process is similar and has a very narrow line width. Such an experiment has already been done by \citet{Pound65}. They measured gravitational effects on 14.4 keV $\gamma$-rays from $^{57}$Fe being sent up and down a vertical path of 22.5 m in helium near room pressure. They found agreement to about 1\% with the predicted fractional redshift of $1.5 \times 10^{-15}$, whereas fractional curvature redshift predicted by Eq.~\ref{ce8} for this density is $1.25 \times 10^{-12}$. Clearly, this is much larger. At   $\gamma$-ray frequencies, the electrons in the helium gas are effectively free and we can use Eq.~\ref{ce24} to compute the refractive index interaction length. For helium at STP, it is $L=0.077$ m, which is much less than curvature-redshift interaction length which for these conditions is $X$=11 m. Hence, we do not expect to see any significant curvature redshift in their results. Pound and Snyder  did observe one-way frequency shifts but they were much smaller than curvature redshift and could be explained by other aspects of the experiment.
However, the Pound and Snyder experiment provides a guide to a possible test for the existence of curvature redshift. Because curvature redshift has a different density variation to that for the inhibiting refractive index it is possible to find a density for which curvature redshift is not inhibited. Although there is a slight advantage in using heavier gases than helium due to their higher atomic number to atomic weight ratio, their increased absorption to $\gamma$-rays rules them out. Hence, we stay with helium and from Eq.~\ref{ce24} we can compute curvature-redshift interaction length to be
\[
X = 10.8\left( {\frac{p_0}{p}} \right)^{1/4} \mbox{\,m},
\]
where $p$ is the pressure and $p_0$ is the pressure at STP. For the same gas the refractive index interaction length is
\[
L = 0.077\left( \frac{p_0}{p} \right){\mbox{\,m}}.
\]
It follows that the curvature redshift will not be inhibited if $X<L$ or in this case, the pressure is less that $0.0014p_0 $ which is about 1 mm of Hg. For this pressure, we find that $X = 57$ m which requires that the apparatus must be much longer than 57 m. For argument let us take the length to be 100 m then the fractional redshift expected is $2.1 \times 10^{-13}$ which is detectable. The experimental method would use a horizontal (to eliminate gravitational redshifts) tube filled with helium and with accurately controlled temperature. Then we would measure the redshift as a function of pressure. The above theory predicts that if it is free of inhibition then the redshift should be proportional to the square root of the pressure.

Alternatively, it may be possible to detect the secondary photons. For helium with a pressure of 1 mm Hg the expected frequency of the secondary radiation is about 100 kHz. The expected power from a 1 Cu source is about $5\times 10^{-22} \mbox{\,W}$. Unfortunately, the secondary radiation could be spread over a fairly wide frequency band which makes its detection somewhat difficult but it may be possible to detect the radiation with modulation techniques.

Another possibility is to use $\gamma$-rays of much shorter wavelength where it may be possible to detect the secondary radiation in an experiment that did not try to measure the redshift. For example consider the passage of keV to Mev gamma rays from radioactive elements or synchrotron sources in air. For air at a density of 1.20 kg m$^{-3}$ and with the $\gamma$-ray energy $E_0$ in keV the frequency of the secondaries is derived from Eq.~\ref{ce14} to be
\[
\lambda  = 0.465 \left( \frac{E_0}{\mbox{\,keV}} \right )^{1/2} \mbox{\,Mhz},
\]
and the gravitational interaction length is
\[
X  = 25.66\left (\frac{E_0}{\mbox{\,keV}} \right )^{1/2} \mbox{\,m}.
\]
Now for there to be no inhibition the gravitational interaction length must be less than the refractive index interaction length ($L$) which from Eq.~\ref{ce24} and for air has the equation
\[
L  = 0.7905 \left (\frac{E_0}{\mbox{\,keV}} \right )  \mbox{\,m}.
\]
In addition the gamma rays must have a path  length greater than $X$. An appropriate measure of this path length is the distance over which the number of $\gamma$-rays have been attenuated to half the original number. Table~\ref{t1a} shows these quantities for a range of primary energies.

\begin{table}
\begin{center}
\caption{Curvature redshift in air.\label{t1a}}
\begin{tabular}{crrcr}
\hline
Energy/keV &$X$\tablenotemark{a}&  $L$\tablenotemark{b} & attn. length\tablenotemark{c} & $ \nu$\tablenotemark{d} \\
\hline
10  & 8.11 &   7.9 &  \ 1.1 & 1.47 \\
20  & 5.74 &  15.8 &  \ 7.4 & 2.08\\
50  & 3.63 &  39.5 &   27.8 & 3.29 \\
100 & 2.57 &  79.1 &   37.4 & 4.65 \\
200 & 1.81 & 158.1 &   44.7 & 6.58\\
500 & 1.15 & 395.2 &   66.1 & 10.4\\
\hline
\end{tabular}
\end{center}
\tablenotetext{a}{Gravitational interaction length in metros}
\tablenotetext{b}{Refractive index interaction length in metros}
\tablenotetext{c}{Distance to halve beam intensity in metros}
\tablenotetext{d}{Secondary frequency in Mhz}
\end{table}

Note that the curvature redshift will be inhibited by the attenuation length until the $\gamma$-rays have an energy a bit less than 20 keV. There is no inhibition from either cause for energies larger than 20 keV. The expected power per gamma ray per meter of path length is given by
\[
\Delta P = 7.24 \times 10^{-21} \left (\frac{E_0}{\mbox{\,keV}} \right )\mbox{\,W m}^{-1}.
\]
Clearly a powerful $\gamma$-ray source with energies greater than about 50 keV is required.

Yet another possible test is to measure the frequency from a spacecraft at two receivers as a function of the differential distance between the receivers and the spacecraft. For example in the analysis of the {\it Pioneer} 10 acceleration anomaly (section (\ref{s7.7})  it was shown that the interplanetary dust density could contribute a measurable frequency shift. Comparison of this frequency shift at the same time at two receivers at different distances would remove most other causes of frequency shifts. One advantage of this test is that it does not require very accurate frequency generation on  the satellite. Typically the two receivers  would be two ground stations. The major problem is the uncertainty and indeed large variation in the density of the exosphere and any other frequency shifts due to earth rotation that cannot be accurately modelled. Note that at the typical X-band frequencies inhibition will prevent the neutral atmosphere showing any curvature-redshift effects.

\subsubsection{Interactions for other particles}
\label{s6.1.6}
Since the focussing due to spacetime curvature applies to the quantum wave, it is expected that electrons and other particles would interact with curved spacetime in a manner similar to photons. The argument is the same up to Eq.~\ref{ce8} but now we have to allow for nonzero mass. The problem (not solved here) is to find a covariant expression that properly describes the energy-momentum loss to secondary particles and yet preserves the correct normalization of the energy-momentum 4-vector. An alternate approach is to consider the motion in a local Minkowskian reference frame. In this case the loss equations (with $ P_0$ denoting the energy component) are
\begin{eqnarray*}
\label{ce28}
\frac{d{ P^0}}{dx}& = &\frac{\beta^2 { P^0}}{a_e} \nonumber \\
\frac{d{ P^j}}{dx}& = &\frac{ P^j }{a_e},\qquad j = 1,2,3 \nonumber
\end{eqnarray*}
where $\beta$ is the usual velocity ratio, $a_e$ is the local radius of curvature for electrons and as required by normalization and the conservation of proper mass, we have from Eq.~\ref{ce28}
\[
\frac{d{ P}^\alpha}{dx} {P}^\alpha = 0.
\]
Noting that for a nonzero rest mass particle ${\bf V}^\alpha {\bf V}_\alpha =-1$.
The radius of curvature $a_e$ can be evaluated for the simple case of a uniform gas (or plasma) using Eq.~\ref{ce5} and Eq.~\ref{ce6} to get
\[
a_e=\left\{{\frac{8\pi G}{c^2}\left[ {\left({\gamma^2-\frac{1}{2}} \right)\rho  + \frac{p}{c^2}\left( {\gamma^2+\frac{1}{2}} \right)} \right]} \right\}^{-1/2},
\]
where $\gamma=1/\sqrt{1-\beta^2}$. Then with the further simplification of negligible pressure and with the material at rest and where $T = \left({\gamma-1}\right)mc^2$ is the kinetic energy, the energy loss rate is
\begin{equation}
\label{ce31}
\frac{1}{T}\frac{dT}{dx} = -\frac{1}{a_e}{\mbox{\, }} = -\left\{\frac{8\pi G\rho \left(\gamma ^2  - \frac{1}{2} \right)}{c^2} \right\}^{1/2} \beta ^2.
\end{equation}
It shows that for nonzero rest mass particles, the energy loss rate has a strong dependence on velocity, and for extreme relativistic velocities, the fractional energy-loss rate is proportional to $\gamma$. Because of the strong velocity dependence, the energy loss rate for electrons will be much higher than that for nuclei in any plasma near thermal equilibrium. In addition, Eq.~\ref{ce31} shows that the energy loss rate has the same square root dependence on density as the energy loss rate for photons.

Since an electron interacts without being absorbed and re-emitted, we do not expect the same type of inhibition that applies to photons. Instead the electron slowly gets excited with the addition of energy  which it releases as low-energy photons when it interacts with some other particle. The need to preserve spin and momentum prevents it from emitting photons without the presence of another particle. In the cosmic medium, the most likely interactions are electro-magnetic scattering off other charged particles and the inverse-Compton effect off 3K background radiation photons. In high temperature plasma the electromagnetic (Rutherford) scattering is probably dominant since there will be many small angle deflections with large impact parameters. Thus the model for curvature redshift of non-photon particles is one in which an excited electron emits most of its excitation energy as a low-energy photon during the scattering off another photon, electron or nucleus.

\subsection{Derivation of curvature pressure}
\label{s6.2}
The hypothesis of curvature pressure is that for moving particles there is a pressure generated that acts back on the matter that causes the curved spacetime. In this case, curvature pressure acts on the matter (plasma) that is producing curved spacetime in such a way as to try to decrease the curvature. In other words, the plasma produces curved spacetime through its density entering the stress-energy tensor in Einstein's field equations. The magnitude of the curvature is an increasing function of the plasma density.

\subsubsection{Gravitation is not a force}
\label{s6.2.1}
The phrase {\em gravitational force} is not only a popular expression but is endemic throughout physics. In particular, gravitation is classified as one of the four fundamental forces with its heritage going back to Newton's law of gravitation. I argue that the formulation of gravitation as a force is a misconception. In both Newtonian theory and general relativity, gravitation is acceleration. To begin let us examine the original Newtonian gravitation equation
\begin{equation}
\label{ce32}
m_{\rm \scriptscriptstyle I} {\bf a} = {\bf F}{\mbox{\, }} = -\frac{GMm_{\rm\scriptscriptstyle G}}
{r^3}{\bf r},
\end{equation}
where (following \citet{Longair91} we identify $m_{\scriptscriptstyle I}$ as the inertial mass of the test object, $M$ as the active gravitational mass of the second object and $m_{\rm\scriptscriptstyle G}$ as the passive gravitational mass of the test object. The vector $\bf a$ is its acceleration and $\bf r$ is its displacement from the second object. This equation is usually derived in two steps: first, the derivation of a gravitational field and second, the force produced by that field on the test mass. By analogy with Coulombs law, the passive gravitational mass has a similar role to the electric charge.

However many experiments by \citet*{Eotvos22}, \citet{Dicke64}, and \citet{Braginskii71} have shown that the passive gravitational mass is equal to the inertial mass to about one part in $10^{12}$. The usual interpretation of the agreement is that they are fundamentally the same thing. However, an alternative viewpoint is that the basic equation is wrong and that the passive gravitational mass and the inertial mass should not appear in the equation. In this case the correct equation is
\begin{equation}
\label{ce33}
{\bf a } = -\frac{{GM}}{{r^3 }}{ \bf r }.
\end{equation}
Thus, the effect of gravitation is to produce accelerations directly; there is no force involved. Some might argue that since the two masses cancel the distinction is unimportant. On the other hand, I would argue that the application of Ockham's razor dictates the use of Eq.~\ref{ce33} instead of Eq.~\ref{ce32}.

The agreement of the inertial mass with the passive gravitational mass is the basis of the weak equivalence principle in that it applies regardless of the composition of the matter used. \citet{Carlip98} Shows that it applies to both the potential and the kinetic energy in the body. The theory of general relativity is based on the principle of equivalence as stated by Einstein: {\em All local, freely falling, non-rotating laboratories are fully equivalent for the performance of physical experiments.} The relevance here is that it is impossible to distinguish between acceleration and a uniform gravitational field.  Thus when gravitation is considered as acceleration and not a force the passive gravitational mass is a spurious quantity that is not required by either theory.

\subsubsection{A Newtonian model}
\label{s6.2.2}
A simple cosmological model using Newtonian physics in four-dimensional space illustrates some of the basic physics subsequently used to derive the features of curvature pressure. The model assumes that the universe is composed of gas confined to the three-dimensional surface of a four-dimensional hypersphere. Since the visualization of four dimensions is difficult let us suppress one of the normal dimensions and consider the gas to occupy the two-dimensional surface of a normal sphere. From Gauss's law (i.e. the gravitational effect of a spherical distribution of particles with radial symmetry is identical to that of a point mass equal in value to the total mass situated at the center of symmetry) the gravitational acceleration at the radius $r$ of the surface is normal to the surface, directed inward and it has the magnitude
\[
\ddot r =  - \frac{GM}{r^2},
\]
where $M$ is the total mass of the particles and the dots denote a time derivative. For equilibrium, and assuming all the particles have the same mass and velocity we can equate the radial acceleration to the gravitational acceleration and get the simple equation from celestial mechanics of
\[
\frac{{v^2 }}{r} = \frac{{GM}}{{r^2 }}.
\]
If there is conservation of energy, this stable situation is directly analogous to the motion of a planet about the sun. When there is a mixture of particles with different masses, there is an apparent problem. In general, particles will have a distribution of velocities and the heavier ones can be expected to have, on average, lower velocities. Thus, equilibrium radii will vary with the velocity of the particles. However, the basis of this model is that all particles are constrained to have the same radius regardless of their mass or velocity with the value of the radius set by the average radial acceleration. Thus for identical particles with a distribution of velocities we average over the squared velocities to get
\begin{equation}
\label{ce36}
\left\langle {v^2 } \right\rangle  = \frac{{GM}}{r}.
\end{equation}
If there is more than one type of particle with different masses then we invoke the precepts of Section~\ref{s6.2.1} and average over the accelerations to get the same result as Eq.~\ref{ce36}.
The effect of this balancing of the accelerations against the gravitational potential is seen within the shell as a curvature pressure that is a direct consequence of the geometric constraint of confining the particles to a shell. If the radius $r$ decreases then there is an increase in this curvature pressure that attempts to increase the surface area by increasing the radius. For a small change in radius in a quasi-equilibrium process where the particle velocities do not change the work done by this curvature pressure (two-dimensions) with an incremental increase of area $dA$ is $p_{\rm c}dA$ and this must equal the gravitational force times the change in distance to give
\[
p_{\rm c} dA = \frac{{GM^2 }}{{r^2 }}\,dr,
\]
where $M = \sum {m_i}$ with the sum going over all the particles. Therefore, using Eq.~\ref{ce36} we can rewrite the previous equation in terms of the velocities as
\[
p_{\rm c} dA = \frac{{M\left\langle {v^2 } \right\rangle }}{r}\,dr.
\]

Now $dA/dr=2A/r$, hence the two-dimensional curvature pressure is
\[
p_{\rm c}  = \frac{{M\left\langle {v^2 } \right\rangle }}{{2A}}.
\]

Thus in this two-dimensional model the curvature pressure is like the average kinetic energy per unit area. This simple Newtonian model provides a guide as to what the curvature pressure would be in the full general relativistic model. The essential result is that there is a curvature pressure that is due to the constraint of requiring all the particles to stay within the two-dimensional surface.

\subsubsection{General relativistic model}
\label{s6.2.3}
In deriving a more general model in analogy to the Newtonian one, we first change $dA/dr=2A/r$ to $dV/dr=3V/r$ and secondly we include the correction  $\gamma^2$ needed for relativistic velocities. The result is
\[
p_{\rm c}  = \frac{{M\left\langle {\gamma ^2 \beta ^2 } \right\rangle c^2 }}
{{3V}} = \frac{{\left\langle {\gamma ^2  - 1} \right\rangle Mc^2 }}
{{3V}}.
\]
In this case the constraint arises from the confinement of all the particles within a three-dimensional hyper-surface. Now we expect to be dealing with fully ionized high temperature plasma with a mixture of electrons, protons, and heavier ions where the averaging is done over the accelerations. Define the average density by $\rho=M/V$ then the cosmological curvature pressure is

\begin{equation}
\label{ce41}
p_{\rm c}  = \frac{1}{3}\left\langle {\gamma^2 -1}\right\rangle \rho c^2.
\end{equation}

In effect, my hypothesis is that the cosmological model must include this curvature pressure as well as thermodynamic pressure. Note that although this has a similar form to thermodynamic pressure it is quite different. In particular, it is proportional to an average over the squared velocities and the thermodynamic pressure is proportional to an average over the kinetic energies. This means that, for plasma with free electrons and approximate thermodynamic equilibrium, the electrons will dominate the average due to their much larger velocities. From a Newtonian point of view, curvature pressure is opposed to gravitational mutual acceleration. In general relativity, the plasma produces curved spacetime through its density entering the stress-energy tensor in Einstein's field equations. Then the constraint of confining the particles to a three-dimensional shell produces a pressure whose reaction is the curvature pressure acting to decrease the magnitude of the curvature and hence decrease the density of the plasma.

For high temperature plasma in equilibrium, the J\"{u}ttner distribution can be used to evaluate the curvature pressure. For a gas with temperature $T$ and for molecules with mass $m$, \citet{deGroot80} showed that
\begin{equation}
\label{ce42}
\gamma ^2 \left(\alpha \right)= 3\alpha K_3 (1/\alpha) / K_2(1 /\alpha ),
\end{equation}
where  $\alpha=kT/mc^2$ and $K_n(1/\alpha)$ are the modified Bessel functions of the second kind \citet{Abramowitz72}. For small, $\alpha$  this has the approximation
\begin{equation}
\label{ce43}
\gamma ^2 (\alpha ) = 1 + 3\alpha  + 152\alpha ^2  + 458\alpha ^3  +  \ldots.
\end{equation}
For a Maxwellian (non-relativistic) distribution, the first two terms are exact and the  $\alpha^2$ term is the first term in the correction for the J\"{u}ttner distribution.

\subsubsection{Local curvature pressure}
\label{s6.2.4}
For the universe, the calculation of curvature pressure is simple because of the constant curvature and homogeneous medium. However, for a localized region such as a star with inhomogeneous medium and curvature the calculation is much more difficult. We start with the premise that it is the  motion of particles that reacts back on the material producing the curvature by producing a pressure that tends to reduce the curvature. The problem is that the calculation of the curvature at any point requires the integration of Einstein's equations of general relativity. Then if the particles'  motion produces a reaction force, the problem is to determine how that reaction force is apportioned amongst the matter that produces the curvature.
One approach that is valid for most astrophysical applications where the spacetime curvature is small is to use the Newtonian approximation. Let $a$, be the effective radius of curvature of the four dimensional space where the particles' are constrained. Then the premise is that this constraint produces an acceleration due to curvature (assuming for the moment that there is only one type of particle) of
\[
g_{\rm c}  = \frac{{\left\langle {v^2 } \right\rangle }}{a},
\]
where the angular brackets denote an averaging over all the velocities. Now consider a spherically symmetric distribution of gas. If the distribution is static, the central gravitational attraction is balanced by some pressure $p_g$, so that
\[
\frac{{dp_{\rm g} }}{{dr}} =  -\rho (r)g(r),
\]
where $\rho \left( r \right)$ is the density at radius $r$ and $g(r)$ is the gravitational acceleration at $r$. Similarly, we define a curvature pressure by
\begin{equation}
\label{ce46}
\frac{{dp_{\rm c} }}{{dr}} =  -\rho (r)g_{\rm c}(r).
\end{equation}
However, if there is a mixture of particles there is an important difference. Because electrons have a much lighter mass than ions the velocity average for mixed particles (provided the gas is ionized) will be dominated by the electrons and the appropriate density to use in Eq.~\ref{ce46} is that for the electrons. Now the curvature radius $a$, is given by Eq.~\ref{ce7}, and for a gas with relativistic particles we put
\[
\left\langle {v^2 } \right\rangle  = \left\langle {\gamma ^2  - 1} \right\rangle c^2.
\]
We need to include a factor of one third because only the velocity component orthogonal to the direction of the acceleration is relevant. Then the curvature pressure acceleration is
\[
g_{\rm c}(r)  = \frac{1}
{3}\left\langle {\left( {\gamma ^2  - 1} \right)\sqrt{\rho (r)} } \right\rangle c^2 \sqrt{\frac{{8\pi G}}
{{c^2 }}},
\]
and
\begin{equation}
\label{ce49}
\frac{{dp_{\rm c} }}
{{dr}} =  - \frac{1}
{3}\left\langle {(\gamma ^2  - 1)\sqrt{\rho (r)}} \right\rangle c^2 \sqrt{\frac{{8\pi G}}
{{c^2 }}} \rho (r).
\end{equation}
Since the hypothesis is that this curvature pressure is a reaction to the accelerations produced by the gas at radius $r$, the averaging over velocities must be over all the gas that is being accelerated. By Gauss's law and symmetry this is the gas with radii greater than $r$ thus we get
\[
\left\langle {\left(\gamma^2-1\right)\sqrt{\rho(r)}} \right\rangle = \frac{\int_r^\infty  {N(\hat r)\hat r^2 (\gamma^2-1)\sqrt{\rho (\hat r)}\,d\hat r}} {\int_r^\infty  {N(\hat r)\hat r^2}\,d\hat r},
\]
where $N(r)$ is the particle number density. Now for plasmas where the temperatures less than about $10^8\,$K we can use Eq.~\ref{ce43} to get
\[
\frac{1}{3}\left\langle {\gamma^2-1} \right\rangle = \frac{kT}{m_{\rm e} c^2}.
\]
Hence the working equation for local curvature pressure is

\[
\frac{dp_{\rm c}}{dr} =  -k\left\langle T(r)\sqrt{\rho (r)} \right\rangle \sqrt{\frac{8\pi G}{c^2}} \rho (r),
\]
where the function in angular brackets is
\[
\left\langle T(r)\sqrt{\rho (r)} \right\rangle = \frac{\int_r^\infty {N_e (\hat r)\hat r^2 T\left({\hat r} \right)\sqrt{\rho (\hat r)}\,d\hat r}} {\int_r^\infty {N_e(\hat r)\hat r^2\,d\hat r}} ,
\]
and $N_{\rm e}(r)$ is the electron number density.

A theory of curvature pressure in a very dense medium where quantum mechanics dominates and where general relativity may be required is needed to develop this model. Nevertheless, without such a theory, we expect the pressure to be proportional to the local gravitational acceleration and an increasing function of the temperature of the particles. Thus, we might expect a curvature pressure that would resist a hot compact object from collapsing to a black hole. Because of the energy released during collapse, it is unlikely for a cold object to stay cold enough to overcome the curvature pressure and collapse to a black hole.
\subsection{The curvature cosmological model}
\label{s6.3}
Curvature cosmology can now be derived by including curvature redshift and curvature pressure  into the equations of general relativity. This is done by using homogeneous isotropic plasma as a model for the real universe. The general theory of relativity enters through the Friedmann equations for a homogeneous isotropic gas. Although such a model is simple compared to the real universe, the important characteristics of CC can be derived by using this model. The first step is to obtain the basic relationship between the density of the gas and the radius of the universe. The inclusion of curvature pressure is not only important in determining the basic equations but it also provides the necessary means of making the solution static and stable. Then it is shown that the effect of curvature redshift is to produce a redshift that is a function of distance, and the slope of this relationship is (in the linear limit of small distances) the Hubble constant.

The first-order model considers the universe to be a gas with uniform density and complications such as density fluctuations, galaxies, and stars are ignored.  In addition, we assume (to be verified later) that the gas is at high temperature and is fully ionized plasma. Because of the high symmetry, the appropriate metric is the one that satisfies the equations of general relativity for a homogeneous, isotropic gas. This metric was first discovered by A. Friedmann  and fully investigated by H. P. Robertson and A. G. Walker. The Robertson-Walker metric   for a space with positive curvature can be written \citep{Rindler77} as
\[
ds^2 = c^2 dt^2 -\left[{R(t)}\right]^2 \left[\frac{dr^2 }
{1 - r^2} + r^2 \left(d\theta ^2 +\sin^2(\theta)d\varphi^2\right) \right]
\]
where $ds$ is the interval between events, $dt$ is time, $R(t)dr$ is the comoving increment in radial distance, $R(t)$ is the radius of curvature and $R_0$ is the value of $R(t)$ at the present epoch.

\subsubsection{The Friedmann equations}
\label{s6.3.2}
Based on the Robertson-Walker metric, the Friedmann equations for the homogeneous isotropic model with constant density and pressure are \citep{Longair91}
\begin{eqnarray*}
\label{ce55}
\ddot R & = & - \frac{4\pi G}{3}\left( \rho+\frac{3p}{c^2} \right)R + \frac{1}{3}\Lambda R, \\
\dot R^2 & = &\frac{8\pi G}{3}\rho R^2-c^2+\frac{1}{3}\Lambda R^2.
\end{eqnarray*}
where $R$ is the radius, $\rho$  is the proper density, $p$ is the thermodynamic pressure, $G$ is the Newtonian gravitational constant, $\Lambda$  is the cosmological constant, $c$ is the velocity of light and the superscript dots denote time derivatives. Working to order of $m_{\rm e}/m_{\rm p} $ thermodynamic pressure may be neglected but not curvature pressure. How to include curvature pressure is not immediately obvious. The thermodynamic pressure appears only as a relativistic correction to the inertial mass density whereas curvature pressure is closer in spirit to the cosmological constant. My solution is to include curvature pressure (with a negative sign) with the thermodynamic pressure and to set the cosmological constant to zero. This is an ad hoc variation to general relativity and its only justification is that it provides sensible equations and show good agreement with observations. Including curvature pressure from Eq.~\ref{ce41} and from Eq.~\ref{ce55} the modified Friedmann equations are
\begin{eqnarray*}
\label{ce56}
\ddot R & = & -\frac{4\pi G\rho}{3} \left[{1 - \left\langle{\gamma^2-1} \right\rangle } \right]R,  \\
\dot R^2 & = & \frac{8\pi G\rho}{3}R^2  - c^2.  \\
\end{eqnarray*}
Clearly there is a static solution if $<\gamma^2-1>=1$, in which case $\ddot R=0$. The second equation, with $\dot R = 0$  provides the radius of the universe which is given by
\begin{equation}
\label{ce57}
R = \sqrt{\frac{3c^2}{8\pi G\rho}} {\mbox{\, }} = \sqrt{\frac{3c^2}
{8\pi GM_{\rm H} N}}.
\end{equation}
Thus, the model is a static cosmology with positive curvature. Although the geometry is similar to the original Einstein static model, this cosmology differs in that it is stable. The basic instability of the static Einstein model is well known \citep{Tolman34,Ellis84}. On the other hand, the stability of CC is shown by considering a perturbation  $\Delta R$, about the equilibrium position. Then the perturbation equation is
\begin{equation}
\label{ce58}
\Delta \ddot R = \frac{3c^2}{4\pi R_0}\left(\frac{d\langle\gamma^2-1\rangle}
{dR} \right)\Delta R.
\end{equation}
For any realistic equation of state for the cosmic plasma, the average velocity will decrease as $R$ increases. Thus the right hand side is negative, showing that the result of a small perturbation is for the universe return to its equilibrium position. Thus, CC is intrinsically stable. Of theoretical interest is that Eq.~\ref{ce58} predicts that oscillations could occur about the equilibrium position.

\subsubsection{Temperature of the cosmic plasma}
\label{s6.3.3}
One of the most remarkable results of CC is that it predicts the temperature of the cosmic plasma from fundamental constants. That is the predicted temperature is independent of the density and independent of any other characteristic of the universe. For a stable solution to Eq.~\ref{ce56} we need that $<\gamma^2-1>=1$, (i.e. $< \gamma^2>=2$) where the average is taken over the electron and nucleon number densities, that is for equal numbers of electrons and protons
\[
\left\langle {\gamma ^2 }\right\rangle \cong 0.5\left\langle{\gamma _{\rm e}^2  + \gamma _{\rm p}^2} \right\rangle,
\]
where the terms on the right are for electrons and protons. Provided the temperatures are small enough for the proton's kinetic energy to be much less than its rest mass energy, we can put $\left\langle {\gamma_{\rm p}^2} \right\rangle=1$  and thus for pure hydrogen, the result is $\left\langle {\gamma_{\rm e}^2} \right\rangle= 3$. Using a more realistic composition that has 8.5\% by number \citep{Allen76} of helium we find that $\left\langle{\gamma _{\rm e}^2 } \right\rangle=2.927$. Hence using Eq.~\ref{ce42} the predicted electron temperature is $2.56 \times 10^9 {\mbox{\,K}}$. For this temperature $\left\langle {\gamma _{\rm p}^2 } \right\rangle= 1.0007$. This shows that the temperature is low enough to justify the assumption made earlier, that the proton's kinetic energy is much smaller than its rest mass energy.

To recapitulate the stability of CC requires that $\ddot R = 0$. This requires that the plasma has the precise temperature that makes $<\gamma^2-1>=1$. The basis for this result is that curvature pressure exists and critical to its derivation is the averaging over accelerations and not over forces. This is where the assertion that gravitation is acceleration and is not a force is important.

\subsubsection{Hubble constant: theory}
\label{s6.3.4}
The Hubble constant is proportional to the local energy loss rate given by Equation(\ref{ce8} which gives
\begin{eqnarray*}
\label{ce60}
H=\frac{c}{E}\frac{{DE}}{{des}} &=& \left( 8\pi GM_{\rm H} N \right)^{1/2} \nonumber \\
 & = & 1.671 \times 10^{ - 18} N^{1/2}\mbox{\,m}^{-1}\nonumber \\
 & = & 51.69 N^{1/2}\mbox{\,kms}^{-1}\mbox{\,Mpc}^{-1}.
\end{eqnarray*}
The usual redshift  parameter $z$ is defined in terms of the wavelengths, frequencies and energies as
\begin{equation}
\label{ce62}
z= \frac{\lambda_0}{\lambda _{\rm e}} - 1 \qquad =\frac{\nu _{\rm e}}{\nu_0}-1
\qquad = \frac{E_{\rm e}}{E_0} - 1.
\end{equation}
If the plasma density is constant then we can integrate the energy loss along the path to get
\begin{equation}
\label{ce63}
z= \exp \left( \frac{Hr}{c}\right )-1,
\end{equation}
where $r$ is the distance travelled.

\subsubsection{Geometry of CC}
\label{s6.3.5}
The Robertson-Walker metric\  shown in Eq.~\ref{ce55} is not in the simplest form that explicitly shows the geometry. Following \citet{D'Inverno92} we can introduce a new variable $\chi$, where $r = R\sin \chi$
and the new metric is
\[
ds^2  = c^2 dt^2  - R^2 \left[ {d\chi^2  + \sin ^2 \chi \left( {d\theta ^2  + \sin ^2 \theta d\phi ^2 } \right)} \right].
\]
In this metric the distance travelled by a photon is $R\chi$ , and since the velocity of light is a universal constant the time taken is $R\chi/c$. There is a close analogy to motion on the surface of the earth with radius R. Light travels along great circles and $\chi$  is the angle subtended along the great circle between two points.
The geometry of this CC is that of a three-dimensional surface of a four-dimensional hypersphere. For this geometry the area of a sphere with radius $R$  is given by
\[
A(r) = 4\pi R^2 \sin ^2 (\chi ).
\]
The surface is finite and  $\chi$ can vary from 0 to $\pi$. Integration of this equation with respect to $\chi$  gives the volume $V$, namely,
\[
V(r) = 2\pi R^3 \left[ {\chi  - \frac{1}{2}\sin (2\chi )} \right].
\]
Clearly the maximum volume is $2\pi^2 R^3$ and we can using Eq.~\ref{ce57} to get $R$ we have
\begin{eqnarray*}
\label{ce68}
R & = & \sqrt{\frac{{3c^2 }}{{8\pi GM_{\rm H} N}}} \nonumber \\
  & = & 3.100 \times 10^{26} N^{-1/2}\mbox{\,m} \nonumber \\
  & = & 10.05N^{-1/2} {\mbox{\,Gpc}}.
\end{eqnarray*}

Examination of Equations (\ref{ce60}) and (\ref{ce68}) shows that there is a simple relationship between $R$ and $H$, namely
\begin{equation}
\label{ce69}
H=\frac{\sqrt{3}c}{R}.
\end{equation}
The next step is to replace $r$ in Eq.~\ref{ce63} with $r=R\chi$ to get
\[
z=\exp(\sqrt{3}\chi)-1,
\]
and
\begin{equation}
\label{ce71}
\chi  = \frac{\ln \left( {1+z} \right)}{\sqrt{3}}.
\end{equation}
This is the fundamental relationship between $z$ and  $\chi$.
Since the geometry of CC does not involve a time coordinate, it is much simpler than that for BB. The key equations define the CC geometry are Eq.~\ref{ce69} which defines the radius of the universe in terms of the Hubble constant and Eq.~\ref{ce71} which defines the distance variable $\chi$ in terms of the redshift parameter $z$. We now examine some topics that are relevant practical applications.

\subsubsection{Luminosities and magnitudes}
\label{s6.3.6}
Let a source have a luminosity $L(\nu)$ (W\,Hz$^{-1}$) at the emission frequency  $\nu$. Then if energy is conserved, the observed flux density (W\,m$^{-2}$\,Hz$^{-1}$) at a distance parameter $\chi$ is the luminosity divided by the area, which is
\[
S(\nu)d\nu = \frac{L(\nu)\,d\nu}{4\pi \left ({R\sin (\chi)}\right )^2}.
\]
However, because of curvature redshift there is an energy loss such that the received frequency $\nu_0$ is related to the emitted frequency  $\nu_{\rm e}$ by Eq.~\ref{ce62}. Including this effect the result is
\[
S(\nu_0)d\nu_0 = \frac{L(\nu _{\rm e} )\,d\nu_{\rm e}}
{4\pi \left( R\sin(\chi)\right)^2 \left({1 + z} \right)}.
\]
The apparent magnitude is defined as $m=-2.5\log(S)$ where the base of the logarithm is 10 and the constant 2.5 is exact. Since the absolute magnitude is the apparent magnitude when the object is at a distance of 10 pc ($3.0857 \times 10^{17}\,$m), the flux density at 10 pc is
\[
S_{10}(\nu_0)\,d\nu_0 = \frac{L(\nu_0)d\nu_0}{2\pi (10pc)^2},
\]
where because 10 pc is negligible compared to $R$, approximations have been made. The flux density ratio is
\[
\frac{S(\nu _0 )}{S_{10}(\nu_0)} = \left\{{\frac{10pc}
{R\sin (\chi )}} \right\}^2 \left\{ {\frac{L(\nu_{\rm e})d\nu_{\rm e}}
{L(\nu_0)d\nu_0}} \right\}\left\{\frac{1}{1 + z} \right\}.
\]
Defining $M$ as the absolute magnitude and putting  $\nu_{\rm e}=(1+z)\nu_0$ we get for ($m-M$)
\begin{eqnarray*}
m - M & = & - 2.5\log \left({\frac{S(\nu _0)}{S_{10}(\nu_0)}}\right) \nonumber\\
&=& 5\log \left\{\frac{R\sin(\chi)}{10pc} \right\}+\mbox{\,K}_z (\nu_0 ) + 2.5\log (1+z),\nonumber
\end{eqnarray*}
where the K-correction \citep{Rowan-Robertson85,Peebles93,Hogg02} is described in section~\ref{s2}
Furthermore, we can use Eq.~\ref{ce69} to replace $R$ by $H$ since
\[
\frac{R}{{\sqrt{3}}} = \frac{c}{H}{\mbox{\, }} = \frac{{2.998}}
{h}{\mbox{\,Gpc}},
\]
where $h$ is the reduced Hubble constant. Hence, we get the distance modulus
\begin{equation}
\label{ce16}
\mu_\CC= 5\log \left[\frac{\sqrt{3} \sin (\chi)}{h} \right] + 2.5\log(1+z)
+ 42.384.
\end{equation}

\section{Application of Curvature Cosmology}
\label{s7}
These topics are relevant to CC but are not part of the comparison of CC with BB. However they are either very important for any cosmology or offer further observational support for CC.
\subsection{Entropy}
\label{s7.1}
Consider a stellar cluster or an isolated cloud of gas in which collisions are negligible or elastic. In either case the virial theorem states that the average kinetic energy $K$, is related to the average potential energy $V$, by the equation $\label{e89}V = V_0  - 2K$ where $V_0$ is the potential energy when there is zero kinetic energy. Let $U$ be the total energy then $U = K + V = V_0  - K$ .
Thus, we get the somewhat paradoxical  situation that since $V_0$ is constant; an increase in total energy can cause a decrease in kinetic energy.  This happens because the average potential energy has increased by approximately twice as much as the loss in kinetic energy. Since the temperature is proportional to (or at the least a monotonic increasing function of) the average kinetic energy it is apparent that an increase in total energy leads to a decrease in temperature. This explains the often-quoted remark that a self-gravitationally bound gas cloud has a negative specific heat capacity. Thus, when gravity is involved the whole construct of thermodynamics and entropy needs to be reconsidered.
One of the common statements of the second law of thermodynamics is that \citep{Longair91}: \emph{The energy of the universe is constant: the entropy of the Universe tends to a maximum}, \citep{Feynman65}: \emph{the entropy of the universe is always increasing} or from Wikipedia \emph{the second law of thermodynamics is an expression of the universal law of increasing entropy, stating that the entropy of an isolated system which is not in equilibrium will tend to increase over time, approaching a maximum value at equilibrium}.

Now the normal proof of the second law considers the operation of reversible and non-reversible heat engines working between two or more heat reservoirs. If we use a self-gravitating gas cloud as a heat reservoir then we will get quite different results since the extraction of energy from it will lead to an increase in its temperature. Thus if the universe is dominated by gravity the second law of thermodynamics needs reconsideration. In addition, it should be noted that we cannot have a shield that hides gravity. To put it another way there is no adiabatic container that is beyond the influence of external gravitational fields. Thus we cannot have an isolated system.

This discussion shows that in a static finite universe dominated by gravity simple discussions of the second law of thermodynamics can be misleading. The presence of gravity means that it is impossible to have an isolated system. To be convincing any proof of the second law of thermodynamics should include the universe and its gravitational interactions in the proof.

\subsection{Olber's Paradox}
\label{s7.2}
For CC, Olber's Paradox is not a problem. Curvature redshift is sufficient to move distant starlight out of the visible band. Visible light from distant galaxies is shifted into the infrared where it is no longer seen.
Of course, with a finite universe, there is the problem of conservation of energy and why we are not saturated with very low frequency radiation produced by curvature redshift. These low-energy photons  are eventually absorbed by the cosmic plasma.  Everything is recycled. The plasma radiates energy into the microwave background radiation and into X-rays. The galaxies develop from the cosmic plasma and pass through their normal evolution. Eventually all their material is returned to the cosmic plasma. Note that very little, if any, is locked up into black holes. Curvature pressure  causes most of the material from highly compact objects to be returned to the surrounding region as high-velocity jets.

\subsection{Black holes and Jets}
\label{s7.3}
The existence of curvature pressure provides a mechanism that could prevent the collapse of a compact object into a black hole. A theory of curvature pressure in a very dense medium where quantum mechanics dominates is needed to develop this model. Nevertheless, without a full theory we can assume that curvature pressure will depend on the local gravitational acceleration and it will be an increasing function of the temperature of the particles. Thus, we might expect a curvature pressure that would resist a hot compact object from collapsing to a black hole. Because of the energy released during collapse it is unlikely for a cold object to stay cold long enough to overcome the curvature pressure and collapse to a black hole.

What is expected is that the final stage of gravitational collapse is a very dense object, larger than a black hole but smaller than a neutron star. This compact object would appear very much like a black hole and would have most of the characteristics of black holes. Such objects could have large masses and be surrounded by accretion discs.  Thus, many of the observations that are thought to show the presence of black hole could equally show the presence of these compact objects. However, there is one observational difference in that many of the mass estimates of black holes come from observations of redshifts from nearby stars. Since part or most of these redshifts may be due to curvature redshift in the surrounding gas, these mass estimates may need to be revised.

If the compact object is rotating there is the tantalizing idea that curvature pressure  may produce the emission of material in two jets along the spin axis. This could be the `jet engine' that produces the astrophysical jets seen in stellar-like objects and in many huge radio sources. Currently there are no accepted models for the origin of these jets.
The postulate here is that the jets are a property of the compact object and do not come from the accretion disk. The spinning object provides the symmetry necessary to generate two jets and curvature pressure provides the force that drives the jets. This mechanism is applicable to both stellar and galactic size structures.

\subsection{Large number coincidences}
\label{s7.4}
It is appropriate to have a brief discussion of famous numerical coincidences in cosmology \citep{Sciama71}. First, however we need the results for the size parameters for the CC universe which are shown in Table~\ref{lnt1} where the $N_{\rm H}$ is the density divided by the mass of a hydrogen atom.
The first large number coincidence is the ratio of the radius of the universe to the classical electron radius ($R/r_0$). The result is $9.49\times 10^{40}$ which is to be compared with the ratio of the electrostatic force to that of the gravitation force between and electron and a proton. This is $4.3\times 10^{38}$ which being about 200 times smaller than $R/r_0$ shows that it is hardly a coincidence and although interesting probably has little physical significance.

\begin{table}
\begin{center}
\caption{Size of CC universe.\label{lnt1}}
\begin{tabular}{lll}
\hline
Quantity    & Value     & SI units \\
\hline
Radius, $R$   & 12.5 Gpc &    $3.86\times 10^{26}\mbox{\,m}$ \\
Volume, $V$   & $2.46\times 10^{31}\mbox{\,pc}^{3}$ &
    $1.14 \times 10^{81}\mbox{\,m}^3 $\\
Density, $N$ & 1.55 m$^{-3}$ & $2.58 \times 10^{-27}{\mbox{\,kg}}{\mbox{\,m}}^{-3} $ \\
Mass, $M$     & $2.94 \times 10^{54}{\mbox{\,kg}}$ &
    $2.94 \times 10^{54}{\mbox{\,kg}}$ \\
$N_{\rm total}=NV$  & $1.77\times10^{81}$ & $1.77\times10^{81}$ \\
\hline
\end{tabular}
\end{center}
\end{table}

\citet{Sciama53,Sciama71} investigated the use of Mach's principle and the role of inertia in general relativity. By direct analogy to Maxwell's equations, he derived {\em for rectilinear motion a combination of Newton's laws of motion and of gravitation, with the inertial frame determined by Mach's principle} (his italics). In effect, there is an acceleration term added to Newton's gravitational equation. The consequence is that the total energy (inertial plus gravitational) of a particle at rest in the universe is zero. He further assumed that matter receding with a velocity greater than that of light makes no contribution. The equivalent distance in CC is the radius, $R$. The implication of his theory is that
\[
\frac{2\pi G\rho R^2}{c^2} \approx 1.
\]
Now using Eq.~\ref{ce68} we get the actual value for the left hand side  to be 3/4 and this value does not depend on the size of the universe. The closeness of this value to unity suggests that Sciama's ideas are worthy of further investigation.

\subsection{Solar neutrino production}
\label{s7.5}
Since the Homestead mine neutrino detector started operation in the late 1960's, its observations have shown a deficiency in the observed intensity of solar neutrinos compared to accurate theoretical calculations. This has led to an enormous activity in the development and testing of solar models. Currently the standard explanation for the deficiency in the arrival rate of solar neutrinos is that it is due to neutrino oscillations. Basically the electron neutrinos produced near the center of the sun are converted into a mix of muon and tau neutrinos by the time they reach the earth. Because of the high densities the matter oscillation as well as vacuum oscillations are important. Although there are several free parameters that must be estimated the most convincing evidence comes from the Sudbury Neutrino Observatory, where the solar neutrino problem was finally solved. There it was shown that only ~34\% of the electron neutrinos (measured with one charged current reaction of the electron neutrinos) reach the detector, whereas the sum of rates for all three neutrinos (measured with one neutral current reaction) agrees well with the expectations. The only reason that I include the following, alternative explanation, is that I was surprised at how accurate were the results predicted by curvature pressure with no additional parameters. Since this is the only place where local curvature pressure is used it is feasible that its derivation in section is flawed. Nevertheless Because it provides accurate predictions and the neutrino oscillation model has to fit several parameters it is worth examination.

The solar model used here is based on that described by  \citet{Bahcall89}. For a local context, curvature pressure is given by Eq.~\ref{ce49}. What was done is to use the tables (for solution BS05) generously provided by Bahcall in his web site and used them to calculate curvature pressure.  It was then assumed that the thermodynamic pressure was reduced by the value of the curvature pressure and then we used the thermodynamic pressure as an index into the same tables to get the temperature. This largely avoids all the complications of equations of state and changing compositions. Naturally, this will only work if the corrections, as they are here, are small. Then this temperature was used as an index into the neutrino production table to get the production rate for each of the eight listed reactions. As a calibration and a check, the same program was used to compute the rates with no curvature pressure. In this test, the maximum discrepancy from the expected rates was 1.3\%.

\begin{table}
\begin{center}
\caption{Computed production rates for solar neutrinos for the standard model including curvature pressure.\label{t22}}
\begin{tabular}{lcccc}
\hline
Reaction&   Relative&   Rate&   Rate/SNU& Rate/SNU \\
    & rate  & /cm$^2$\,s$^{-1}$   & for $^{37}$Cl&   for $^{71}$Ga \\
\hline
pp  &0.829  &$4.93\times 10^{10}$ & 0.0   & 57.8 \\
pep &0.767  &$1.07\times 10^8$    & 0.17  & 2.15 \\
$_{7}$Be &0.537  &$2.56\times 10^9$    & 0.64  & 18.4 \\
$_{8}$B  &0.288  &$1.45\times 10^6$    & 1.67  & 3.48 \\
$_{13}$N &0.503  &$2.76\times 10^8$    & 0.045 & 1.71 \\
$_{15}$O &0.349  &$1.68\times 10^8$    & 0.115 & 1.91 \\
$_{17}$F &0.318  &$1.79\times 10^8$    & 0.0   & 0.03 \\
hep &0.905  &$8.42\times 10^3$    & 0.036 & 0.09 \\
Totals&     &       & $2.66\pm0.42$ &$85.6\pm5.4$ \\
\hline
\end{tabular}
\end{center}
\end{table}

At a radius of 0.1 solar radii, the reduction in thermodynamic pressure was 12.5\% and the reduction in temperature was 4.1\%. The computed rates with curvature pressure included in the solar model are shown in Table~\ref{t22}. The standard rates are from \citet*{Bahcall01, Bahcall89}. The solar neutrino unit (SNU) is a product of the production rate times the absorption cross section and has the units of events per target atom per second and one SNU is defined to be $10^{-36}$ s$^{-1}$. For example for each $^{71}$Ga target atom in the detector the expected event rate due to solar neutrinos for the pp reaction would be $57.7\times 10^{-36}$ s$^{-1}$. The last row shows the expected event rates for $^{37}$Cl and $^{71}$Ga target atoms where the uncertainties are proportional to those provided by \citet{Bahcall01}. Another type of detector uses Cherenkov light from the recoiling electron that is scattered by the neutrino. Because this electron requires high-energy neutrinos to give it enough energy to produce the Cherenkov light this type of experiment is essentially sensitive only to the $^8$B neutrinos.

\citet{McDonald04} provides a list of recent observational results and they are compared with the predictions in Table~\ref{t23}. The columns show the name of the experiment, the type of detector, the unit, the predicted rate (with curvature pressure), the observed rate, and the $\chi^2$ of the difference from the predicted value.
The statistical and systematic uncertainties have been added in quadrature to get the observed uncertainty. The result in the last row from SNO is from the charged current reaction ( $\nu_{\rm e}$+d$\rightarrow$p+p+e) that is the expected rate if there are no neutrino oscillations.
The agreement is excellent. However, there may be some biases that could be either theoretical or experimental in origin. The crucial test requires computation with a solar model that includes curvature pressure so that the more subtle effects are properly handled. The benefit of this agreement is that it gives very strong support for curvature pressure in a non-cosmological context.

\begin{table}
\begin{center}
\caption{Comparison of predicted and observed solar neutrino production rates.\label{t23}}
\begin{tabular}{llcccc}
\hline
Experiment&   Unit&   Predicted&  Observed&  $\chi^2$ \\
Homestead&    SNU&    $2.66\pm0.42$&  $2.56\pm0.23$&     0.04 \\
GALLEX+GNO&     SNU&    $85.6\pm5.4$&   $70.8\pm5.9$&     3.42 \\
SAGE &         SNU&    $85.6\pm5.4$&   $70.9\pm6.4$ &    3.08 \\
Kamiokande & \tablenotemark{a}&  $1.45\pm0.26$ & $2.8\pm0.38$ &   8.60 \\
Super-Kamiokande &   \tablenotemark{a}&   $1.45\pm0.26$&  $2.35\pm0.08$&  10.95 \\
SNO  ( e+d)&  \tablenotemark{a}&   $1.45\pm0.26$&   $1.76\pm0.10$ &       0.25 \\
\hline
\end{tabular}
\end{center}
\tablenotetext{a}{$10^6\mbox{\,cm}^{-2}\mbox{\,s}^{-1}$}
\end{table}

\subsection{Heating of the solar corona}
\label{s7.6}
For over fifty years, astrophysicists have been puzzled by what mechanism is heating the solar corona. Since the corona has a temperature of about $2\times 10^6\,$K and lies above the chromosphere that has a temperature of about 6000K, the problem is where the energy comes from to give the corona this high temperature. Let us consider whether curvature redshift due to the gas in the corona can heat the corona via the energy loss from the solar radiation. \citet{Aschwanden04} quotes the number distribution of electrons in the corona to be
\begin{equation}
\label{sce1}
N_{\rm e} =2.99\times 10^{14} r^{-16}  + 1.55 \times 10^{14} r^{-6} + 3.6 \times 10^{12} r^{-1.5} \mbox{\,m}^{-3},
\end{equation}
where $r$ is the distance from the solar center in units of solar radii. If we assume spherical symmetry then all the radiation leaving the sun must pass through a shell centrad on the sun and we can use Eq.~\ref{ce8} and Eq.~\ref{sce1} to compute the fractional energy loss in that shell. To the accuracy required, we can also assume that the hydrogen number density is the same as the electron density and then the integration of Eq.~\ref{sce1} from the solar surface to 4 solar radii above the surface gives a total fractional energy loss of $1.32\times 10^{-11}$. Thus with a solar power output of $3.83\times 10^{26}$ W the total energy loss to the solar corona by curvature redshift is $5.1\times 10^{15}$ W which is equivalent to $8.3\times 10^{-4}$ W\,m$^{-2}$ at the surface of the sun. This may be compared with the energy losses from the corona to conduction, solar wind and radiation. The total loss rates are quoted by \citet{Aschwanden04} to be $8\times 10^2$ W\,m$^{-2}$ for coronal holes, $3\times 10^3$ W\,m$^{-2}$ for the quiet corona and $10^4$ W\,m$^{-2}$ for an active corona. Since these are about seven magnitudes larger than the predicted loss, curvature redshift is not important in the inner corona. Although it is not pursued here, there is a similar problem in that the Milky Way has a corona with a high temperature. It is intriguing to speculate that curvature redshift may explain the high temperature of the galactic halo.

\subsection{Pioneer 10 acceleration}
\label{s7.7}
Precise tracking of the {\it Pioneer} 10/11, {\it Galileo} and {\it Ulysses} spacecraft \citep{Anderson98a, Anderson02} have shown an anomalous constant acceleration for {\it Pioneer} 10 with a magnitude $(8.74\pm 1.55)\times 10^{-10}$ m\,s$^{-2}$ directed towards the sun. The major method for monitoring {\it Pioneer} 10 is to measure the frequency shift of the signal returned by an active phase-locked transponder. These frequency measurements are then processed using celestial mechanics in order to get the spacecraft trajectory. The simplicity of this acceleration and its magnitude suggests that {\it Pioneer} 10 could be a suitable candidate for investigating the effects of curvature redshift. There is a major problem in that the direction of the acceleration corresponds to a blue shift whereas curvature redshift predicts a redshift. Nevertheless, we will proceed, guided by the counter-intuitive observation that a drag force on a satellite actually causes it to speed up. This is because the decrease in total energy makes the satellite change orbit with a redistribution of kinetic and potential energy.

The crucial point of this analysis is that the only information available that can be used to get the {\it Pioneer} 10 trajectory is Doppler shift radar. There is no direct measurement of distance. Thus the trajectory is obtained by applying celestial mechanics and requiring that the velocity matches the observed frequency shift. Since the sun produces the dominant acceleration we can consider that all the other planetary perturbations and know drag effects have been applied to the observations and the required celestial mechanics is to be simple two body motion. If the observed velocity (away from the sun) is increased (in magnitude) by an additional apparent velocity due to curvature redshift the orbit determination program will compensate by assuming that the spacecraft is closer to the sun than its true distance. It will be shown that this distance discrepancy produces an extra apparent acceleration that is directed towards the sun. The test of this model is whether the densities required by curvature redshift agree with the observed densities.

Let the actual velocity of {\it Pioneer} 10 at a distance $r$, be denoted by $v(r)$, then since the effect of curvature redshift is seen as an additional velocity, $\Delta v(r)$  where from Eq.~\ref{ce8} it is given by
\begin{equation}
\label{pe1}
\Delta v(r) = 2\sqrt{8\pi G} \int_0^r\sqrt{\rho (r)}\,dr
\end{equation}
where the factor of 2 allows for the two-way trip and the density at the distance $r$ from the sun is $\rho(r)$. Since {\it Pioneer} 10 has a velocity away from the sun this redshift shows an increase in the magnitude of its velocity.
We will assume that all the perturbations and any other accelerations that may influence the {\it Pioneer} 10 velocity have been removed as corrections to the observed velocity and the remaining velocity, $v(r)$,  is due to the gravitational  attraction of the sun. In this case the energy equation is
\begin{equation}
\label{pe3}
v(r)^2= v^2_\infty +\frac{2\mu}{r},
\end{equation}
where  $\mu=GM$ is the gravitational constant times the mass of the sun ($\mu=1.327\times 10^{20} \, \mbox{m}^3 \, \mbox{s}^{-2} $) and $v_{\infty}$ is the velocity at infinity.
The essence of this argument is that the tracking program is written to keep energy conserved so that an anomalous change in velocity, $\Delta v(r)$, will be interpreted as a change in radial distance which is
\[
\Delta r=-\sqrt{\frac{2r^3}{\mu}}\,\Delta v(r).
\]
Thus an increase in magnitude of the velocity will be treated as a decrease in radial distance which, in order to keep the total energy constant, implies an increase in the magnitude of the acceleration. Either by using Newton's gravitational equation or by differentiating Eq.~\ref{pe3} the acceleration $a(r)$ is given by
\begin{equation}
\label{pe4}
a(r)=  -\frac{\mu}{r^2}.
\end{equation}
Hence with $v_\infty=0$ and therefore $v(r) =\sqrt{2\mu }/r$ we get
\[
\Delta a(r)=  \frac{2\mu }{r^3}\,\Delta r =\sqrt{\frac{8\mu}{r^3}}\,\Delta r
\]
and then to the first order an increase in velocity of $\Delta v(r)$ will produce an apparent decrease in acceleration of $\Delta a(r)$, and
\begin{eqnarray*}
\label{pe5}
\Delta a&=& 8\sqrt{\pi \mu G}\, r^{-3/2} \int_0^r{\sqrt{\rho(r)}}\,dr \nonumber \\
 & = & 16\sqrt{\pi \mu G}\, r^{-1/2}<\sqrt{\rho(r)}> \nonumber \\
 & = & 6.90 R^{-1/2}<\sqrt{\rho(r)}>
\end{eqnarray*}
where for the last equations we measure the distance in AU so that  $r=1.496\times 10^{11}R$ and the angle brackets show an average value. Now fig.~7 from \citep{Anderson02} shows that after about 20 AU the anomalous acceleration is essentially constant. The first step is to get an estimate of the required density and see if  is feasible. Using the observed acceleration of $a_{\rm P}=8.74\times 10^{-10}$ m\,s$^{-2}$  the required average density for the two-way path is $1.60\times 10^{-20}R\,$kg\,m$^{-3}$ and for R=20 it is $3.21\times 10^{-19}\,$kg\,m$^{-3}$.

The only constituent of the interplanetary medium that approaches this density is dust. One estimate by \citet{D'Hendecourt80} of the interplanetary dust density at 1 AU is $1.3\times 10^{-19}\,$kg\,m$^{-3}$ and more recently, \citet{Grun99} suggests a value of $10^{-19}\,$kg\,m$^{-3}$ which is consistent with their earlier estimate of $9.6\times 10^{-20}\,$kg\,m$^{-3}$ \citep*{Grun85}. Although the authors do not provide uncertainties it is clear that their densities could be in error by a factor of two or more. The main difficulties are the paucity of information and that the observations do not span the complete range of grain sizes.  The meteoroid experiment on board {\it Pioneer} 10 measures the flux of grains with masses larger than $10^{-10}$ g. The results show that after it left the influence of Jupiter the flux \citep{Anderson98b} was essentially constant (in fact there may be a slight rise) out to a distance of 18 AU. It is thought that most of the grains are being continuously produced in the Kuiper belt. As the dust orbits evolve inwards due to Poynting-Robertson drag and planetary perturbations, they achieve a roughly constant spatial density.  The conclusion is that interplanetary dust could  provide the required density to explain the  anomalous acceleration by a frequency shift due to curvature redshift.

\citet{Anderson02} also reports a annual velocity variation of $(1.053\pm0.107)\times 10^{-4} \mbox{\,m\,s}^{-1}$ with a phase angle relative to conjunction of $5\arcdeg.7\pm 1\arcdeg.7$. The cause of this variation is the changing path length through the dust at about 1 AU as the earth cycles the sun. However if this annual variation is due to curvature redshift it cannot be easily distinguished from a position displacement in the plane of the ecliptic: for example this anomalous velocity corresponds to a position shift of about $5\times 10^{-3}$ arcsec. From Eq.~\ref{pe1} and a density of $10^{-19}\,$kg\,m$^{-3}$ the predicted curvature-redshift velocity is $3.9 \times 10^{-3}\mbox{\,m\,s}^{-1}$ which is an order of magnitude larger than the reported anomalous diurnal velocity. Clearly most of the predicted velocity could have been interpreted by the  orbit determination program as a very small angular displacement. This could also explain the phase angle. The predicted phase angle is $90\arcdeg$ from conjunction, whereas the observed phase angle is very close to the line of conjunction.

Finally \citet{Anderson02} reports a diurnal component. Reading from their fig.~18 the diurnal velocity amplitude is about $1.4 \times 10^{-4}\mbox{\,m\,s}^{-1}$. Note that due to inhibition  there is no curvature redshift to be expected from the atmosphere. The major redshift will come from the inter-planetary dust. Then using the earths radius and a density of $10^{-19}\,$kg\,m$^{-3}$  the expected diurnal velocity amplitude due to curvature redshift is $1.7 \times 10^{-7}\mbox{\, m\,s}^{-1}$ which is three orders of magnitude too small. The average density that is needed is about $7.2 \times 10^{-14}\,$kg\,m$^{-3}$. Unless there is such a density it is unlikely that curvature redshift could explain the diurnal velocity effect. Note a critical test would be to compare the simultaneous observation of the {\it Pioneer} 10 velocity from two tracking stations as a function of their different distances from {\it Pioneer} 10.

Overall, this analysis has shown that it is possible to explain the acceleration anomaly of {\it Pioneer} 10 but that a more definitive result requires curvature redshift to be included in the fitting program and more accurate estimates of the dust density are certainly needed. Subject to the caveat about the dust density, curvature redshift could explain the anomaly in the acceleration of {\it Pioneer} 10 (and by inference other spacecraft).

\section{Conclusion}
\label{s8}
The major conclusion from this evaluation of Big Bang cosmologies is that all of the topics covered in Section~\ref{s4} namely Tolman surface brightness, type 1a supernova, angular size, gamma ray bursts, galaxy distribution, quasar distribution, radio source counts, quasar variability in time and the Butcher--Oemler effect are in excellent agreement with a static universe. Furthermore quantitative estimates of evolution derived within the BB paradigm are very close to what would be predicted by BB with a time dilation term of $(1+z)$ removed from the equations.

Results for the topics of the Hubble redshift, X-ray background radiation, the cosmic background radiation and dark matter show strong support for curvature cosmology. In particular CC predicts that the Hubble constant is $64.4\pm 0.2\mbox{\,km\,s}^{-1} \mbox{\,Mpc}^{-1}$ whereas the value estimated from the type 1a supernova data is $63.8\pm0.5\,$kms$^{-1}$ Mpc$^{-1}$ and the result from the Coma cluster (Section~\ref{s5.14}) is $65.7\,$kms$^{-1}$ Mpc$^{-1}$. In CC the theoretical cosmic temperature is $2.56\times 10^9\,$K for the cosmic gas and the temperature estimated from fitting the X-ray data is $(2.62\pm 0.04)\times10^9\,$K. The predicted temperature for the CMBR is 3.18 K. whereas  \citet{Mather90} measured the temperature to be 2.725 K. This prediction does depend on the nuclei mix in the cosmic gas and could vary from this value by several tenths of a degree. Curvature cosmology does not need dark matter to explain the velocity dispersion in clusters of galaxies or the shape of galactic rotation curves. Nor does it need dark energy to explain type 1a supernovae observations.

Other topics in Section~\ref{s5} namely the Sunyaev--Zel'dovich effect, gravitational lens, the Gunn--Peterson trough, redshifts in our Galaxy and  voids can be explained by CC or are fully compatible with CC. Currently CC provides a qualitative explanation for the abundances of light elements (the 'primordial' abundances) but not a quantitative predictions. The remaining topics in Section~\ref{s5} namely  Lyman-$\alpha$ forest, galaxy rotation and anomalous redshift are compatible but with problems.

Curvature pressure can explain the non-cosmological topic of solar neutrino production but since this already explained by neutrino oscillations it must remain a curiosity. The explanation of the {\it Pioneer} 10 anomalous acceleration is feasible if the inter-planetary dust density is a little larger than current estimates.

\section*{Acknowledgments}
This research has made use of the NASA/IPAC Extragalactic Database (NED) that is operated by the Jet Propulsion Laboratory, California Institute of Technology, under contract with the National Aeronautics and Space Administration. The calculations have used Ubuntu Linux and the graphics have used the DISLIN plotting library provided by the Max-Plank-Institute in Lindau.

\appendix
\section{Analytic methods}
\label{appendix}
Traditional least squares regression provides two lines, the regression of $y$ on $x$ and the regression of $x$ on $y$. In general, these regression lines do not have reciprocal slopes. If the data includes values for the uncertainties in both coordinates, the traditional minimization of $\chi^2$ methods would ignore the uncertainty in the independent variable. A better solution is obtained by the subroutine {\sc fitexy} provided by \citet{Press07} in their book Numerical Recipes, which also gives the basic references. This solution uses both sets of uncertainties in a completely symmetric manner that has a single solution regardless of which variable is deemed to be the independent. The method for estimating $a$ and $b$ in $y=a+bx$ is to minimize the $\chi^2$ function defined by
\[
\chi^2 = \sum\limits_i {\frac{(y_i-a-bx_i)^2}{\sigma _i^2+b^2 \varepsilon _i^2+\eta^2}},
\]
where  $\sigma_i$ is the uncertainty in $y_i$,  $\varepsilon _i$ is the uncertainty in $x_i$ and $\eta$ is described below. It is apparent that this equation has the basic symmetry in that if $\eta$ is zero then interchanging $x$ and $y$ will give the reciprocal of the slope $b$. If all the $\varepsilon$ are zero it has an analytic solution otherwise the minimization can be done numerically.

A problem arises when the $\chi^2$ value is significantly different from the number of degrees of freedom. The case where the value of $\chi^2$ is much less than the number of degrees of freedom is handled by multiplying the  uncertainties by the square root of the ratio of $\chi^2$ to the degrees of freedom. If the $\chi^2$ is larger than the number of degrees of freedom the approach used here is to assume that there is an additional, unknown contribution which is uncorrelated with the existing uncertainties.  This is a common occurrence when the uncertainties are measurement errors and may be precise but some other factor  makes a significant contribution the scatter of values. the solution is to add a quadratic term, $\eta^2$, to the denominator and estimate its value by making the $\chi^2$  equal to the degrees of freedom.  The prime advantage of this technique is that it provides a smooth transition from the situation where the given uncertainties dominate to the alternative where they are negligible. The main effect is a change in the relative weights.

\end{document}